\documentclass[lettersize,journal]{IEEEtran}
\usepackage{amsmath,amsthm,amssymb,amscd}
\usepackage{array}
\usepackage{graphicx}
\usepackage{url}
\usepackage{verbatim}
\usepackage{cite}
\usepackage[table]{xcolor}
\usepackage{cuted}
\usepackage{algorithm}
\usepackage{algpseudocode}
\usepackage{longtable}
\usepackage{booktabs}
\usepackage{balance}
\usepackage{diagbox}
\usepackage{tikz}
\usepackage{preview}
\usepackage{xcolor}
\usepackage{subfig}
\usepackage{booktabs}  
\usepackage{multirow}  
\usepackage{pifont}
\begin{document}

\title{``X of Information'' Continuum: A Survey on AI-Driven Multi-dimensional Metrics for Next-Generation Networked Systems}

\author{Beining Wu,~\IEEEmembership{Member,~IEEE,} 
         Jun Huang,~\IEEEmembership{Senior Member,~IEEE,} 
            and 
         Shui Yu,~\IEEEmembership{Fellow,~IEEE}
\thanks{Beining Wu and Jun Huang are with the Department of Electrical Engineering and Computer Science, South Dakota State University. Email: Wu.Beining@jacks.sdstate.edu., Jun.Huang@sdstate.edu.}
\thanks{S. Yu is with the School of Computer Science, University of Technology Sydney, Ultimo, NSW 2007, Australia. E-mail: Shui.Yu@uts.edu.au.}}

\markboth{IEEE Communications Surveys \& Tutorials,~Vol.~14, No.~8, August~2021}{Shell \MakeLowercase{\textit{et al.}}: A Sample Article Using IEEEtran.cls for IEEE Journals}

\maketitle

\begin{abstract}
The development of next-generation networking systems has inherently shifted from throughput-based paradigms towards intelligent, information-aware designs that emphasize the quality, relevance, and utility of transmitted information, rather than sheer data volume. While classical network metrics, such as latency and packet loss, remain significant, they are insufficient to quantify the nuanced information quality requirements of modern intelligent applications, including autonomous vehicles, digital twins, and metaverse environments. In this survey, we present the first comprehensive study of the ``X of Information'' continuum by introducing a systematic four-dimensional taxonomic framework that structures information metrics along temporal, quality/utility, reliability/robustness, and network/communication dimensions. We uncover the increasing interdependencies among these dimensions, whereby temporal freshness triggers quality evaluation, which in turn helps with reliability appraisal, ultimately enabling effective network delivery. Our analysis reveals that artificial intelligence technologies, such as deep reinforcement learning, multi-agent systems, and neural optimization models, enable adaptive, context-aware optimization of competing information quality objectives. In our extensive study of six critical application domains, covering autonomous transportation, industrial IoT, healthcare digital twins, UAV communications, LLM ecosystems, and metaverse settings, we illustrate the revolutionary promise of multi-dimensional information metrics for meeting diverse operational needs. Our survey identifies prominent implementation challenges, including metric integration, information-driven resource allocation, semantic communication, federated learning optimization, and the sustainable design of networks. In conclusion, the survey highlights critical research agendas in formulating unified theoretical models, AI-augmented dynamic optimization, and cross-layer orchestration mechanisms, laying the foundation for intelligent, value-aware communication systems that will redefine the future of network infrastructures.
\end{abstract}


\section{Introduction}
\begin{figure*}
    \centering
    \subfloat[\footnotesize Time-oriented Metrics]{
        \includegraphics[width=0.22\textwidth]{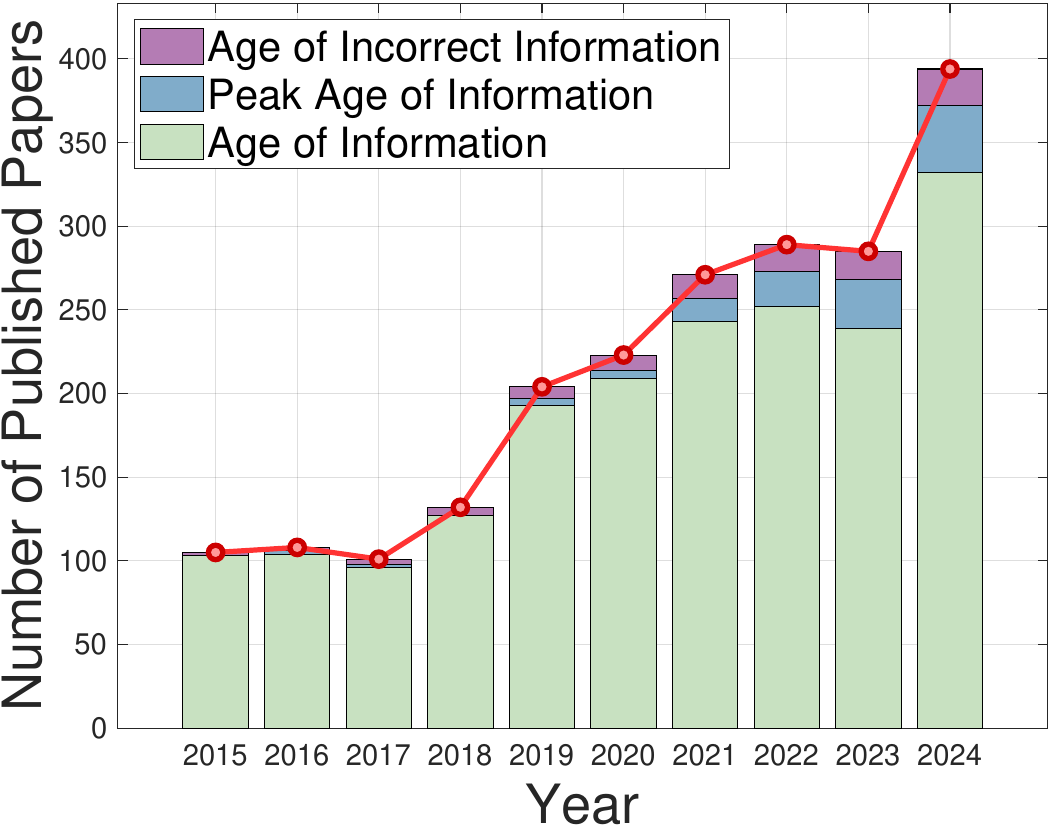}
    }
    \hspace{0mm}
    \subfloat[\footnotesize Quality/Utility-oriented Metrics]{
        \includegraphics[width=0.22\textwidth]{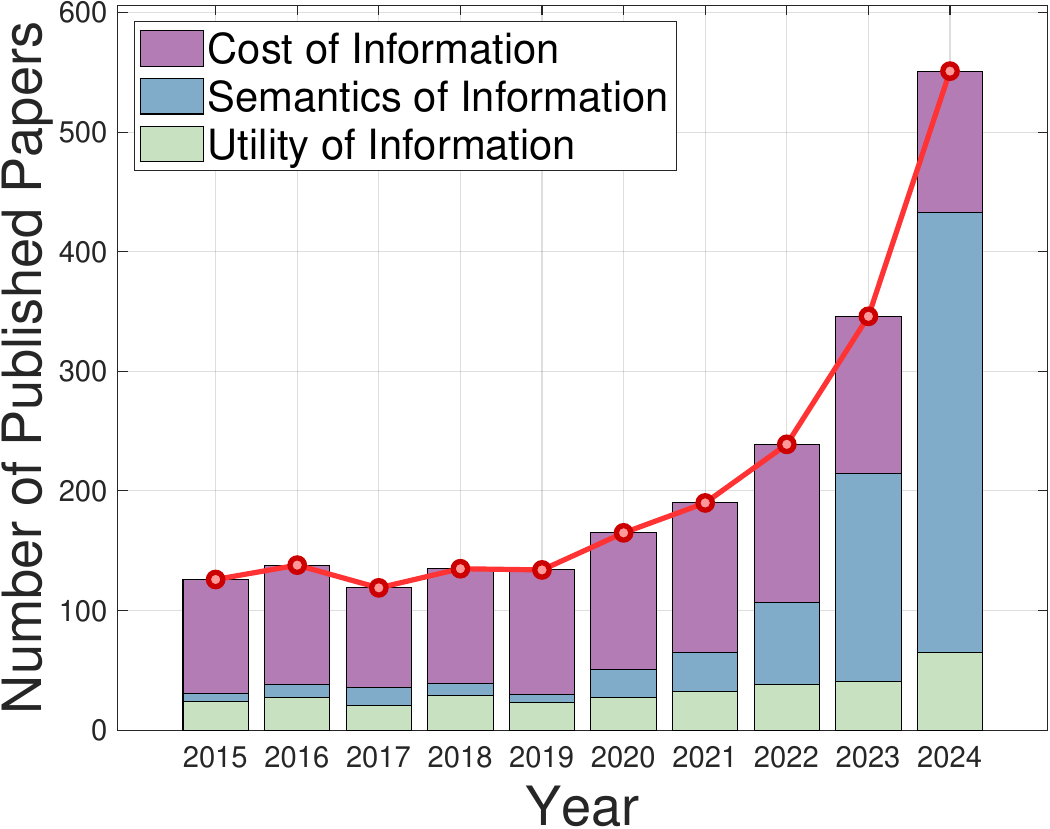}
    }
    \hspace{0mm}
    \subfloat[\footnotesize Reli/Robust-oriented Metrics]{
        \includegraphics[width=0.22\textwidth]{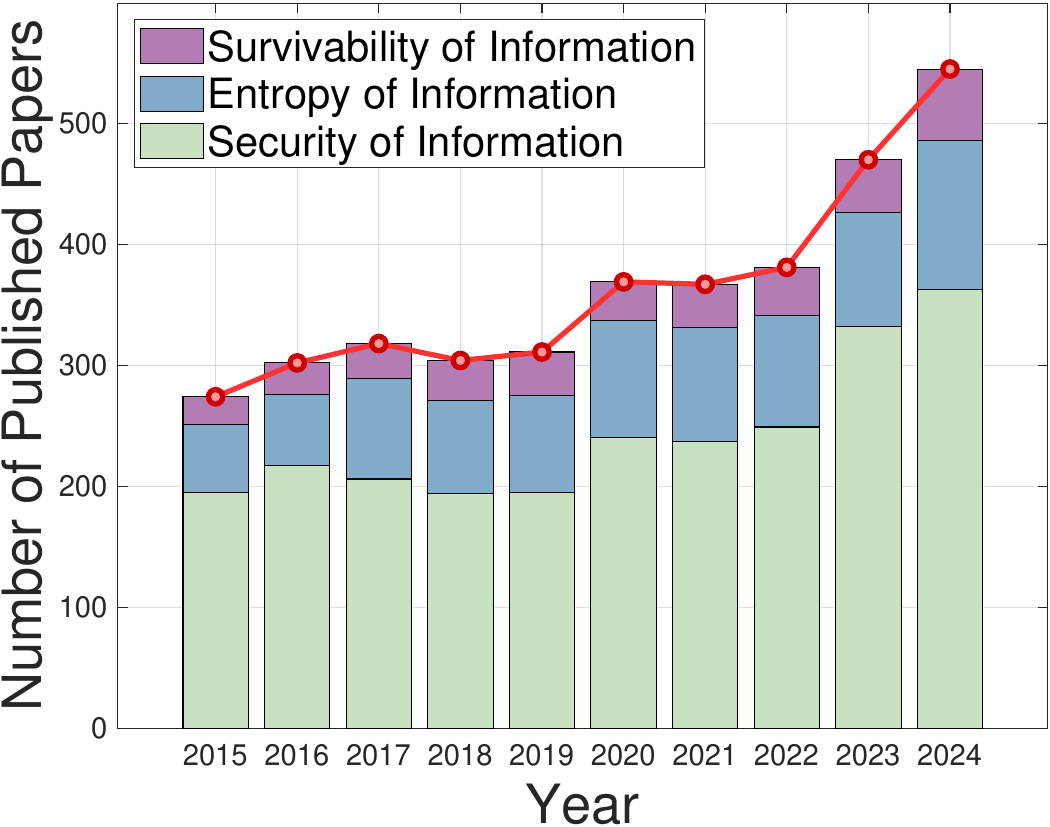}
    }
    \hspace{0mm}
    \subfloat[\footnotesize Net/Comm-oriented Metrics]{
        \includegraphics[width=0.22\textwidth]{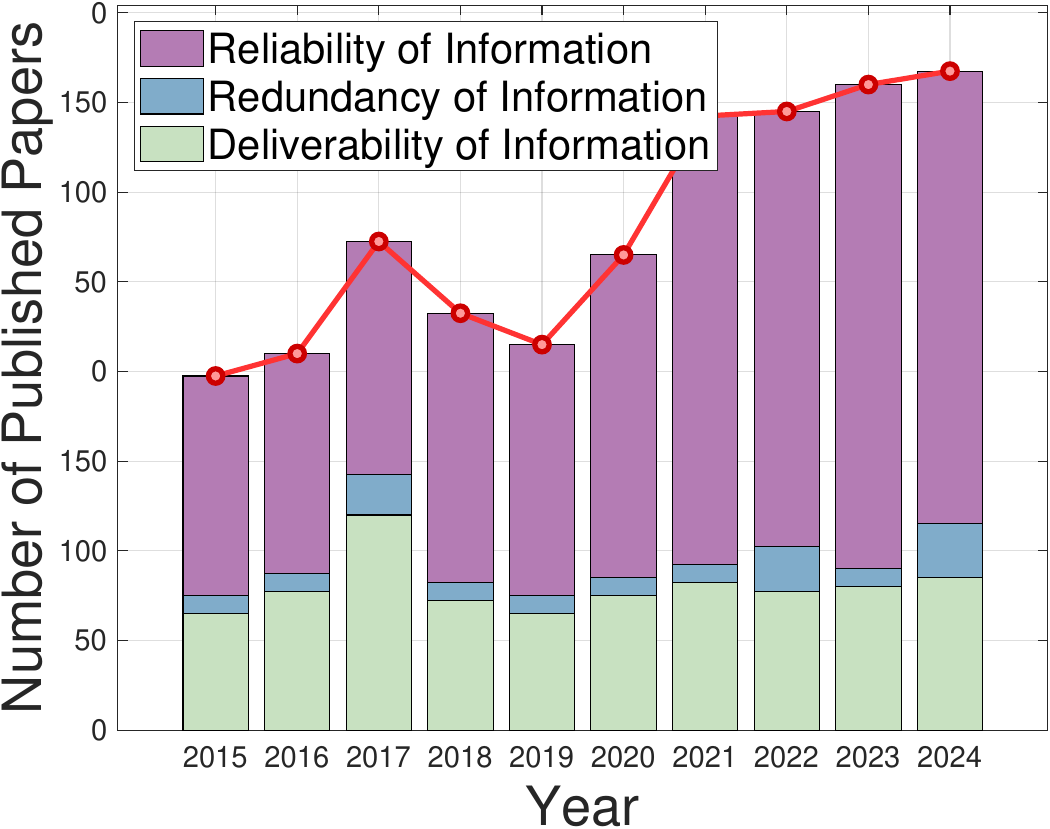}
    }
    \caption{Number of published papers by searching representative keywords for each information metric category in Web of Science (Access date: July 22, 2025).}
    \label{fig:metric_trends}
\end{figure*}

\subsection{Background}


Modern communication networks have experienced a fundamental transformation from traditional data transmission infrastructures toward intelligent, information-centric systems. These systems now prioritize the quality, relevance, and utility of transmitted information over mere throughput optimization \cite{Lin2024CM,Chaccour2025COMST,Zhong2024MNET,Sun2025OJCOMS}. Next-generation networks, including 6G systems, Internet of Things (IoT) deployments, autonomous transportation networks, and digital twin architectures, exhibit unprecedented complexity in their information processing requirements \cite{Lin2024CM,Jiang2024MWC,Chen2024IOTJ,Benedictis2023JBHI}. These intelligent systems demand sophisticated evaluation frameworks that extend beyond conventional network performance indicators to embrace semantic understanding, temporal relevance, security, and adaptive decision-making capabilities \cite{Das2023CIC,Hu2024TMC}. The convergence of artificial intelligence technologies with networking infrastructure has created new paradigms where information value, rather than data volume, becomes the primary optimization objective for system design and operation \cite{Jayanetti2024TPDS,Fu2022TNET}.

Traditional network performance metrics, although foundational to communication system design, have been inadequate for capturing the information quality requirements of modern intelligent applications. Traditional metrics such as throughput, latency, and packet loss rate presuppose that all transmitted information holds the same value and significance. They overlook the contextual relevance and semantic importance that characterise next-generation networked applications \cite{Adil2023TITS,He2024TYCB,Xu2025JSAC,Fu2023INFOCOM}. Cyber-physical applications, autonomous vehicles, and industrial IoT deployments require evaluation regimes that distinguish between safety-critical information and routine status reports. These applications demand metrics that integrate temporal freshness and content usefulness \cite{Lin2024CM,Li2025IOTJ,Sorkhoh2022TITS}. Moreover, the increasing ubiquity of adversarial threats and system vulnerabilities calls for security-aware evaluation of information, which traditional metrics cannot adequately address \cite{Wang2025TVT,Kumar2023TITS}.

The recognition of these limitations has stimulated the emergence of specialised ``X of Information'' metrics, each of which addresses specific dimensions of information quality in networked systems. Age of Information (AoI) and its variants have established temporal freshness as a fundamental quality dimension. These frameworks provide mathematical tools for optimizing information timeliness in real-time applications \cite{Costa2014ISIT,Yates2019TIT,Yates2021JSAC,Zheng2025IOTJ}. Utility-oriented metrics have evolved to capture the decision-making value of information. They incorporate economic and contextual considerations into network optimization strategies \cite{Howard1966TSSC,Simon2023CM,Wang2023TFS}. Security-oriented information metrics have emerged to address trustworthiness and integrity requirements in adversarial environments \cite{Ferrag2022JAS,Ramos2017COMST}. Network-centric metrics have adapted to evaluate deliverability, redundancy, and relevance characteristics specific to dynamic communication infrastructures \cite{Sourlas2018TNSM,ArafatJSEN,Anisetti2022TNSM}. These specialised metrics have demonstrated significant impact in diverse application domains, from healthcare monitoring systems to vehicular networks and industrial automation \cite{Wu2023access,Wu2025MNET,Xu2022INFOCOM}. As shown in Fig.~\ref{fig:metric_trends}, the research interest in all four metric categories has grown substantially over the past decade, with particularly rapid growth observed since 2020.

However, the proliferation of these individual metrics has created a critical challenge that current research has yet to address adequately. While each ``X of Information'' metric provides valuable insights into specific quality dimensions, real-world intelligent systems require holistic optimization across multiple, often conflicting objectives simultaneously \cite{Fei2017COMST,Mazloomi2022AHN}. The complex interdependencies between temporal freshness, semantic utility, security robustness, and network efficiency demand unified theoretical frameworks that can capture cross-dimensional trade-offs and enable systematic optimization strategies \cite{Zeng2023CVPR,Iyer2022TIT}. Moreover, the integration of artificial intelligence techniques into information metric evaluation presents opportunities for adaptive, context-aware optimization that existing fragmented approaches cannot fully exploit \cite{Natarajan2005POTICOML,Ji2014VTCSpring}. This fundamental gap between specialised metric development and holistic system optimisation represents a critical barrier to realising the full potential of next-generation intelligent networks.

\subsection{Motivation}

Individual information metrics have shown tremendous potential for fine-tuning specific aspects of networked system operations. However, intelligent applications that subscribe to multi-dimensional information services fail to address end-to-end, adaptive quality assessment that can perceive complex interdependencies likely to occur in next-generation networks \cite{Jiang2024MWC, Das2023CIC, Wu2025MNET}. First, specialized metrics such as AoI and Utility of Information (UoI) can be used to evaluate temporal timeliness and semantic merit separately. Still, they provide insight toward particular quality aims only \cite{Costa2014ISIT,Howard1966TSSC,Yates2021JSAC}. Next, network managers can call upon security-centric metrics and deliverability assessment to address individual reliability and communication requirements via specialized assessment infrastructures \cite{Adil2023TITS,Sourlas2018TNSM,Kumar2023TITS}. Nevertheless, since they are standalone, individual metric methods possess inherent flaws in perceiving inter-dimensional trade-offs and dynamic quality requirements. Accordingly, integrated multi-dimensional information assessment infrastructures that employ artificial intelligence for adaptive improvement present a more viable strategy for next-generation intelligent networks, as shown in Fig.~\ref{fig:framework_Hierarchical} \cite{Chaccour2025COMST,Fu2023INFOCOM,Wang2023ICDCS}. When embedded in multi-dimensional information systems, AI-enhanced infrastructures coordinate simultaneous sophisticated optimization of temporal, quality, reliability, and network aspects, which requires high-order machine learning algorithms and adaptive coordination infrastructures. Beyond that, there are also issues of real-time metric consolidation, trade-off decision-making, and contextual adaptation that are delegated to the intelligent layer, which necessitates skilled neural infrastructures and distributed learning mechanisms. Finally, there exists a need for the application layer to maintain dynamic quality requirements of information with minimal overhead, which represents unique challenges that are solvable through goal-driven networking and semantic communication methodologies. Specifically, motivations for designing AI-driven multi-dimensional information metrics include:

\begin{figure*}[ht]
    \centering
    \includegraphics[width=0.9\textwidth]{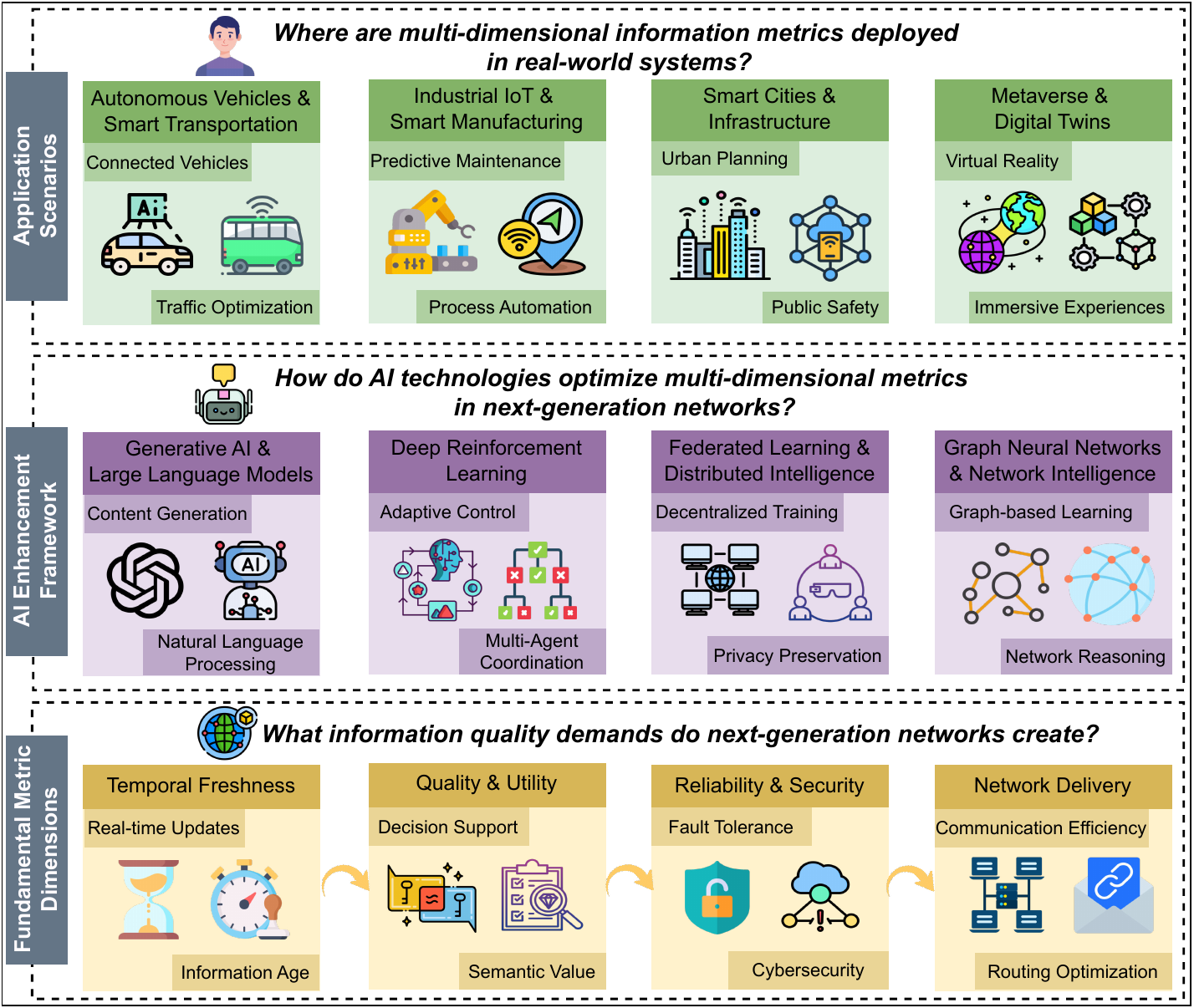} 
    \caption{Hierarchical framework for AI-driven multi-dimensional information metrics in next-generation networked systems. The framework consists of three tiers: fundamental metric dimensions with progressive dependencies (bottom), AI enhancement technologies (middle), and real-world application scenarios (top).}
    \label{fig:framework_Hierarchical}
\end{figure*}

\begin{itemize}
    \item \textit{Cross-dimensional optimization:} Instead of optimizing individual metrics in isolation, intelligent systems can achieve holistic performance improvements through joint optimization across temporal, utility, security, and network dimensions \cite{Li2025IOTJ,Wang2025TVT,Mazloomi2022AHN}. For example, autonomous vehicles can obtain optimal information quality by dynamically balancing freshness requirements with security constraints and communication efficiency, which supports safety-critical decision-making processes.
    \item \textit{Adaptive intelligence and context-awareness:} In AI-enhanced information systems, machine learning algorithms can continuously adapt metric weighting and optimization strategies based on evolving application requirements and environmental conditions \cite{He2024TMC,Zhao2024TASE,Abegaz2023TNSM}. Furthermore, contextual factors can serve as input for dynamic metric prioritization, which addresses specific operational demands while maintaining system-wide performance objectives.
    \item \textit{Theoretical unification and systematic evaluation:} Unified mathematical frameworks can establish fundamental relationships between different information quality dimensions and enable systematic performance analysis across diverse application scenarios \cite{Iyer2022TIT,Fei2017COMST}. On one hand, theoretical foundations can provide rigorous optimization guarantees and performance bounds for multi-dimensional systems. On the other hand, practitioners can leverage standardized evaluation methodologies for comparing different approaches and validating system designs.
\end{itemize}

When intelligent systems access multi-dimensional information services through AI-enhanced optimization frameworks, the complex interdependencies between different quality dimensions pose significant challenges for achieving optimal performance while maintaining computational efficiency, as illustrated in Fig.~\ref{fig:framework_Hierarchical}. First, resource allocation algorithms must balance trade-offs among competing objectives such as temporal freshness, semantic accuracy, security robustness, and energy consumption across distributed network elements \cite{Wu2013ECC, Liu2025TWC, Dong2023TWC}. Second, dynamic optimization requirements necessitate real-time adaptation capabilities that can respond to changing application semantics, environmental conditions, and threat landscapes without compromising system stability. Third, theoretical frameworks that capture cross-dimensional dependencies must be developed to enable principled optimization strategies and performance guarantees in complex operational scenarios. Finally, standardized evaluation methodologies and implementation guidelines should be established to facilitate practical deployment and fair comparison of different approaches across diverse application domains. Compared to traditional single-metric optimization, multi-dimensional information assessment requires comprehensive theoretical foundations, adaptive learning capabilities, and systematic integration methodologies for effective implementation in next-generation intelligent networks.

\subsection{Related Works and Contributions}

\begin{table*}[!t]
\centering
\caption{Summary of related survey works versus our survey.}
\label{tab:related_works}
\renewcommand{\arraystretch}{1}
\begin{tabular}{>{\centering\arraybackslash}m{1.0cm}>{\centering\arraybackslash}m{1.2cm}p{7cm}>{\centering\arraybackslash}m{2.4cm}>{\centering\arraybackslash}m{1.8cm}>{\centering\arraybackslash}m{2.0cm}}
\toprule
\textbf{Year} & \textbf{Ref.} & \textbf{Contributions} & \textbf{Multi-dimensional Framework} & \textbf{AI-driven Integration} & \textbf{Progressive Dependencies} \\
\midrule
\multirow{2}{1.0cm}{\centering\textbf{2021}} & \cite{Yates2021JSAC} & Summarize recent contributions in Age of Information (AoI) design and optimization, present general AoI evaluation methods applicable to increasingly complex systems from single-server queues to wireless networks & \textcolor{red}{\ding{55}} & \textcolor{red}{\ding{55}} & \textcolor{red}{\ding{55}} \\
\cmidrule(l){2-6}
& \cite{Syu2021TNSM} & Provide a comprehensive survey of time series-based QoS modeling and forecasting for Web services, classify research into four essential components and discuss future directions & \textcolor{red}{\ding{55}} & \textcolor{red}{\ding{55}} & \textcolor{red}{\ding{55}} \\
\midrule
\multirow{3}{1.0cm}{\centering\textbf{2022}} & \cite{Fadlullah2022COMST} & Survey practices, challenges, and opportunities in emerging 6G networks where AI-based techniques meet classical optimization to balance service performance and security levels & \textcolor{red}{\ding{55}} & \textcolor{green}{\ding{51}} & \textcolor{red}{\ding{55}} \\
\cmidrule(l){2-6}
& \cite{Ghafouri2022TSC} & Introduce QoS prediction methods for Web services and divide them into memory-based, model-based, and collaborative filtering combined approaches & \textcolor{green}{\ding{51}} & \textcolor{red}{\ding{55}} & \textcolor{red}{\ding{55}} \\
\cmidrule(l){2-6}
& \cite{Alazzawe2022TBDATA} & Propose a taxonomy for data-driven storage management techniques and discuss research challenges in identifying valuable data items with minimal overhead and latency & \textcolor{red}{\ding{55}} & \textcolor{green}{\ding{51}} & \textcolor{red}{\ding{55}} \\
\midrule
\multirow{4}{1.0cm}{\centering\textbf{2023}} & \cite{Yang2023COMST} & Provide a holistic review of semantic communication fundamentals, organize system design into semantic information extraction, transmission, and metrics dimensions & \textcolor{red}{\ding{55}} & \textcolor{green}{\ding{51}} & \textcolor{red}{\ding{55}} \\
\cmidrule(l){2-6}
& \cite{Xu2023JSTSP} & Present an overview of practical distributed edge learning techniques and their interplay with advanced communication optimization from dual perspectives of communications for learning and learning for communications & \textcolor{green}{\ding{51}} & \textcolor{red}{\ding{55}} & \textcolor{red}{\ding{55}} \\
\cmidrule(l){2-6}
& \cite{Bai2023COMST} & Provide a comprehensive review of reinforcement learning-based approaches in autonomous multi-UAV wireless networks, summarize RL applications including resource allocation and trajectory planning & \textcolor{red}{\ding{55}} & \textcolor{green}{\ding{51}} & \textcolor{red}{\ding{55}} \\
\cmidrule(l){2-6}
& \cite{DeKeersmaeker2023COMST} & Survey existing datasets containing IoT network traffic, classify them according to features for traffic classification and intrusion detection research & \textcolor{green}{\ding{51}} & \textcolor{green}{\ding{51}} & \textcolor{red}{\ding{55}} \\
\midrule
\multirow{3}{1.0cm}{\centering\textbf{2024}} & \cite{Liu2024COMST} & Present a comprehensive review of federated learning, meta learning, and federated meta learning methodologies and their applications over wireless networks & \textcolor{green}{\ding{51}} & \textcolor{green}{\ding{51}} & \textcolor{red}{\ding{55}} \\
\cmidrule(l){2-6}
& \cite{Khalek2024COMST} & Provide an in-depth examination of machine learning-based cognitive radio integration in emerging wireless networks including IoT, vehicular communications, and UAV networks & \textcolor{red}{\ding{55}} & \textcolor{green}{\ding{51}} & \textcolor{red}{\ding{55}} \\
\cmidrule(l){2-6}
& \cite{Mao2024TIM} & Provide a comprehensive survey on UAV channel sounder design including hardware scheme, sounding signal, time synchronization, calibration, and data postprocessing & \textcolor{red}{\ding{55}} & \textcolor{green}{\ding{51}} & \textcolor{green}{\ding{51}} \\
\midrule
\multirow{3}{1.0cm}{\centering\textbf{2025}} & \cite{Pan2025JIOT} & Present a comprehensive survey of digital twin network reference architecture and enabling technologies, propose a five-level evolution taxonomy framework for digital twin models & \textcolor{green}{\ding{51}} & \textcolor{red}{\ding{55}} & \textcolor{red}{\ding{55}} \\
\cmidrule(l){2-6}
& \cite{Zhang2025COMST} & Provide a comprehensive survey of Industrial Metaverse enabling technologies including blockchain, privacy-preserving computing, digital twin, 5G/6G, XR, and artificial intelligence & \textcolor{red}{\ding{55}} & \textcolor{green}{\ding{51}} & \textcolor{red}{\ding{55}} \\
\cmidrule(l){2-6}
\rowcolor[HTML]{e5e5f7}
& \textbf{Ours} & Investigate the deployment of unified multi-dimensional information quality optimization frameworks for next-generation intelligent networks via comprehensive four-dimensional taxonomic analysis, AI-driven cross-layer integration, and extensive application scenarios across six major domains & \textcolor{green}{\ding{51}} & \textcolor{green}{\ding{51}} & \textcolor{green}{\ding{51}} \\
\bottomrule
\end{tabular}
\end{table*}

In this survey, we provide a comprehensive analysis of information quality metrics research across next-generation networked systems. Given the increasing complexity of intelligent networks and the growing demand for AI-driven applications, several surveys and research efforts have explored various aspects of information quality assessment. Table~\ref{tab:related_works} presents a comparison of these surveys with our work.

The existing literature on information quality metrics can be systematically categorized into two primary research streams. The first stream focuses on individual metric dimension surveys, where researchers have extensively studied specific aspects of information quality in isolation. The study in \cite{Yates2021JSAC} provided a foundational examination of AoI metrics. It established theoretical frameworks for timeliness assessment in cyber-physical systems and examined AoI evaluation methods across various system configurations from elementary queues to complex wireless networks. Based on traditional AoI concepts, the authors \cite{Maatouk2020TON} introduced the Age of Incorrect Information (AoII) as a novel performance metric that extends the notion of fresh updates to fresh ``informative'' updates. This approach addressed limitations of conventional AoI frameworks through Markov Decision Process formulations. In \cite{Fu2023INFOCOM}, the authors analyzed Peak Age of Information (PAoI) strategies in cache-enabled Industrial IoT networks, which demonstrated advanced timeliness optimization techniques. The study \cite{Chaccour2025COMST} presented a comprehensive examination of semantic communication paradigms. It detailed the fundamentals of semantic information representation and entropy, while exploring how deep learning-enabled semantic communication serves as a promising enabler of 6G technology. In \cite{Fu2023INFOCOM}, the authors explored content-aware semantic communication for goal-oriented wireless systems. The study in \cite{Xu2025JSAC} established the foundational \emph{Information Value Theory} framework for quantitative assessment. The authors \cite{Ramos2017COMST} provided a comprehensive survey of model-based quantitative network security metrics, which examined various approaches for security assessment in networked systems. In \cite{Sourlas2018TNSM}, the study investigated information resilience enhancement in disruptive information-centric networks. The survey \cite{Fei2017COMST} offered a comprehensive tutorial on multi-objective optimization techniques for wireless sensor networks. It examined challenging tradeoffs among conflicting optimization criteria including energy dissipation, coverage, and network lifetime.

\begin{figure*}[!t]
    \centering
    \includegraphics[width=0.9\textwidth]{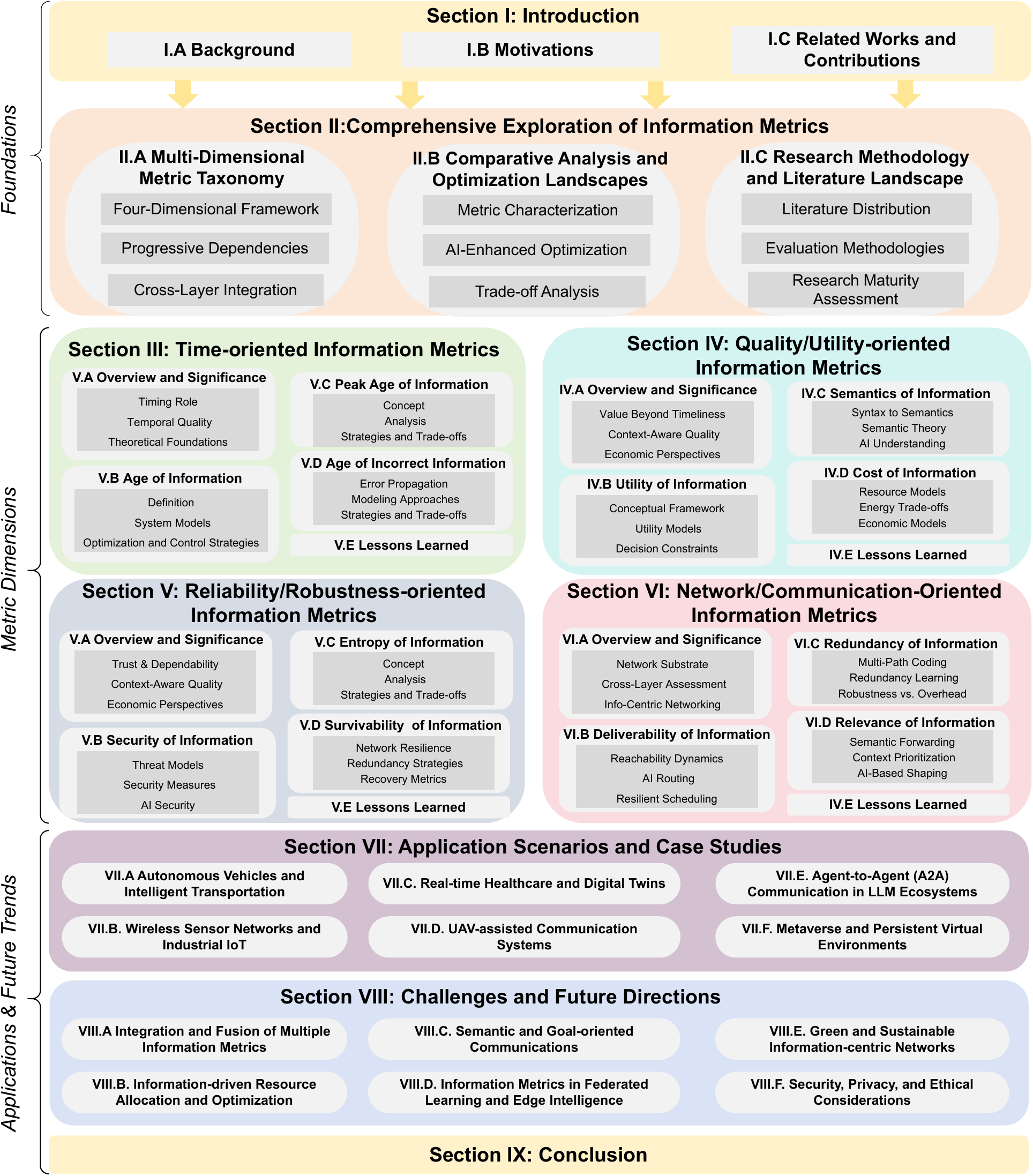} 
    \caption{Structure of our survey. We begin with foundational concepts and taxonomy of information metrics, with emphasis on the evolution toward AI-driven, context-aware evaluation in networked systems. The taxonomy is followed by a detailed exposition of core metric families, timeliness, utility, semantics, robustness, and communication, and their interrelationships. The survey further explores how these metrics are applied in intelligent systems such as UAV networks, LLM-based agent interactions, and digital twins. We conclude by identifying key challenges and future research directions to guide the evolution of multi-dimensional ``X of Information'' frameworks.}
    \label{fig:framework_overall}
\end{figure*}

The second research stream includes cross-cutting technology and application-specific surveys that bridge multiple information quality dimensions. The study in \cite{Jiang2024MWC} explored the integration of large language models with multi-agent systems for 6G communications. It proposed CommLLM as a framework that enhanced LLM capabilities through retrieval, planning, memory, and evaluation agents to address communication-related tasks using natural language interfaces. In \cite{Hu2024TMC}, the authors presented AutoFL, a Bayesian game approach for autonomous client participation in federated edge learning, which demonstrated AI-driven optimization in distributed systems. The study \cite{Jayanetti2024TPDS} investigated multi-agent deep reinforcement learning frameworks for renewable energy-aware workflow scheduling in distributed systems. The authors \cite{He2024TYCB} provided a comprehensive review of cascading failure modeling and vulnerability analysis in cyber-physical systems, which examined reliability and robustness considerations. In \cite{Fu2022TNET}, the study explored generative-AI-driven human digital twin applications in IoT healthcare systems. The survey \cite{Adil2023TITS} examined cross-layer security metrics and optimization strategies for semantic communications. The authors \cite{Chen2019TOIT} investigated dynamic service migration mechanisms in edge cognitive computing environments. In \cite{Li2025IOTJ}, the study presented multiobjective optimization approaches for AoI and rate in secure dual-functional networks. The research \cite{Sorkhoh2022TITS} focused on optimizing information freshness for MEC-enabled cooperative autonomous driving applications.

Distinct from existing surveys and tutorials, our survey concentrates on the deployment of unified multi-dimensional information quality optimization frameworks for next-generation intelligent networks. While individual metrics have been extensively studied in isolation, and cross-cutting technologies have been explored within specific domains, no comprehensive survey has systematically examined the progressive dependencies among time-oriented, quality/utility-oriented, reliability/robustness-oriented, and network/communication-oriented information metrics. Furthermore, existing work lacks a cohesive analysis of how AI-driven integration strategies can enable joint optimization across these competing dimensional objectives.

The contributions of our survey are as follows:
\begin{itemize}
\item We introduce the first comprehensive four-dimensional taxonomic framework that systematically organizes information quality metrics into progressive dependency hierarchies, which establishes fundamental relationships between temporal, semantic, security, and network delivery dimensions.

\item We analyze cross-dimensional interdependencies and multi-objective optimization strategies. We provide unified theoretical foundations and AI-enhanced methodologies for resolving competing information quality objectives through deep reinforcement learning and multi-agent coordination frameworks.

\item We present extensive application scenarios and case studies across six critical domains—autonomous vehicles, industrial IoT, healthcare digital twins, UAV communications, LLM ecosystems, and metaverse environments. These demonstrate practical deployment of integrated multi-dimensional information quality frameworks.

\item We identify key research challenges and future directions. These encompass metric integration and fusion, information-driven resource allocation, semantic and goal-oriented communications, federated learning optimization, green sustainable networking, and privacy-preserving quality assessment for next-generation information-centric network architectures.
\end{itemize}

This survey is organized as shown in the outline illustrated in Fig.~\ref{fig:framework_overall}. Section~\ref{explore} establishes the theoretical foundations of multi-dimensional information quality assessment and introduces our four-dimensional taxonomic framework. Section~\ref{Time} systematically examines time-oriented metrics including AoI, PAoI, and AoII. Section~\ref{Quality} explores quality and utility-oriented metrics encompassing UoI, Semantics of Information, and Cost of Information. Section~\ref{Reliability} investigates reliability and robustness-oriented metrics including Security of Information, Entropy of Information, and Survivability of Information measures. Section~\ref{Network} analyzes network and communication-oriented metrics covering Deliverability of Information, Redundancy of Information, and Relevance of Information. Section~\ref{Application} presents comprehensive application scenarios and case studies across six major domains with detailed implementation guidelines and performance analysis. Section~\ref{future} discusses future research challenges and emerging directions encompassing metric integration, resource optimization, and privacy-preserving assessment. Section~\ref{conclus} provides the conclusions.

\section{A Comprehensive Exploration of Information Metrics}\label{explore}
\begin{figure*}[ht]
    \centering
    \includegraphics[width=0.9\textwidth]{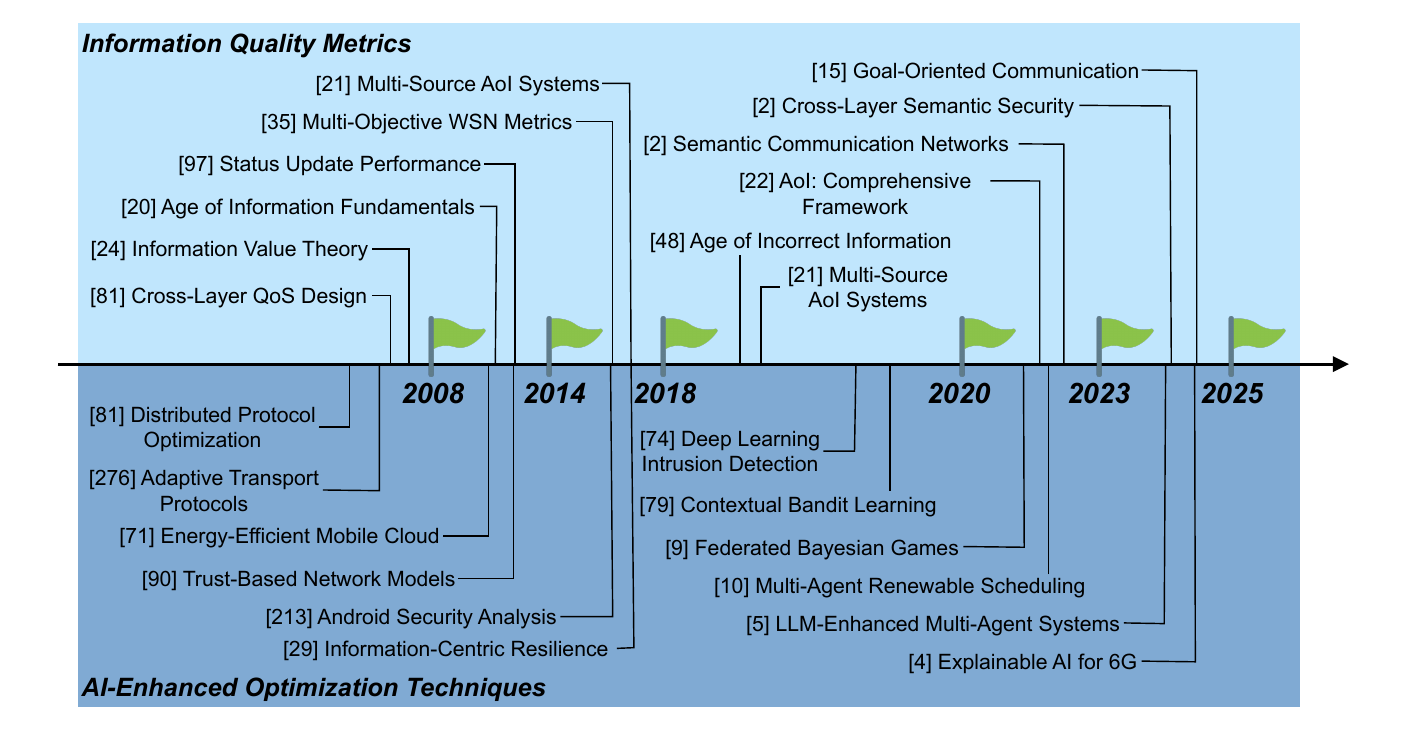}
    \caption{Evolution of information quality metrics and AI-enhanced optimization techniques in next-generation networks.}
    \label{fig:information_metrics_evolution}
\end{figure*}

In this section, we present an in-depth study of information metrics for next-generation network systems. It provides both the theoretical foundations and practical frameworks necessary for understanding multi-dimensional information quality assessment. We begin by constructing a comprehensive taxonomic framework that classifies information metrics into four core dimensions: temporal, quality/utility, reliability/robustness, and network/communication. We then examine how artificial intelligence enables the transformation of traditional static metrics into adaptive and intelligent evaluation systems. We also discuss the interdependencies among these dimensions and offer practical guidelines for applying such metrics in real-world networked environments. Fig.~\ref{fig:information_metrics_evolution} illustrates the evolution of information quality metrics alongside AI-driven optimization techniques, which together form the analytical basis of this discussion.

\subsection{Multi-Dimensional Metric Taxonomy for Next-Generation Networks}

Application proliferation across autonomous vehicles \cite{Guo2023MNET}, smart cities \cite{Lin2024CM}, industrial IoT \cite{Adil2023TITS}, and augmented reality necessitates a redefinition of traditional single-metric evaluation paradigms. Conventional metrics such as throughput, latency, and packet loss inadequately capture the information quality requirements of intelligent systems that demand semantic interpretation, temporal awareness, security assurance, and adaptive communication capabilities.

\subsubsection{Four-Dimensional Information Quality Space}

Next-generation networks require evaluation across four core dimensions that together define information value in intelligent applications. Temporal freshness utilizes Age of Information (AoI) \cite{Costa2014ISIT} and Peak Age of Information (PAoI) \cite{Yates2019TIT} to quantify information currency for real-time decision-making \cite{Wu2023access}. Quality and utility assessment employs Utility of Information (UoI) and Semantics of Information (Sem-oI) metrics \cite{Howard1966TSSC} to capture semantic relevance and determine information contribution to specific application goals \cite{Chaccour2025COMST}. Information integrity relies on Security of Information (Sec-oI) and Entropy of Information (EoI) measurements \cite{Adil2023TITS} to maintain resilience against threats and failures in cyber-physical systems \cite{He2024TYCB}. Network delivery performance utilizes Deliverability of Information (DoI) and Redundancy of Information (RoI) metrics \cite{Sourlas2018TNSM} to assess deliverability and redundancy. These metrics optimize infrastructure effectiveness through routing \cite{Liu2022IOTJ} and adaptive resource allocation \cite{Jayanetti2024TPDS}.

\subsubsection{AI-Driven Metric Evolution and Enhancement}

Artificial intelligence integration transforms evaluation requirements in networked systems beyond traditional performance metrics. Large Language Models (LLMs) \cite{Jiang2024MWC} and generative AI models \cite{Chen2024IOTJ} introduce content creation demands, while semantic communication frameworks \cite{Chaccour2025COMST} enable meaning-aware data transmission. Digital twins \cite{Benedictis2023JBHI} and metaverse applications \cite{Lotfi2024JSAC} necessitate real-time synchronization and high-fidelity representation. Multi-agent systems \cite{Das2023CIC} require coordinated decision-making based on shared information quality, while federated learning paradigms \cite{Hu2024TMC} distribute training across edge networks with novel freshness requirements. UAV-assisted networks \cite{Wang2025IOTJ1}, mobile edge computing \cite{Chen2019TOIT}, 6G communications \cite{Zhong2024MNET}, blockchain-enabled AI systems \cite{Li2024TGCN}, and privacy-preserving collaborative learning \cite{Gao2021CVPR} introduce semantic richness, temporal dependencies, security constraints, and adaptive allocation challenges that single-dimensional metrics cannot address.

These technologies impose quality requirements that span multiple evaluation domains simultaneously. Generative AI applications \cite{Xu2025JSAC} demand temporal freshness for real-time generation and semantic utility for meaningful output. Autonomous systems \cite{Guo2023MNET} require reliability guarantees and network deliverability for safety-critical operations. Semantic-aware communication systems \cite{Fu2023INFOCOM} prioritize contextual relevance over throughput, while goal-oriented networking \cite{Chen2024GLOBECOM} requires dynamic metric adaptation. Multi-objective optimization frameworks \cite{Li2025IOTJ} and intelligent resource management systems \cite{Jayanetti2024TPDS} balance competing objectives across temporal, semantic, security, and network dimensions. Privacy-preserving mechanisms \cite{Wang2024JIOT}, energy-efficient protocols \cite{Chen2015TCC}, and sustainable computing paradigms add complexity layers that require comprehensive multi-dimensional evaluation.

\subsubsection{Cross-Dimensional Dependencies in Networked Systems}

Information quality dimensions in next-generation networks exhibit hierarchical interdependencies. Temporal freshness through AoI \cite{Costa2014ISIT} and PAoI \cite{Yates2019TIT} establishes baseline requirements that directly affect quality and utility evaluations in real-time decision-making \cite{Wu2023access}. Quality and utility assessment utilizes semantic communication \cite{Chaccour2025COMST} and goal-oriented networking \cite{DiLorenzo2023IOTM} to determine information value based on temporal currency. Reliability and robustness dimensions maintain trustworthiness through security mechanisms \cite{Adil2023TITS} and fault-tolerant design \cite{He2024TYCB}. Network and communication performance transmits processed information through optimized routing \cite{Liu2022IOTJ}, cross-layer protocols \cite{Wu2024TCOMM}, and adaptive resource allocation \cite{Wang2025TVT}, where modifications at one layer impact others throughout the stack.

Dimensional interdependencies require integrated optimization strategies that balance multiple objectives \cite{Li2025IOTJ} rather than isolated approaches. UAV-assisted networks \cite{Wang2025IOTJ1}, federated learning environments \cite{Hu2024TMC}, and edge computing systems \cite{Chen2019TOIT} demonstrate requirements for intelligent coordination mechanisms \cite{Jayanetti2024TPDS} to manage cross-dimensional trade-offs. Advanced optimization frameworks employ machine learning \cite{Fu2022TNET} and reinforcement learning \cite{He2024TMC} to adapt to complex dependencies in real time. Multi-agent systems \cite{Das2023CIC} enable distributed decision-making for system-wide performance maintenance. Network architectures integrate privacy-preserving techniques \cite{Wang2024JIOT}, energy-efficient protocols \cite{Chen2015TCC}, and adaptive service management to sustain multi-dimensional performance.

\subsection{Comparative Analysis and Optimization Landscapes}

This subsection examines the fundamental characteristics and optimization challenges inherent in multi-dimensional information metrics for next-generation networks. We analyze the computational complexity, measurement precision, and real-time feasibility differences across temporal, quality, reliability, and network dimensions. The section then explores AI-enhanced optimization strategies that enable intelligent trade-off management and conflict resolution in complex networked environments where multiple objectives compete for limited resources.

\subsubsection{Multi-Dimensional Metric Characterization and Trade-offs}

Information metrics across four dimensions exhibit distinct computational complexity and measurement characteristics. Temporal metrics like AoI \cite{Costa2014ISIT} and PAoI \cite{Yates2019TIT} demonstrate low computational overhead with polynomial complexity \cite{Yates2021JSAC}, which enables real-time monitoring in resource-constrained environments. Quality metrics require higher computational demands due to semantic processing \cite{Howard1966TSSC} and machine learning-based utility functions \cite{Fu2022TNET}. Security metrics introduce exponential complexity through cryptographic operations \cite{Adil2023TITS} and threat detection \cite{Otoum2019LNET}. Network metrics scale with topology size and dynamic routing requirements \cite{Sourlas2018TNSM}, which produces precision variations from quantitative temporal assessments \cite{sun2022age} to probabilistic utility evaluations and categorical security measurements.

Dimensional differences generate fundamental trade-offs in next-generation networks. Key conflicts include security overhead versus temporal freshness \cite{Li2025IOTJ}, semantic accuracy versus communication efficiency \cite{Wu2024TCOMM}, and reliability versus energy consumption \cite{Wang2025TVT}. Resolution mechanisms employ multi-objective optimization \cite{Fei2017COMST}, reinforcement learning \cite{He2024TMC}, federated decision-making \cite{Hu2024TMC}, and multi-agent collaboration \cite{Jayanetti2024TPDS} for priority management. Optimization approaches integrate semantic communication principles \cite{Chaccour2025COMST} and goal-oriented networking \cite{DiLorenzo2023IOTM} to enable application-specific prioritization.

\subsubsection{AI-Enhanced Multi-Objective Optimization Strategies}

AI techniques enable multi-dimensional optimization space exploration through Pareto frontier navigation and adaptive strategies. Evolutionary algorithms \cite{Fei2017COMST} and genetic programming identify trade-off solutions across temporal, quality, reliability, and network dimensions through population-based search. Deep reinforcement learning \cite{He2024TMC} coordinates multi-agent systems \cite{Jayanetti2024TPDS} for real-time objective balancing. Transformer architectures \cite{Jiang2024MWC} and attention mechanisms adjust priorities based on application requirements and network context. Neural network-based optimization \cite{Dang2025MCI} integrates federated learning \cite{Hu2024TMC} for distributed decision-making across edge-cloud systems \cite{Gort2025TMLCN} under latency and resource constraints. Edge computing solutions \cite{Bakhrani2025IOTJ} balance local processing with global optimization for scalable network designs.

AI-driven conflict resolution mechanisms address optimization trade-offs through adaptive learning and distributed coordination. Multi-agent reinforcement learning \cite{Jayanetti2024TPDS} enables collaborative decision-making for resource allocation and priority management based on performance feedback. Federated optimization \cite{Wang2024JIOT} coordinates decentralized learning while it preserves privacy and reduces overhead for semantic communication \cite{Chaccour2025COMST} and goal-oriented networking \cite{DiLorenzo2023IOTM,Chen2024GLOBECOM}. Online learning methods \cite{Fu2022TNET} adapt to dynamic network conditions and user behavior. Contextual bandit algorithms \cite{NgoICDCS} manage exploration-exploitation trade-offs in uncertain environments. Zero-touch network management \cite{Sun2025OJCOMS} employs explainable AI for autonomous optimization in critical systems.

\subsection{Research Methodology and Literature Landscape}

\begin{figure*}[ht]
    \centering
    \includegraphics[width=0.9\textwidth]{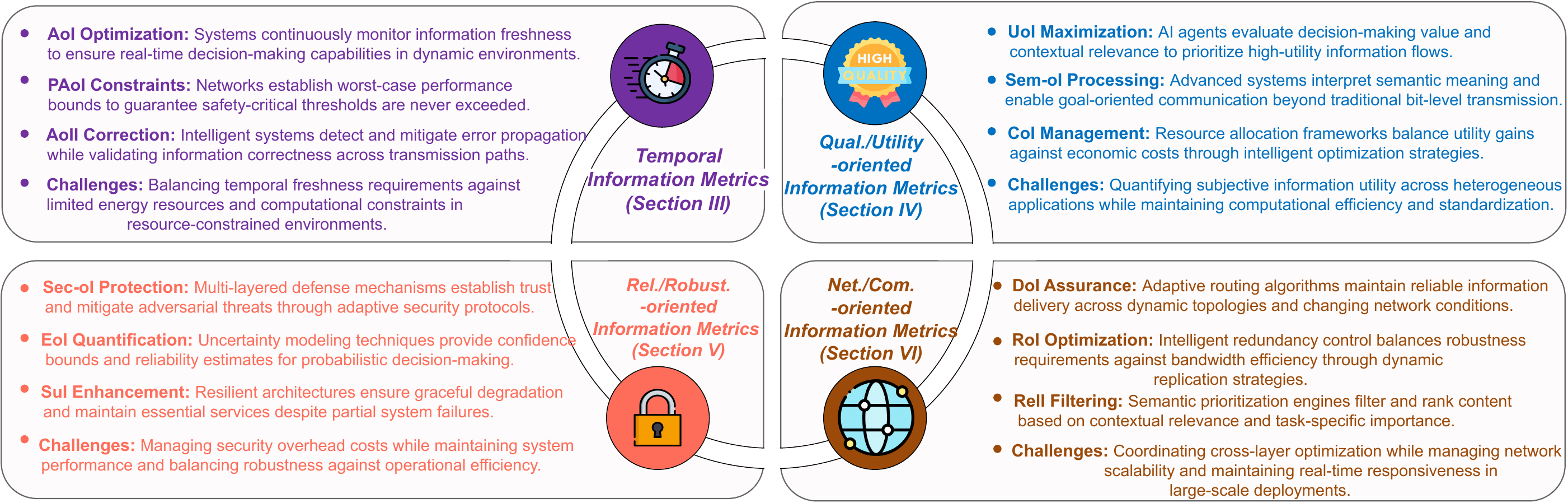} 
    \caption{Multi-dimensional information metrics framework for Next-generation networks.}
    \label{ummary4}
\end{figure*}

Fig.~\ref{ummary4} illustrates our comprehensive four-dimensional framework that organizes these diverse research areas into temporal freshness, quality/utility assessment, reliability/robustness assurance, and network/communication delivery optimization categories. Understanding these patterns provides essential context for navigating the detailed metric exploration and helps identify both well-established areas and emerging research frontiers that shape the field's evolution.

\subsubsection{Research Maturity and Literature Distribution Across Dimensions}

Research maturity across the four dimensions exhibits significant variation. Temporal metrics are the most extensively studied area with established theoretical foundations. AoI demonstrates high maturity, supported by seminal contributions \cite{Costa2014ISIT} and comprehensive surveys \cite{Yates2021JSAC}. PAoI has developed as a well-established extension \cite{Zheng2025IOTJ,Sinha2024ISIT}. Quality and utility metrics demonstrate fragmented advancement. They evolve from classical information theory \cite{Howard1966TSSC} toward semantic communication \cite{Chaccour2025COMST} and goal-oriented networking \cite{DiLorenzo2023IOTM} without reaching the theoretical consistency found in temporal metrics. Reliability and security metrics maintain domain-specific focus through cybersecurity research \cite{Adil2023TITS,Kumar2023TITS} and cyber-physical system resilience studies \cite{He2024TYCB}. Machine learning-based security detection receives growing attention \cite{Otoum2019LNET,Ferrag2022JAS}. Network and communication metrics exhibit greatest fragmentation across information-centric networking \cite{Sourlas2018TNSM}, vehicular networks \cite{Liu2022IOTJ}, and cross-layer optimization \cite{Papandriopoulos2008TNET}. Artificial intelligence integration varies across dimensions: multi-agent deep reinforcement learning achieves broad adoption in resource management \cite{Jayanetti2024TPDS}, federated learning emerges in distributed optimization \cite{Hu2024TMC}, and edge computing AI adoption continues to expand \cite{He2024TMC}.

\subsubsection{Evaluation Methodologies and Validation Approaches}

Evaluation methodologies in information metric research exhibit clear preferences that are shaped by dimensional characteristics and practical limitations. Simulation analysis dominates the field, particularly in multi-objective optimization \cite{Fei2017COMST} and complex system modeling \cite{Yang2024IOTJ}. It is supported by specialized tools for IoT applications \cite{Chen2023IOTJ} and the evaluation of distributed systems \cite{Lu2023TPDS}. Theoretical modeling remains central in foundational studies, particularly for uncertainty quantification \cite{Wang2023TNNLS} and reliability assessment \cite{Zhang2025TR}, while Bayesian game-theoretic approaches \cite{Hu2024TMC} and network utility optimization \cite{Fu2022TNET} offer mathematical rigor in performance bound analysis. Testbed implementations continue to mature, with industrial IoT deployments \cite{Karamyshev2024TII} and wireless experiments that demonstrate feasibility; however, real-world validation remains limited to domains such as healthcare \cite{Wu2025MNET} and vehicular systems \cite{Sorkhoh2022TITS}. Benchmarking remains underdeveloped, with few standardized frameworks beyond specific areas, such as network traffic classification \cite{Aceto2014MCOM} and quality assessment \cite{Thi2017arxiv}. This highlights the need for systematic performance comparisons. Recent validation methods increasingly incorporate trust modeling \cite{Cho2015ACS} and explainable AI \cite{Sun2025OJCOMS}, while privacy-preserving evaluation \cite{Wang2024JIOT} addresses concerns in distributed metric analysis. This indicates a shift toward more advanced, AI-supported validation approaches that lay the foundation for a detailed study of individual metric categories.

\section{Time-oriented Information Metrics} \label{Time}
\begin{figure}[ht]
    \centering
    \includegraphics[width=1\linewidth]{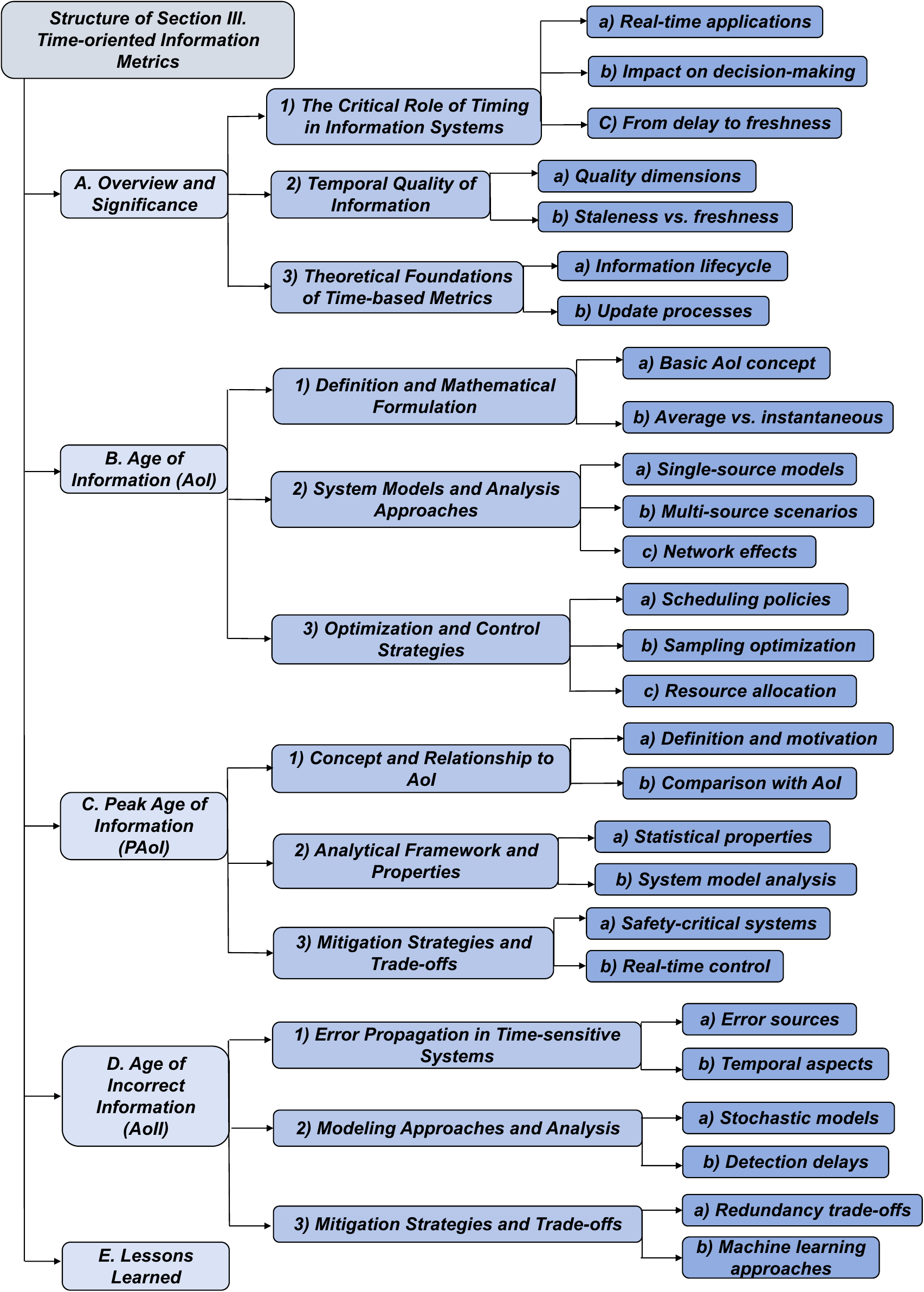}
    \caption{The roadmap of Section III.}
    \label{fig:RMIII}
\end{figure}

In this section, we provide an overarching framework of time-oriented information metrics that emphasizes their definitions, analytical foundations, practical implications, and strategies that facilitate network optimization. Metrics such as Age of Information (AoI), Peak Age of Information (PAoI), and Age of Incorrect Information (AoII) are explored, each with distinctive features and roles that optimize information freshness and correctness within real-time and critical application environments. The roadmap of Section \ref{Time} is illustrated in Fig. \ref{fig:RMIII}.

\subsection{Overview and Significance}
This subsection presents an overview of time-oriented information metrics, which includes their applications, underlying quality dimensions, theoretical foundations, and extensions that facilitate network optimization. It also provides insights into the role of these metrics that optimize information timeliness within various practical scenarios.

Time-oriented information metrics assess how fresh and timely information is retained for effective decision-making. AoI measures the time elapsed since the generation of the most recently received update, providing direct assessment of information freshness. PAoI captures the maximum AoI value before successful update reception, offering crucial insights into worst-case freshness scenarios for safety-critical applications. AoII extends these concepts by incorporating correctness, measuring the duration during which received information remains invalid.

These metrics find applications in autonomous vehicle networks for Vehicle-to-Everything (V2X) communication timing, healthcare systems for patient monitoring data freshness, and Industrial Internet of Things (IoT) networks for detecting sensor malfunctions. Practical implementation requires timestamp-based tracking systems, statistical analysis of update patterns, and sophisticated optimization algorithms that balance freshness requirements against energy consumption and bandwidth constraints in resource-limited environments.

\subsubsection{The Critical Role of Timing in Information Systems}

Timing critically affects the performance and reliability of information systems. The following paragraphs delve into real-time applications, the impact of timing on decision-making processes, and the evolution from delay-centric to freshness-centric perspectives.

\paragraph{Real-time applications}Real-time performance is increasingly becoming a fundamental requirement in modern networked systems, which face stringent timing demands across various fields. For instance, Lin {\sl et al.}~\cite{Lin2024CM} demonstrated that intelligent transportation systems critically depend on real-time data processing to prevent accidents and enhance traffic efficiency. Similarly, the precision of time-critical actions is paramount in Internet of Things (IoT) applications, which include industrial robotics, autonomous systems, and interactive human-machine experiences~\cite{Cavalcanti2022IOTJ}.

\paragraph{Impact on decision-making}Timeliness directly influences the effectiveness of data-driven decision-making processes. Guan {\sl et al.}~\cite{Guan2025CM} emphasized that meeting deterministic latency requirements, particularly in 5G integrated time-sensitive networking (TSN), is crucial for both industrial and consumer wireless applications. Furthermore, delays in critical information can significantly impair timely and accurate decisions in various scenarios such as healthcare diagnostics and epidemic disease management~\cite{Lin2022TMBMC}.

\paragraph{From delay to freshness} The shift from focusing merely on network delay to explicitly considering information freshness marks a significant conceptual evolution. AoI emerges as a critical metric that presents this paradigm shift and reflects the time elapsed since the most recently received update. The focus on freshness rather than delay enables enhanced precision in managing data timeliness, which ensures reliability and effectiveness in information systems~\cite{Costa2014ISIT}.

\subsubsection{Temporal Quality of Information}

To better understand the practical implications of timing, we explore various dimensions of temporal information quality that focus specifically on the complementary metrics of freshness and staleness.

\paragraph{Quality dimensions}Temporal dimensions of information quality, such as freshness, staleness, and timeliness, are key metrics in diverse application contexts. Freshness, which is quantified primarily by AoI, plays a vital role in assessing the utility of information in real-time systems~\cite{Yates2019TIT}.

\paragraph{Staleness vs. freshness}PAoI, which captures the worst-case freshness scenarios, highlights critical moments where stale information could cause safety hazards, particularly in applications such as healthcare monitoring and intelligent vehicular networks~\cite{Wu2023access,Guo2023MNET}. Recent developments like Ultra-Low AoI (ULAoI) address scenarios where traditional metrics fail to adequately handle extreme freshness deviations, particularly in digital twin and power grid scenarios~\cite{Liao2023JSAC}.

\subsubsection{Theoretical Foundations of Time-based Metrics}

Understanding the theoretical foundation of temporal information metrics is crucial for their effective implementation and further innovation. The following paragraphs outline the lifecycle of information and describe advanced update processes that ensure the freshness and accuracy of information.

\paragraph{Information lifecycle}Time-oriented metrics build upon rigorous theoretical foundations that derive from classical information lifecycle models and queueing theories. Costa {\sl et al.}~\cite{Costa2014ISIT} initially provided analytical insights into how packet management influences temporal freshness within information systems.

\paragraph{Update processes}Advanced modeling techniques such as stochastic mixed-integer programming and mean-field games have been introduced to handle complex update processes. Qin {\sl et al.}~\cite{Qin2024TNSE} employed deep reinforcement learning to optimize age-sensitive tasks in UAV-edge-computing scenarios. Additionally, Emami {\sl et al.}~\cite{Emami2024TVT} proposed a hybrid proximal policy optimization framework that integrates predictive AI methods to effectively manage the AoI of multi-agent UAV systems, which further enriches the theoretical toolkit available for managing information freshness.

In summary, timing as a critical dimension of information quality significantly impacts real-time system performance across various application domains. Metrics such as AoI, PAoI, and AoII are essential indicators for evaluating and optimizing systems to ensure reliability and accuracy in diverse operational environments. The integration of advanced AI techniques and theoretical innovations continues to push research forward, which promises enhanced efficiency in next-generation networked systems. The roadmap of Section \ref{Time} is illustrated in Fig. \ref{fig:RMIII}.

\begin{figure}[t!]
    \centering
    \includegraphics[width=1\linewidth]{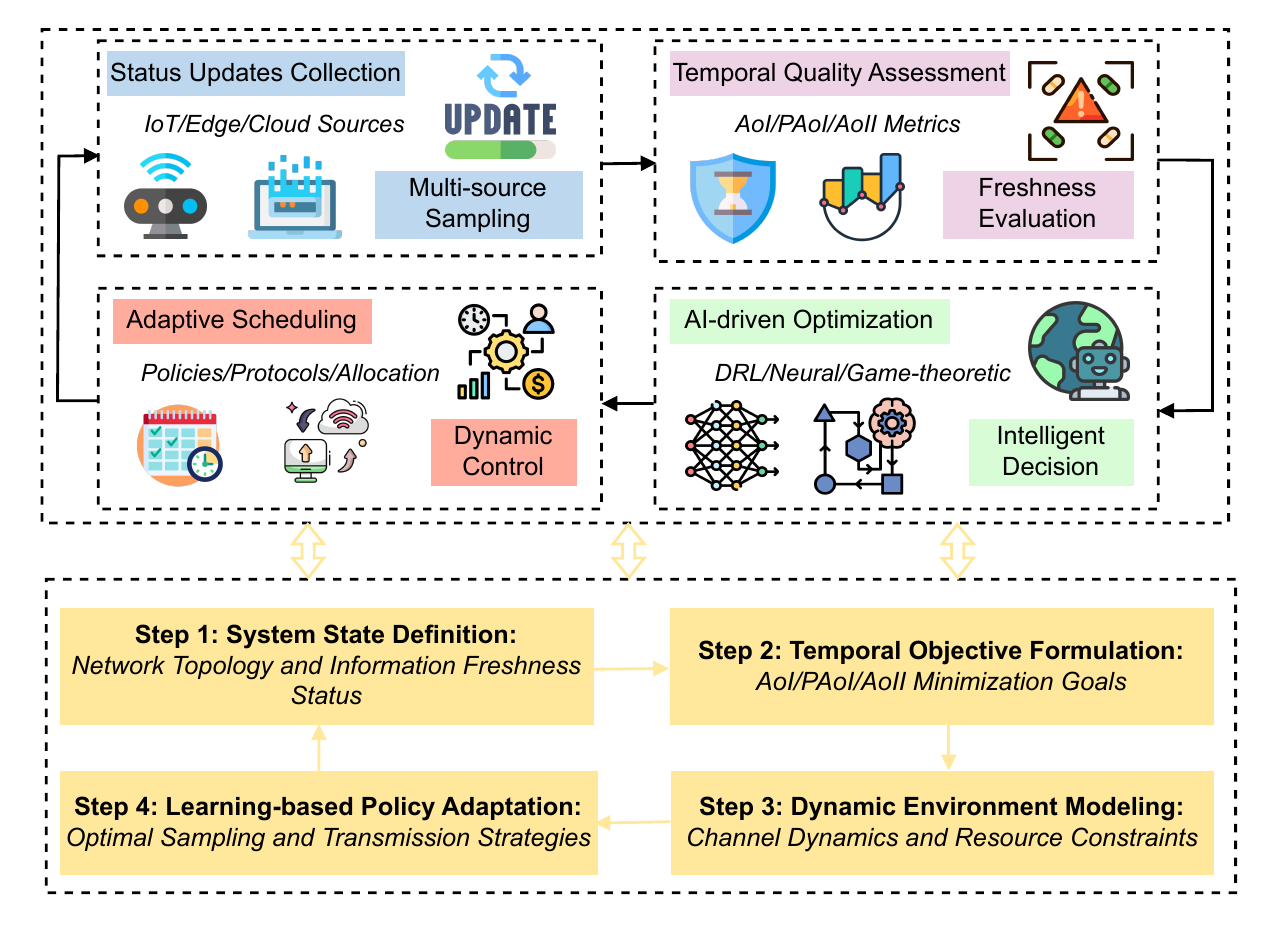}
    \caption{AI-enhanced time-oriented information metrics optimization framework for next-generation networks.}
    \label{fig:ai_time_optimization}
\end{figure}

As illustrated in Fig.~\ref{fig:ai_time_optimization}, the integration of artificial intelligence technologies enables systematic optimization of time-oriented information metrics through intelligent decision-making and adaptive control strategies. This framework demonstrates how modern AI approaches, particularly deep reinforcement learning and multi-agent systems, can dynamically balance temporal objectives with network constraints to achieve optimal information freshness in next-generation networked systems.

\subsubsection{Definition and Mathematical Formulation}

Information freshness assessment requires precise mathematical definitions of AoI, distinguishing between instantaneous and average metrics.

\paragraph{Basic AoI concept}
AoI quantifies the freshness of the latest received information update. Sun {\sl et al.}~\cite{sun2022age} defined AoI as the time elapsed since the generation of the latest successfully received packet, emphasizing applications in autonomous driving and financial trading where information freshness impacts performance. Costa {\sl et al.}~\cite{Costa2016TIT} extended this definition through packet management strategies within status update systems for networked scenarios. The authors in \cite{Kam2018TIT} analyzed AoI under packet deadline constraints, demonstrating deadline optimization for enhanced freshness and system performance.

\paragraph{Average vs. instantaneous AoI}
Practical implementations distinguish between instantaneous and average AoI metrics. Instantaneous AoI provides real-time freshness measures for immediate responsiveness in dynamic environments. Average AoI evaluates long-term system performance. The authors in \cite{Soysal2021TIT} developed analytical frameworks for computing average AoI in G/G/1/1 queueing systems under different operational disciplines. Farazi {\sl et al.}~\cite{Farazi2018INFOCOM} investigated AoI in energy harvesting scenarios, developing analytical expressions for energy constraint impacts on AoI. Krikidis {\sl et al.}~\cite{Krikidis2019WCL} examined AoI in wireless powered sensor networks, providing closed-form results for optimal operational parameters. Both instantaneous and average AoI understanding proves essential for comprehensive freshness management across diverse real-time systems.

\subsubsection{System Models and Analysis Approaches}

AoI analysis employs systematic modeling approaches from single-source scenarios to complex multi-source environments and network topology considerations.

\paragraph{Single-source models}
Single-source AoI modeling establishes foundational freshness management principles. The authors in \cite{Akar2025TCOM} introduced a dual-server, single-source system model, analyzed transmission policies and their impacts on AoI distributions and performance. Inoue {\sl et al.}~\cite{Inoue2019TIT} derived a general stationary AoI distribution formula for single-server queues, applying it across different service disciplines and demonstrating queue management influence on AoI performance.

\paragraph{Multi-source scenarios}
Multi-source analysis addresses practical network complexity beyond single-source foundations. Sun {\sl et al.}~\cite{Sun2024JSAC} analyzed multi-source networks under stochastic energy harvesting conditions, proposing approximation policies. The authors in \cite{Zakeri2024TWC} developed dynamic and learning-based scheduling algorithms to minimize AoI in resource-constrained multi-source relaying systems. Pan {\sl et al.}~\cite{Pan2022TNSE} introduced optimization formulations for IoT networks handling multiple sources through computationally efficient solutions.

\paragraph{Network effects}
Network topology impacts AoI management by influencing system performance and optimization strategies. AoI optimization under topology constraints spans multiple scenarios. The authors in \cite{Sun2024JSAC,Zakeri2024TWC,Pan2022TNSE,Saurav2021INFOCOM} demonstrated how network structures and source interactions determine AoI performance, requiring topology-specific optimization strategies.

\subsubsection{Optimization and Control Strategies}

Effective management of AoI requires sophisticated optimization and control strategies, including scheduling, sampling, and resource allocation techniques for real-world scenarios.

\paragraph{Scheduling policies}
Effective scheduling policies minimize AoI \cite{Fang2022JSAC,Wang2024TCOM,Wang2025TIT}. Specifically, Fang {\sl et al.}~\cite{Fang2022JSAC} developed adaptive scheduling policies for massive multiple access networks, incorporating energy harvesting constraints to reduce AoI. Furthermore, Wang {\sl et al.}~\cite{Wang2024TCOM} formulated optimization methods for opportunistic channel access, leveraging advanced iterative solutions to enhance AoI performance in practical environments.

\paragraph{Sampling optimization}
Sampling strategies play a critical role in AoI optimization, particularly under resource constraints. The authors in \cite{He2024TMC} introduced reinforcement learning-based methods to optimize sampling schedules in Mobile Edge Computing (MEC) contexts, reducing AoI through adaptive learning frameworks. Meanwhile, Wang {\sl et al.}~\cite{Wang2025TIT} utilized grouping-based cyclic scheduling methods, optimizing sampling intervals to manage AoI effectively across correlated information sources.

\paragraph{Resource allocation}
Resource allocation mechanisms must be intelligently managed to ensure AoI minimization and enhanced network performance ~\cite{Dong2024TMC,Picano2025IOTJ,He2024TMC,Wang2025TWC}. Dong {\sl et al.}~\cite{Dong2024TMC} proposed FedAoI, a federated learning-based resource management approach, enhancing model training efficiency by optimizing client selection based on AoI metrics. Similarly, Picano {\sl et al.}~\cite{Picano2025IOTJ} proposed a matching-game-based algorithm to optimize resource assignments in computationally intensive IoT applications, aiming to minimize AoI. Moreover, Wang {\sl et al.}~\cite{Wang2025TWC} developed an age-weighted federated learning strategy to mitigate gradient divergence, improving convergence and reducing energy consumption.

\subsection{Peak Age of Information (PAoI)}

We explore the PAoI, a critical extension of AoI that focuses on worst-case performance in time-sensitive information systems. We also examine its theoretical foundations, analytical properties, and applications in safety-critical environments.

\subsubsection{Concept and Relationship to AoI}

Understanding PAoI requires clear differentiation from traditional AoI metrics, which establishes its unique value in ensuring reliable performance bounds for time-critical applications.

\paragraph{Definition and motivation}
The PAoI quantifies the maximum value that instantaneous age reaches before dropping due to successful update reception and provides crucial insights into worst-case freshness scenarios \cite{Liu2021INFOCOM, Champati2021JSAC, Sinha2024ISIT, Zhao2022VTC}. The authors in \cite{Liu2021INFOCOM} introduced a robust queueing approach to analyze PAoI that enabled approximation of expected PAoI performance for general arrival and service processes that exhibit heavy-tailed behaviors or correlations, where traditional probabilistic approaches proved insufficient. Meanwhile, the authors in \cite{Champati2021JSAC} developed a theoretical framework to determine the minimum achievable average PAoI in single-server-single-source queuing systems under service preemptions and request delays, which characterized optimal fixed-threshold policies and established conditions for beneficial preemptions. Furthermore, Sinha {\sl et al.}~\cite{Sinha2024ISIT} pioneered a recursive framework to characterize mean PAoI in multi-server tandem queue systems with different service rates and compared preemptive and non-preemptive policies to enhance time-sensitive information delivery.

\paragraph{Comparison with AoI}
While AoI measures average information freshness, PAoI specifically targets worst-case performance bounds critical for systems with stringent timeliness requirements \cite{Li2022TCOM, Hu2021CL, Zou2021TON, akar2021ARXIV}. Li {\sl et al.}~\cite{Li2022TCOM} provided unified closed-form expressions for both average AoI and average PAoI in Hybrid-ARQ systems and illustrated how PAoI offered complementary insights into freshness-critical status update systems. Specifically, the authors in \cite{Hu2021CL} introduced analyses of optimal arrival rates based on hard AoI and PAoI constraints and proved that for either loose or tight constraints, the optimal arrival rate converged to half the service rate. Furthermore, Zou {\sl et al.}~\cite{Zou2021TON} examined the trade-off between average AoI and average PAoI in edge computing scenarios with computation-transmission dependencies and showed how system parameters differently impacted these metrics. Meanwhile, Akar {\sl et al.}~\cite{akar2021ARXIV} developed exact distributions for both AoI and PAoI in probabilistic scheduling systems with two sources and provided analytical tools to optimize weighted average AoI/PAoI for service differentiation.

\subsubsection{Analytical Framework and Properties}

Comprehensive analysis of PAoI requires sophisticated statistical tools and system modeling approaches to characterize its behavior across various network architectures and update policies.

\paragraph{Statistical properties}
Statistical characteristics of PAoI provide essential insights for system design and optimization \cite{Akar2025TCOM, Hui2024VTC, Zheng2025IOTJ, Xu2021TIT}. The authors in \cite{Akar2025TCOM} introduced an absorbing Markov chain method to exactly obtain the distributions of both AoI and PAoI processes in a single-source dual-server status update system with phase-type distributed service times, which facilitated comprehensive performance evaluation beyond simple mean analysis. Furthermore, Hui {\sl et al.}~\cite{Hui2024VTC} derived closed-form expressions for mean PAoI in energy-aware computation offloading scenarios and showed how PAoI and energy consumption related to task generation rates and local computation capabilities in industrial IoT networks. Additionally, Zheng {\sl et al.}~\cite{Zheng2025IOTJ} developed PAoI cumulative distribution functions to analyze violation probabilities in cache-enabled Industrial IoT networks and established closed-form expressions for multicast-aware and hierarchical-aware status update strategies. Meanwhile, the authors in \cite{Xu2021TIT} provided exact derivations of the expected PAoI for priority queueing systems with different buffer configurations and service disciplines and investigated how priority ordering affected PAoI performance.

\paragraph{System model analysis}
Diverse system models require specific analytical approaches to characterize PAoI behavior effectively \cite{Qin2025TVT, Liu2021INFOCOM, Champati2021JSAC, Zou2021TON}. Qin {\sl et al.}~\cite{Qin2025TVT} pioneered a velocity-aware statistical framework for PAoI analysis that accommodated both ground and aerial users and employed dominant interferer-based approximation to analyze spatio-temporal correlation effects on PAoI distributions. Meanwhile, the authors in \cite{Liu2021INFOCOM} developed uncertainty modeling techniques for stochastic arrival and service processes and derived new bounds for PAoI in both single-source and multi-source scenarios. Furthermore, Champati {\sl et al.}~\cite{Champati2021JSAC} established theoretical characterizations of minimum achievable PAoI under service preemptions and showed improved performance for heavy-tailed service time distributions. Additionally, Zou {\sl et al.}~\cite{Zou2021TON} analyzed various queue management schemes in tandem queue systems with computation-transmission dependencies and obtained closed-form expressions for average PAoI under different buffer and service configurations.

\subsubsection{Mitigation Strategies and Trade-offs}

Effective PAoI management requires sophisticated control strategies and scheduling mechanisms designed for application requirements that balance PAoI minimization with other system objectives.

\paragraph{Safety-critical systems}
PAoI minimization proved important in safety-critical applications where timeliness directly impacted system reliability and operational safety \cite{Khorsandmanesh2021TCCN, Ling2021ICC, Jiang2021ICC, Jiang2023ICC, Jiang2022ICC}. Khorsandmanesh {\sl et al.}~\cite{Khorsandmanesh2021TCCN} investigated PAoI in wireless powered cooperative networks for time-critical IoT applications and showed how relaying strategies could improve PAoI despite increased delay through enhanced reliability. Meanwhile, the authors in \cite{Ling2021ICC} developed distributionally robust optimization techniques for PAoI minimization in E-Health IoT systems with uncertain channel state information and showed PAoI-energy tradeoffs and established superior performance of CVaR-based methods over non-robust approaches. Furthermore, Jiang {\sl et al.}~\cite{Jiang2021ICC} proposed UAV-aided sensing and communication schemes for wireless sensor networks that minimized average PAoI through joint optimization of UAV trajectory, communication power, and service time. Additionally, the authors in \cite{Jiang2023ICC} analyzed PAoI violation probability for multi-connectivity short-packet transmission in industrial IoT and derived closed-form expressions for PAoI distributions and showed how performance gains saturated at large connection numbers.

\paragraph{Real-time control}
Effective real-time control required sophisticated PAoI-aware scheduling and resource allocation mechanisms \cite{An2024VTC, Fang2022TVT, Song2024IOTJ, Zhu2024TON, Chen2024TON}. An {\sl et al.}~\cite{An2024VTC} introduced joint sensing, transmission, and computation frameworks for mobile edge computing systems and derived closed expressions for average PAoI across different computation scenarios to optimize packet sensing rates and server computation allocations. Specifically, Fang {\sl et al.}~\cite{Fang2022TVT} proposed active queue management policies for underwater wireless sensor networks that beneficially compressed packets with large waiting times and derived closed-form expressions for average PAoI and energy costs. Furthermore, the authors in \cite{Song2024IOTJ} developed deep-reinforcement-learning-based AoI-aware queue management for IoT sensor networks and formulated Markov decision processes to optimize cluster head actions for reduced power consumption while maintaining acceptable PAoI. Meanwhile, Zhu {\sl et al.}~\cite{Zhu2024TON} investigated optimal update generation timing in MEC systems and established optimality of fixed threshold policies in non-preemptive systems and transmission-aware threshold policies in preemptive systems. Additionally, Chen {\sl et al.}~\cite{Chen2024TON} examined joint scheduling of data transmission and energy replenishment to minimize maximum PAoI at wireless-powered network edges and derived theoretical bounds and proposed approximation algorithms with performance guarantees.

\subsection{Age of Incorrect Information (AoII)}

We explore AoII, an advanced information freshness metric that extends beyond timeliness to incorporate correctness. This exploration examines error propagation, modeling approaches, and mitigation strategies for next-generation networked systems.

\subsubsection{Error Propagation in Time-sensitive Systems}

Understanding error propagation in time-sensitive systems requires analyzing error sources and their temporal evolution, which impact information reliability in networked applications.

\paragraph{Error sources}
Information errors in networked systems originate from diverse sources that compromise data reliability and accuracy \cite{munari2023ARXIV, Maatouk2020TON, Maatouk2023TWC, Bonagura2023CSR}. The authors in \cite{Maatouk2020TON} introduced AoII as a performance metric that addressed the shortcomings of both AoI and conventional error penalty functions by extending fresh updates to fresh "informative" updates, which brought new and correct information to the monitor side. Meanwhile, Maatouk {\sl et al.}~\cite{Maatouk2023TWC} positioned AoII as an enabler for semantics-empowered communication. They demonstrated how traditional communication paradigms were vulnerable to performance bottlenecks. Furthermore, the authors in \cite{munari2023ARXIV} analyzed AoII in random access channels without feedback. They considered three ALOHA variations (random, reactive, and hybrid) to capture the receiver's ability to maintain accurate perception of tracked processes. Additionally, Bonagura {\sl et al.}~\cite{Bonagura2023CSR} examined AoII in adversarial scenarios where false data injection compromised information accuracy. They modeled system transitions between correct and incorrect knowledge states.

\paragraph{Temporal aspects}
Temporal dynamics of information errors influence system performance and reliability across diverse application domains \cite{Kam2020INFOCOM, Chen2021GLOBECOM, Chen2024TON, Kriouile2021ISIT}. Kam {\sl et al.}~\cite{Kam2020INFOCOM} investigated AoII in monitoring a symmetric binary information source over a delay system with feedback. They compared the metric with real-time error and AoI to show that optimal policies for real-time error and AoII were equivalent to the sample-at-change policy. Meanwhile, the authors in \cite{Chen2021GLOBECOM} considered AoII in systems with unreliable channels and power constraints. They formulated a Constrained Markov Decision Process (CMDP) and proved that the optimal policy was a mixture of two deterministic threshold policies. Furthermore, Chen {\sl et al.}~\cite{Chen2024TON} analyzed AoII over channels with random delays. They examined scenarios where updates received guaranteed delivery within certain time slots or immediate discard when transmission time exceeded predetermined values. Additionally, Kriouile {\sl et al.}~\cite{Kriouile2021ISIT} addressed monitoring Markovian remote sources without knowing if the information at the monitor was correct. They modeled the problem as a Partially Observable Markov Decision Process and developed a Whittle index-based scheduling policy.

\subsubsection{Modeling Approaches and Analysis}

Effective AoII management requires sophisticated analytical frameworks and detection mechanisms to accurately characterize information correctness across diverse network scenarios.

\paragraph{Stochastic models}
Stochastic modeling establishes fundamental frameworks for characterizing AoII behavior across diverse system architectures and information dynamics \cite{kriouile2022ARXIV, Saurav2021INFOCOM, Kam2020INFOCOM, Chen2021GLOBECOM}. The authors in \cite{kriouile2022ARXIV} considered the challenge of monitoring unknown Markovian sources. They balanced exploration and exploitation to simultaneously estimate source parameters and minimize Mean AoII. They proved that optimal solutions followed threshold-based policies. Meanwhile, Saurav {\sl et al.}~\cite{Saurav2021INFOCOM} examined stochastic arrival models for minimizing the combined objective of AoI and transmission cost. They derived an explicit optimal online policy for Poisson arrivals and proposed a simple randomized algorithm for arbitrary distributions. Furthermore, Kam {\sl et al.}~\cite{Kam2020INFOCOM} formulated optimal sampling problems as Markov decision processes across different performance metrics (real-time error, AoI, and AoII). They applied dynamic programming algorithms to compute optimal performance and policies. Additionally, Chen {\sl et al.}~\cite{Chen2021GLOBECOM} cast AoII minimization under power constraints as a CMDP. They used relative value iteration and structural properties of threshold policies to develop efficient optimization algorithms.

\paragraph{Detection delays}
Detection delays substantially impact information correctness and system performance in time-sensitive applications \cite{Nayak2023ICC, Chen2023JSTSP, Zhang2024ICC, Maatouk2023TWC}. Nayak {\sl et al.}~\cite{Nayak2023ICC} presented a decentralized policy for AoII minimization in slotted ALOHA systems. They developed a heuristic that started from a threshold-based policy and used dual methods to converge to a better solution. This approach achieved an efficient balance between frequent updates and low packet collisions. Meanwhile, the authors in \cite{Chen2023JSTSP} constructed a multi-user uplink non-orthogonal multiple access system to analyze semantic communication performance using AoII. They derived closed-form expressions for packets' AoI and formulated a non-convex optimization problem that combined both error and AoI-based performance. Furthermore, Zhang {\sl et al.}~\cite{Zhang2024ICC} introduced AoII for measuring traffic data freshness in UAV-aided vehicular Metaverse applications. They jointly optimized UAV trajectory, vehicle scheduling, and semantic data extraction under the Markov decision process framework. Additionally, Maatouk {\sl et al.}~\cite{Maatouk2023TWC} demonstrated how AoII captured the purpose of data more meaningfully than both AoI and error-based metrics. They proved that optimal transmission strategies followed randomized threshold policies.

\subsubsection{Mitigation Strategies and Trade-offs}

Effective AoII minimization requires balancing multiple objectives through sophisticated redundancy management and machine learning approaches designed for application requirements.

\paragraph{Redundancy trade-offs}
Effective redundancy management is crucial for balancing AoII performance with resource constraints in networked systems \cite{chiariotti2025ARXIV, Munari2024ICC, Ayik2023INFOCOM, Zakeri2024TCOM}. The authors in \cite{chiariotti2025ARXIV} presented the Dynamic Epistemic Logic for Tracking Anomalies (DELTA) protocol that limited collisions and minimized AoII in anomaly reporting over dense networks. This protocol used local AoII knowledge and beliefs about other sensors based on their AoI. Meanwhile, Munari {\sl et al.}~\cite{Munari2024ICC} analyzed three ALOHA variations for IoT networks. They provided closed-form analytical expressions for average AoII, missed detection probability, and incorrect knowledge duration. Their work highlighted non-trivial trade-offs for protocol design. Furthermore, Ayik {\sl et al.}~\cite{Ayik2023INFOCOM} derived closed-form Whittle Index formulations for push-based multi-user networks with AoII-dependent cost functions. They also proposed the Query AoII (QAoII) metric for pull-based systems that quantified AoII at specific query instants. Additionally, Zakeri {\sl et al.}~\cite{Zakeri2024TCOM} addressed the joint optimization of sampling and transmission policies for partially observable Markov sources in energy harvesting systems. They expressed belief as a function of AoI to effectively truncate the corresponding belief space.

\paragraph{Machine learning approaches}
Machine learning techniques offer powerful solutions for optimizing AoII in complex and dynamic networked environments \cite{Bonagura2025TCNS, Zhang2024ICC, Zakeri2024TCOM, kriouile2022ARXIV}. Bonagura {\sl et al.}~\cite{Bonagura2025TCNS} modeled strategic interactions between controllers and adversaries in cyber-physical systems using game theory. They analyzed how adversaries that injected false data affected AoII and demonstrated the existence of Nash equilibrium solutions. Meanwhile, the authors in \cite{Zhang2024ICC} used proximal policy optimization-based deep reinforcement learning to minimize expected long-term system AoII in UAV-aided vehicular Metaverse applications without requiring statistical knowledge of physical-world randomness. Furthermore, Zakeri {\sl et al.}~\cite{Zakeri2024TCOM} proposed a deep reinforcement learning policy for the AoII metric in partially observable Markov decision processes with unreliable channels. They demonstrated a non-monotonic switching-type structure of real-time optimal policies with respect to AoI. Additionally, Kriouile {\sl et al.}~\cite{kriouile2022ARXIV} developed a learning algorithm that balanced exploration and exploitation for unknown Markovian sources. They proved that the regret bound at time $T$ was $\log(T)$ compared to genie solutions that had perfect future information.

\subsection{Lessons Learned}

\begin{table*}[ht]
\centering
\caption{Comparative analysis of AoI, PAoI, and AoII.}
\label{table:comparison_time_metrics}
\renewcommand{\arraystretch}{1.5}
\begin{tabular}{|p{3cm}|p{4cm}|p{4cm}|p{4.3cm}|}
\hline
\rowcolor[gray]{0.85} \textbf{Key Features} & \textbf{Age of Information (AoI)} & \textbf{Peak Age of Information (PAoI)} & \textbf{Age of Incorrect Information (AoII)} \\
\hline
Definition & 
Time elapsed since the generation of the most recently received update \cite{sun2022age, Costa2016TIT, Kam2018TIT}. & 
Maximum value of AoI, capturing worst-case freshness performance \cite{Liu2021INFOCOM, Champati2021JSAC, Sinha2024ISIT}. & 
Time duration where the received information remains incorrect or invalid \cite{Maatouk2020TON, Maatouk2023TWC, munari2023ARXIV}. \\
\hline
Primary Applications & 
General freshness-sensitive scenarios in IoT, sensor networks, and UAV systems \cite{Fang2022JSAC, Sun2024JSAC, Pan2022TNSE}. & 
Safety-critical systems requiring worst-case guarantees for vehicular safety and emergency response \cite{Khorsandmanesh2021TCCN, Ling2021ICC, Jiang2021ICC}. & 
Systems sensitive to both freshness and correctness in autonomous driving and healthcare monitoring \cite{Bonagura2023CSR, Zhang2024ICC, Chen2023JSTSP}. \\
\hline
Analytical Techniques & 
Queueing theory, stochastic processes, and Markovian analysis \cite{Akar2025TCOM, Inoue2019TIT, Soysal2021TIT}. & 
Extreme value theory, probabilistic bounds, and queueing analysis \cite{Qin2025TVT, Akar2025TCOM, Zheng2025IOTJ}. & 
Stochastic modeling, Bayesian methods, and error propagation analysis \cite{kriouile2022ARXIV, Kam2020INFOCOM, Zakeri2024TCOM}. \\
\hline
Typical Optimization Approaches & 
Scheduling algorithms, sampling optimization, and reinforcement learning \cite{Wang2024TCOM, He2024TMC, Dong2024TMC}. & 
Priority-based scheduling, worst-case optimization, and AI-driven prediction methods \cite{An2024VTC, Fang2022TVT, Song2024IOTJ}. & 
Redundancy techniques, predictive machine learning, and error correction strategies \cite{chiariotti2025ARXIV, Munari2024ICC, Bonagura2025TCNS}. \\
\hline
Main Limitations & 
Does not capture peak deviations; mainly concerned with average freshness \cite{Wang2025TIT, Picano2025IOTJ, Saurav2021INFOCOM}. & 
Conservative metric; optimizing for PAoI might lead to resource inefficiency \cite{Zou2021TON, Li2022TCOM, Hu2021CL}. & 
Complex modeling requirements; accuracy heavily depends on error detection capabilities \cite{Kriouile2021ISIT, Nayak2023ICC, Maatouk2023TWC}. \\
\hline
\end{tabular}
\end{table*}

Examining time-oriented information metrics reveals theoretical and practical insights for networked system optimization. AoI, PAoI, and AoII complement each other as time-oriented metrics evolve from simple delay measurements to sophisticated frameworks that incorporate both timeliness and correctness \cite{Kaswan2025TCOM, Kahraman2024IOTJ, Yates2021JSAC}. AoI provides fundamental assessment of information freshness, PAoI captures worst-case performance for safety-critical applications, and AoII extends these concepts to address information correctness within temporal dimensions. The literature on these metrics shows their critical importance across diverse domains including vehicular networks, healthcare monitoring, industrial automation, and emerging metaverse applications.

Time-oriented information metrics face several practical challenges. Resource constraints represent the primary challenge that calls for trade-offs between monitoring frequency, transmission power, and computational resources. Continuous monitoring to minimize AoI may exhaust energy resources in IoT deployments. Meanwhile, stringent PAoI requirements can lead to excessive resource consumption without proportional benefits. In addition, the integration of these metrics into existing protocols and standards requires substantial modification of conventional network architectures. Successful implementations have employed machine learning techniques, adaptive sampling, and context-aware scheduling to optimize performance under specific application constraints.

Future research in time-oriented information metrics should focus on several promising directions \cite{Yates2021JSAC, Maatouk2023TWC, chiariotti2025ARXIV}. The integration of semantic awareness with temporal freshness metrics has great potential for next-generation systems that enable goal-oriented communication. This approach prioritizes information value rather than raw data transmission. Cross-layer optimization approaches that jointly consider sensing, transmission, and computation are another critical research area, particularly for edge computing with heterogeneous devices and varying resource constraints. Furthermore, the development of standardized testing methodologies and benchmarks would facilitate fair comparison of different optimization strategies across diverse application scenarios. As networked systems continue to grow in complexity and scale, these time-oriented metrics will remain essential measures for reliable, efficient, and timely information delivery.

\section{Quality/Utility-oriented Information Metrics}\label{Quality}
\begin{figure}[h]
    \centering
    \includegraphics[width=1\linewidth]{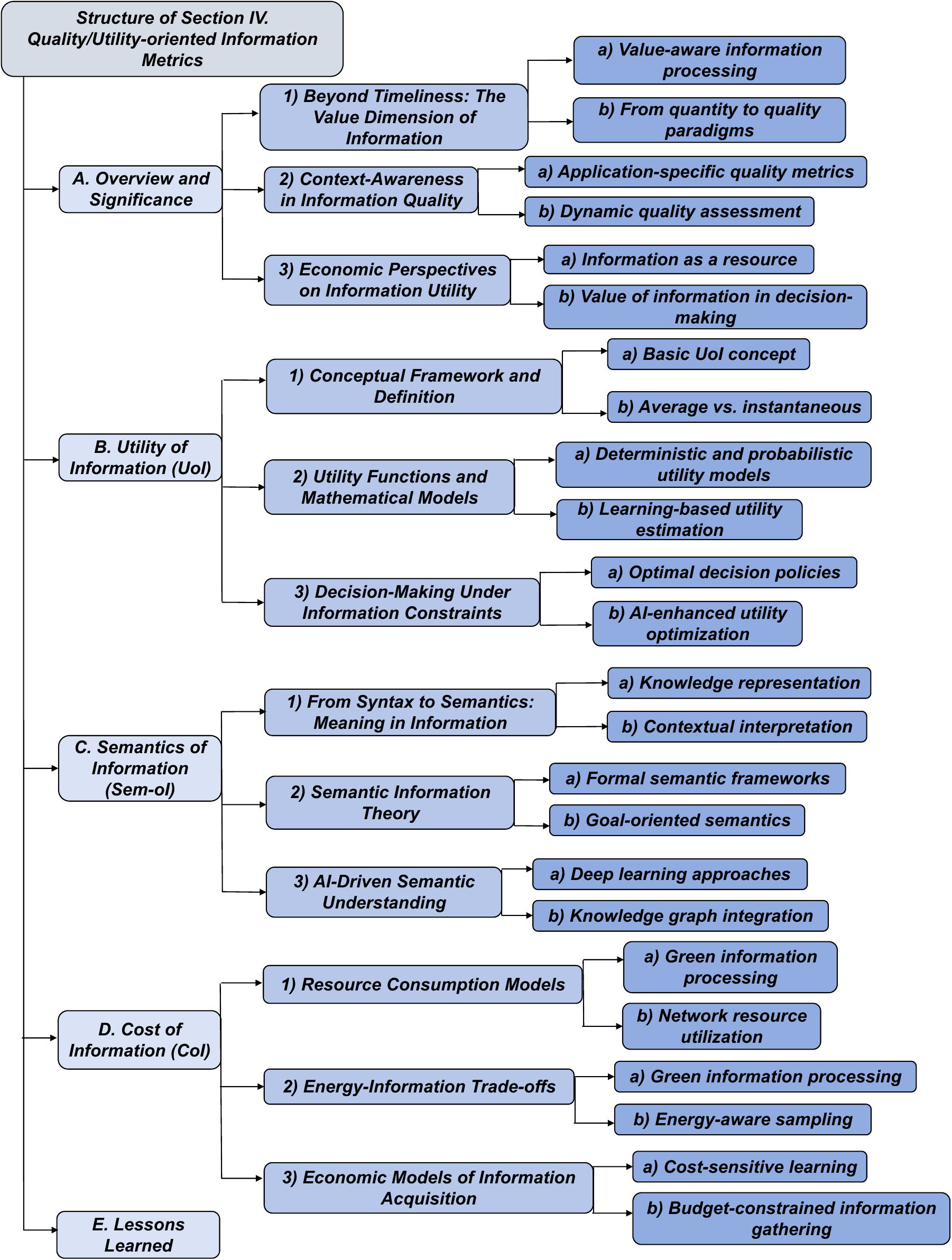}
    \caption{The roadmap of Section IV.}
    \label{fig:RMIV}
\end{figure}

\begin{figure}[ht]
    \centering
    \includegraphics[width=0.9\linewidth]{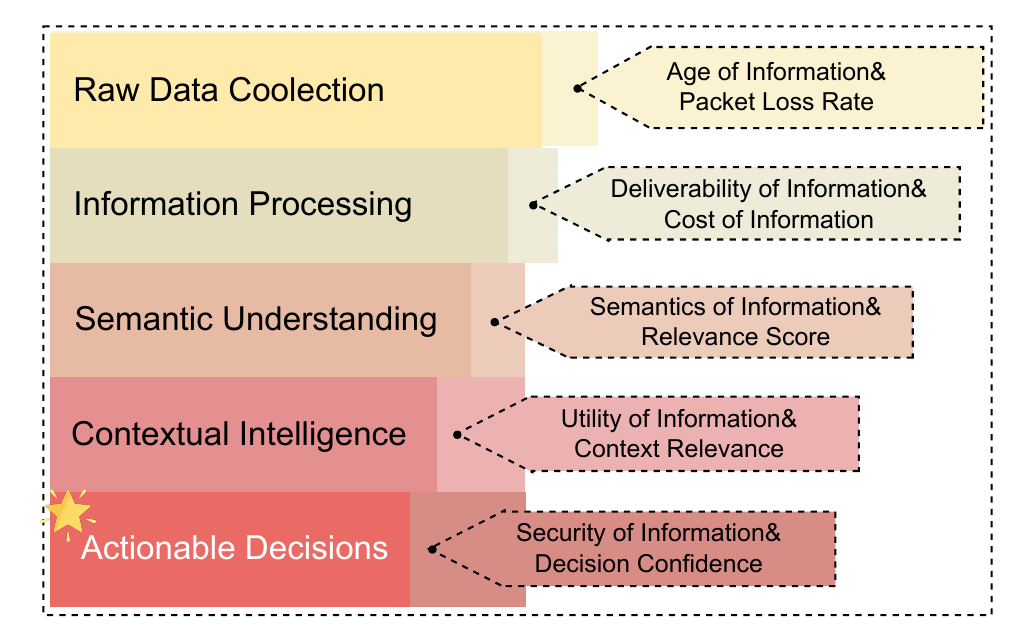}
    \caption{Information value enhancement chain in next-generation networks.}
    \label{fig:information_value_chain}
\end{figure}

This section provides a comprehensive framework for quality/utility-oriented information metrics that focus on their fundamental principles, contextual dimensions, and economic perspectives. Metrics such as Utility of Information (UoI), Semantics of Information (Sem-oI), and Cost of Information (CoI) are systematically explored, each that illustrates unique roles in information value optimization across diverse application scenarios and constrained resource environments. Fig.~\ref{fig:information_value_chain} illustrates the hierarchical value enhancement process that provides the theoretical foundation for understanding how these metrics contribute to transforming raw data into actionable intelligence through AI-driven processing stages. The detailed roadmap of Section \ref{Quality} is illustrated in Fig. \ref{fig:RMIV}.

\subsection{Overview and Significance}
Quality and utility-oriented metrics exceed traditional time-based evaluation and quantify the real-world value and practical usefulness of information in decision-making contexts. These metrics provide indispensable insights for next-generation information systems optimization, exceptionally where resources are constrained and contextual relevance directly influences effectiveness.

These metrics fundamentally answer the question: "What is the actual value and meaning of the information we process?" Unlike time-oriented metrics that focus on freshness, quality-oriented metrics evaluate the worth and applicability of information. UoI quantifies the practical decision-making value of information for achieving specific goals, measured through utility functions that assess improvement in decision outcomes. Sem-oI evaluates the meaning and contextual relevance of information content, moving beyond syntax to assess interpretability and goal alignment. CoI measures the economic and resource expenditures associated with information acquisition, processing, and utilization throughout its lifecycle.

These metrics have critical applications in federated learning systems, where UoI guides client selection for model training efficiency. Autonomous systems employ Sem-oI for goal-oriented communication, transmitting only task-relevant data. Energy-constrained Internet of Things (IoT) networks utilise CoI frameworks to balance information value against power consumption. Practical implementation involves developing utility assessment algorithms, deploying semantic filtering mechanisms, and establishing resource accounting systems that collectively enable networks to prioritize valuable information while optimizing resource allocation under operational constraints.

\subsubsection{Beyond Timeliness: The Value Dimension of Information}
Information value transcends mere timeliness. Recent advancements emphasize the integration of context and utility into metric frameworks, which represents a fundamental transformation toward value-oriented evaluation approaches within next-generation networks.

\paragraph{Value-aware information processing}
Value-aware information processing frameworks are critical for informed decision-making in environments that are characterized by resource constraints. Early theoretical foundations that were laid by Howard \cite{Howard1966TSSC} established that numerical values assigned to uncertainty reductions directly enhance decision-making efficacy and formed a cornerstone for subsequent research. Simon \textit{et al.} \cite{Simon2023CM} proposed a techno-economic multi-stakeholder modeling framework that explicitly embeds stakeholder preferences into utility functions, which bridges economic theories and practical decision-making scenarios. Wang \textit{et al.} \cite{Wang2023TFS} developed a sophisticated framework that converts heterogeneous information into structured interval-valued belief models and enriched utility in complex, multi-criteria environments. Zhan \textit{et al.} \cite{Zhan2023CAA} introduced a behavioral decision-making model for hesitant fuzzy systems that explicitly captures human cognitive and social decision dynamics under uncertainty. Ferguson \textit{et al.} \cite{Ferguson2023TCSS} addressed network-wide implications of information valuation and employed incentive mechanisms that are designed for congestion networks that quantify strategic informational value under worst-case scenarios, which established the broader applicability and robustness of value-oriented metrics.

\paragraph{From quantity to quality paradigms}
The paradigm shift from data volume prioritization to information quality emphasis underscores a fundamental transformation in networked system operations and reinforces the importance of actionable and relevant information over mere data abundance. Howard's seminal work \cite{Howard1966TSSC} articulated the critical interdependence between information quality and decision outcomes and argued that pure probability-based metrics prove insufficient without accounting for decision consequences. Wang \textit{et al.} \cite{Wang2023TFS} provided practical mechanisms for multi-granular linguistic and heterogeneous data integration and enhanced decision quality in complex, criteria-rich scenarios. Zhan \textit{et al.} \cite{Zhan2023CAA} illustrated the practical benefits of this evolution toward quality-centric paradigms and demonstrated how hesitant fuzzy information models effectively represent uncertainties when precise membership values are difficult to determine, which enhanced decision-making quality and robustness.

\subsubsection{Context-Awareness in Information Quality}
Information relevance varies inherently across different contexts. Quality metrics must dynamically adapt to specific environmental and application requirements. Unlike universally applicable time-oriented metrics, these adaptive approaches ensure meaningful evaluations that are customized for diverse operational contexts.

\paragraph{Application-specific quality metrics}
Recent studies highlight the emergence of context-aware metrics that are specifically designed for distinct application scenarios and enhance utility in diverse network environments. Liu \textit{et al.} \cite{Liu2023TII} established a learning-based system for advanced manufacturing and employed bandit learning techniques to integrate contextual insights into quality prediction tasks effectively. Zhang \textit{et al.} \cite{Zhang2023TKDE} emphasized the indispensable role of local context through their novel sequence labeling framework (ConPhrase) and improved phrase mining quality. Sang \textit{et al.} \cite{Sang2025TIP} introduced an energy-adaptive perceptual model that is specifically designed for quality assessment of point clouds and emphasized the interplay between content diversity and perceptual distortions. Zhou \textit{et al.} \cite{Zhou2023TNSM} created a forecasting model that integrates spatio-temporal contextual data and outperformed traditional QoS prediction techniques. Staelens \textit{et al.} \cite{Staelens2014TBC} provided empirical insights into subjective perceptions of video quality and highlighted that user evaluations are heavily influenced by contextual factors like content dynamics and segment quality variability, which reinforced the need for context-specific evaluations.

\paragraph{Dynamic quality assessment}
Dynamic quality assessment methodologies adaptively respond to evolving environmental conditions and changing user needs and ensure consistent utility across operational contexts. Zhou \textit{et al.} \cite{Zhou2023TNSM} introduced a dynamic similarity measure and incorporated real-time QoS fluctuations into adaptive quality evaluations, which effectively leveraged spatio-temporal context to enhance predictive accuracy. Zhu \textit{et al.} \cite{Zhu2025TCSVT} developed specialized dynamic convolutional architectures for underwater image assessments and adapted kernel structures that are uniquely suited to each image's intrinsic content and distortions. Staelens \textit{et al.} \cite{Staelens2014TBC} highlighted dynamic quality perception in video playback scenarios, where quality perceptions shift with content motion intensity and segment quality changes, which illustrated the importance of dynamic quality assessment adjustments based on real-time contextual feedback.

\subsubsection{Economic Perspectives on Information Utility}
Economic frameworks critically inform information valuation and furnish methodologies for cost and benefit quantification associated with information acquisition, processing, and utilization under stringent resource constraints.

\paragraph{Information as a resource}
The conceptualization of information as a valuable resource has fundamentally reshaped network resource management strategies. Esposito \textit{et al.} \cite{Esposito2025IOTJ} established a practical IoT system that is customized for earthquake early warning scenarios and reduced computational overhead and improved efficiency on resource-limited devices. Jia \textit{et al.} \cite{Jia2024MNET} expanded resource utilization principles into pervasive intelligence frameworks within integrated space-air-ground networks and showcased enhanced computational and communication efficiency for microservices. Rehman \textit{et al.} \cite{Rehman2025IOTJ} integrated adaptive security measures into resource-constrained IoT environments and reduced computational burdens and improved energy efficiency, which exemplified optimal resource-aware information management.

\paragraph{Value of information in decision-making}
The relationship between information quality and decision-making effectiveness remains a pivotal theme in contemporary network design and is explicitly addressed through innovative methodologies. Howard \cite{Howard1966TSSC} articulated the economic considerations that are critical to information value quantification in decision-making contexts. Ferguson \textit{et al.} \cite{Ferguson2023TCSS} leveraged informational insights to optimize societal-level network behaviors through customized incentive mechanisms. Zou \textit{et al.} \cite{Zou2025TITS} developed a ``Cognitive Tree" prioritization framework for autonomous driving and improved decision outcomes by critical environmental factor emphasis. Lin \textit{et al.} \cite{Lin2025TAI} introduced privacy-preserving recommendation methods and optimized decision utility by balancing information quality, privacy protection, and computational efficiency.

\begin{figure*}[ht]
    \centering
    \includegraphics[width=0.9\textwidth]{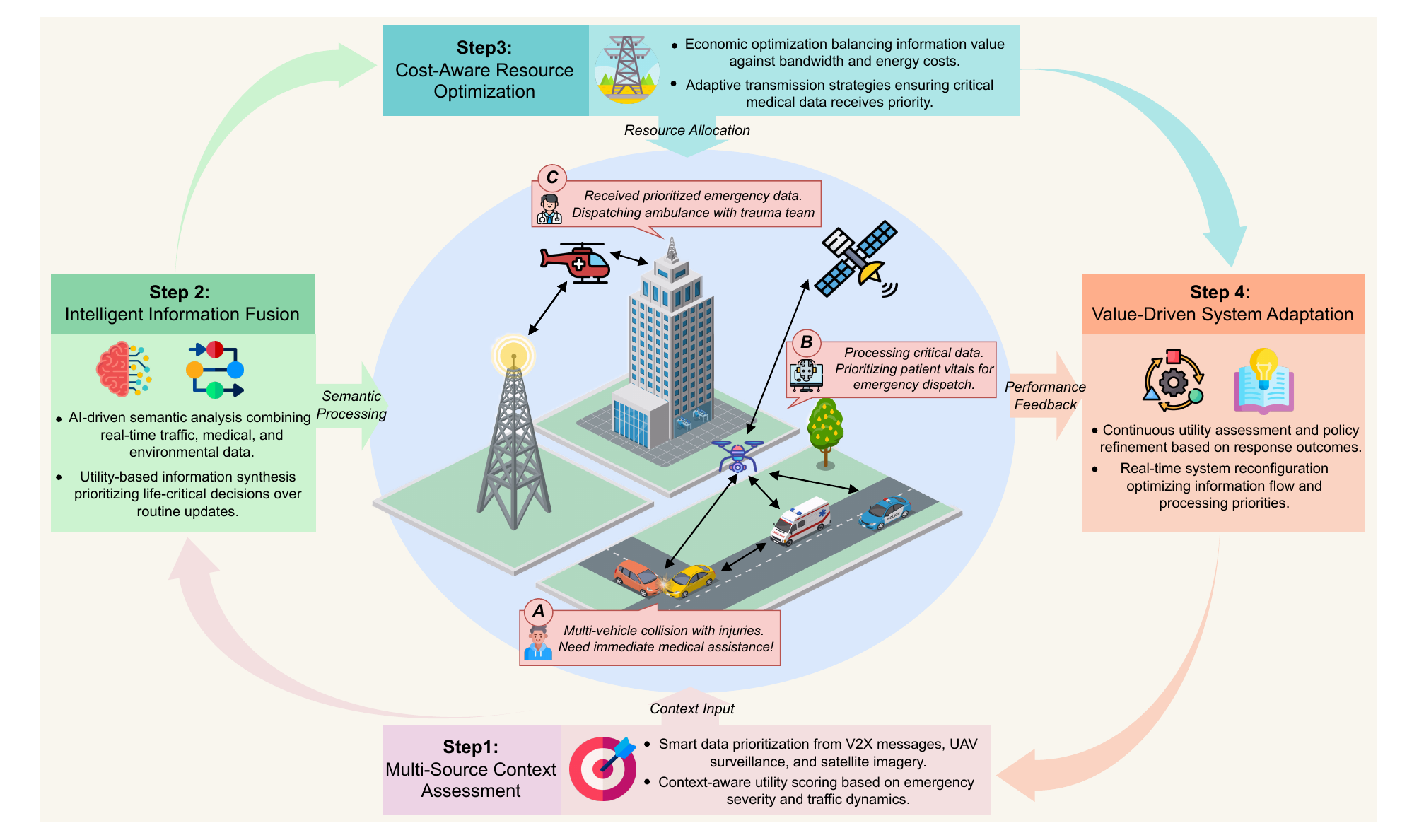}
     \caption{AI-enhanced quality/utility-oriented information metrics optimization framework for emergency medical response systems. The framework demonstrates a four-step process: (1) Multi-source context assessment prioritizing V2X messages, UAV surveillance, and satellite imagery; (2) Intelligent information fusion combining real-time data through semantic analysis; (3) Cost-aware resource optimization balancing information value against bandwidth and energy costs; (4) Value-driven system adaptation providing continuous utility assessment and policy refinement based on emergency response outcomes.}
     \label{fig:quality_utility_framework}
\end{figure*}

As illustrated in Fig.~\ref{fig:quality_utility_framework}, the integration of value-aware processing, context-awareness, and economic perspectives culminates in comprehensive optimization frameworks for emergency response systems. This AI-enhanced framework demonstrates how quality and utility-oriented metrics, including UoI, Sem-oI, and CoI, work synergistically to transform raw emergency data into actionable intelligence and prioritize life-critical decisions while resource allocation optimization and cost-effective information transmission assurance.

\begin{table*}[!t]
\centering
\caption{Information metrics optimization methods and performance.}
\label{tab:metrics_optimization}
\renewcommand{\arraystretch}{1}
\begin{tabular}{>{\centering\arraybackslash}p{2.8cm}>{\centering\arraybackslash}p{2.2cm}>{\centering\arraybackslash}p{0.8cm}>{\centering\arraybackslash}p{2.8cm}>{\centering\arraybackslash}p{3.2cm}>{\centering\arraybackslash}p{3.2cm}}
\toprule
\textbf{Metric Category} & \textbf{Specific Metric} & \textbf{Paper} & \textbf{Application Scenario} & \textbf{Optimization Method} & \textbf{Key Results/Performance} \\
\midrule
\multirow{6}{2.8cm}{\centering\textbf{Time-oriented}} & \multirow{2}{1.8cm}{\centering AoI} & \cite{Fang2022JSAC} & B5G massive energy harvesting networks & Sleep-scheduling with linear search and CCP & Lower peak AoI, reduced power consumption \\
\cmidrule(l){3-6}
& & \cite{Yin2021VTC} & 6G multiple-UAV networks & RL-based trajectory design with AoI reward & 21.7\% and 7.7\% AoI performance gains \\
\cmidrule(l){2-6}
& \multirow{2}{1.8cm}{\centering PAoI} & \cite{Khorsandmanesh2021TCCN} & Wireless powered IoT networks & Decode/amplify and forward relaying & Improved age metrics, lower average PAoI \\
\cmidrule(l){3-6}
& & \cite{Jiang2021ICC} & UAV-relayed wireless sensor networks & Successive convex approximation with SOCP & Better convergence, much smaller peak AoI \\
\cmidrule(l){2-6}
& \multirow{2}{1.8cm}{\centering AoII} & \cite{Zhang2024ICC} & UAV-aided vehicular Metaverse systems & Proximal policy optimization with deep RL & Minimized expected long-term AoII \\
\cmidrule(l){3-6}
& & \cite{Chen2023JSTSP} & Multi-user uplink NOMA for XR & Linear search algorithm with DeepSC & Beneficial error and AoI performance evaluation \\
\midrule
\multirow{6}{2.8cm}{\centering\textbf{Quality/Utility-\\oriented}} & \multirow{2}{1.8cm}{\centering UoI} & \cite{Hu2024TMC} & Federated edge learning with autonomous clients & Bayesian game with AutoFL framework & 15.5-24.5\% time, 82.2-86.8\% energy savings \\
\cmidrule(l){3-6}
& & \cite{NgoICDCS} & Distributed hierarchical edge IoT systems & Contextual bandit with RL policy network & 71.4\% detection delay reduction \\
\cmidrule(l){2-6}
& \multirow{2}{1.8cm}{\centering Sem-oI} & \cite{Wu2025TII} & Industrial knowledge graph construction & Concept-aware entity alignment network & Outperforms SOTA alignment methods \\
\cmidrule(l){3-6}
& & \cite{Wang2024TNNLS} & Weakly supervised semantic segmentation & WS-FCN with FCA and SF2 modules & 65.02\% and 64.22\% mIoU on PASCAL VOC \\
\cmidrule(l){2-6}
& \multirow{2}{1.8cm}{\centering CoI} & \cite{Agarwal2024TGCN} & IoT-enabled WSNs for real-time queries & Mobile sinks with set cover optimization & 41.26\% energy, 39.84\% delay improvements \\
\cmidrule(l){3-6}
& & \cite{Wang2023TSMC} & Cost-sensitive learning with missing data & CALS with softmax regression strategy & Superior to existing active learning methods \\
\midrule
\multirow{6}{2.8cm}{\centering\textbf{Reliability/\\Robustness-\\oriented}} & \multirow{2}{1.8cm}{\centering Sec-oI} & \cite{Adil2023TITS} & UAV-aided IoT systems security threats & Comprehensive survey with comparative analysis & Theoretical taxonomy and countermeasures \\
\cmidrule(l){3-6}
& & \cite{Kumar2023TITS} & IoT-enabled maritime transportation systems & DLTIF with DFE, CTIDD, and CTIATI schemes & Up to 99\% threat detection accuracy \\
\cmidrule(l){2-6}
& \multirow{2}{1.8cm}{\centering EoI} & \cite{Chen2024TIT} & Finite-state Markov sources scheduling & Restless bandit with universal index policy & Near-optimal optimal performance \\
\cmidrule(l){3-6}
& & \cite{Li2024Neuro} & CNN deployment on edge devices & Graph-based pruning with k-core decomposition & 49.39\% FLOPs reduction, 0.01\% accuracy loss \\
\cmidrule(l){2-6}
& \multirow{2}{1.8cm}{\centering SuI} & \cite{Alam2016TNSM} & Network virtualization with failure tolerance & SiMPLE multi-path link embedding & Near-optimal survivability results \\
\cmidrule(l){3-6}
& & \cite{Rubin2009ADH} & Mobile ad hoc wireless networks & RFAR with robust throughput measure & Enhanced route lifetime and reliability \\
\midrule
\multirow{6}{2.8cm}{\centering\textbf{Network/\\Communication-\\oriented}} & \multirow{2}{1.8cm}{\centering DoI} & \cite{ArafatJSEN} & Heterogeneous healthcare systems & ARMR with link lifetime and adaptive Q-learning & Improved PDR, E2E delay, network lifetime \\
\cmidrule(l){3-6}
& & \cite{Sourlas2018TNSM} & Dynamic fragmented ICN/NDN networks & Satisfied interest table information resilience & Serves most requests after disruptive events \\
\cmidrule(l){2-6}
& \multirow{2}{1.8cm}{\centering RoI} & \cite{Zhang2015ICCC} & CCN high-bandwidth file sharing networks & LT code aided multipath forwarding & Reduced download time, improved load balance \\
\cmidrule(l){3-6}
& & \cite{Ramasubramonian2010LSP} & Video multicast in lossless networks & Network coding with multiple description & Significant control over rate allocation \\
\cmidrule(l){2-6}
& \multirow{2}{1.8cm}{\centering RelI} & \cite{Zhao2024TASE} & Dynamic job shop with random arrivals & PPO with attention-based policy network & Better performance and training efficiency \\
\cmidrule(l){3-6}
& & \cite{Liu2006TIT} & Wireless network downlink scheduling & U/R-rule gradient-based algorithm & Optimal in many TDM-type capacity cases \\
\bottomrule
\end{tabular}
\end{table*}
\subsection{Utility of Information (UoI)}

We provide an in-depth exploration of the Utility of Information (UoI) that details its conceptual foundations, mathematical frameworks, and practical decision-making applications. UoI quantifies the tangible value of information and facilitates optimal resource allocation and enhances decision quality across diverse networked scenarios.

\subsubsection{Conceptual Framework and Definition}

The Utility of Information explicitly links the value of information to improved decision outcomes and changes the focus from mere availability or freshness to the practical impacts on decision-making effectiveness.

\paragraph{Basic UoI concept}
The foundational work by Howard \cite{Howard1966TSSC} introduced a seminal concept whereby uncertainty reduction is quantitatively tied to decision improvement and integrated probabilistic and economic considerations. Recent developments have expanded on this by introducing privacy-aware information utility models, such as Mutual Information Autoencoders, achieving precise trade-offs between utility and privacy through rate-distortion optimization \cite{Yang2024TIFS}. Moreover, sophisticated theoretical advancements have emerged, including controlled privacy leakage frameworks \cite{Zamani2024TIT} and strategic models for multiple information providers interacting with decision-makers, which highlight competitive dynamics that benefit the decision receiver \cite{Velicheti2025TAC}. Complementing these technical frameworks, Zhan \textit{et al.} \cite{Zhan2023CAA} have integrated behavioral insights into utility assessment and illustrated how cognitive factors shape information valuation under uncertainty.

\paragraph{Average vs. instantaneous}
UoI metrics can be evaluated either instantaneously or averaged over time, each that offers unique insights for decision-making scenarios. Alawad \textit{et al.} \cite{Alawad2022JSEN} differentiate observed instantaneous value from expected long-term utility and emphasize distinct application requirements from immediate response systems to strategic planning contexts. Further enriching this discussion, Chen \textit{et al.} \cite{Chen2024TIT} develop an entropy-based framework to quantify information uncertainty dynamically and illustrated how instantaneous observations and long-term averages are both critical in comprehensive decision support systems.

\subsubsection{Utility Functions and Mathematical Models}

Formal utility functions and mathematical modeling offer structured approaches to accurately quantify information value across multiple contexts and operational constraints.

\paragraph{Deterministic and probabilistic utility models}
Utility models have transitioned from initial deterministic frameworks to sophisticated probabilistic approaches that address real-world complexities and uncertainties. Arrigoni \textit{et al.} \cite{Arrigoni2023TON} enhanced deterministic models by introducing Bayesian frameworks that progressively optimize information selection based on expected utility. Zhang \textit{et al.} \cite{Zhang2024LCOMM} further advanced this by employing fuzzy probabilistic modeling to address uncertainty in dynamic channel state information, which significantly outperformed traditional deterministic approaches. Chen \textit{et al.} \cite{Chen2024TIT} contributed an entropy-based universal index policy applicable to restless multi-armed bandit problems and highlighted the effectiveness of probabilistic utility approaches. Additionally, Hu \textit{et al.} \cite{Hu2024TMC} illustrated practical benefits of Bayesian game-theoretic modeling and showcased higher efficiency and convergence to stable equilibria in federated learning contexts.

\paragraph{Learning-based utility estimation}
Learning-based techniques have revolutionized utility estimation and enable adaptive, context-aware valuation across dynamic environments. Early frameworks like the contextual-bandit anomaly detection system by Ngo \textit{et al.} \cite{NgoICDCS} reduced detection delays significantly through adaptive modeling strategies. Building on these advancements, Bongole \textit{et al.} \cite{Bongole2025ICASSP} established theoretical foundations for robust decision making using reinforcement learning with proven minimax optimality. Further practical implementations, such as the Adaptive Contextual Bandit algorithm by Ghosh \textit{et al.} \cite{Ghosh2024TIT}, dynamically refine model selection to match the true underlying environments effectively. Similarly, Li \textit{et al.} \cite{Li2017LCOMM} demonstrated stable relay selection in challenging underwater acoustic networks through minimal contextual information and underscored learning-based methods' practical utility.

\subsubsection{Decision-Making Under Information Constraints}

The application of UoI principles practically involves sophisticated decision frameworks capable of balancing informational value against resource and operational limitations.

\paragraph{Optimal decision policies}
Recent frameworks, such as the three-way group decision method by Pan \textit{et al.} \cite{Pan2024TFS}, incorporated regret theory and consensus-reaching processes to handle uncertainty and decision ambiguity effectively. Further extending these concepts, Xiao \textit{et al.} \cite{Xiao2024TETCI} introduced sequential multi-scale decision systems and optimized decision consistency through comprehensive matrix-theory-based evaluations. In parallel, Xiao \textit{et al.} \cite{Xiao2024TFS} integrated prospect theory and fuzzy modeling to capture psychological influences in decision-making and demonstrated improved stability and practical effectiveness in real-world applications.

\paragraph{AI-enhanced utility optimization}
Artificial intelligence has significantly advanced utility optimization methodologies and enables dynamic decision-making under diverse operational constraints. Contextual-bandit-based approaches, as demonstrated by Ngo \textit{et al.} \cite{NgoICDCS}, allow adaptive model selection for IoT and edge environments and considerably outperform traditional static methods. Further theoretical enhancements, including minimax regret bounds for reinforcement learning by Bongole \textit{et al.} \cite{Bongole2025ICASSP}, ensure robust policy formation in uncertain environments. Practical algorithmic refinements by Ghosh \textit{et al.} \cite{Ghosh2024TIT} and sophisticated self-learning frameworks by Xiao \textit{et al.} \cite{Xiao2024TETCI} demonstrate how AI transforms utility optimization and significantly enhance system resilience and adaptability to complex information constraints.

\subsection{Semantics of Information (Sem-oI)}

This subsection provides an in-depth examination of Sem-oI, which extends beyond structural encoding and temporal ordering to explore how meaning is extracted, represented, and leveraged in intelligent networked systems. When systems move from transmission focus to interpretation, Sem-oI forms the foundation for decision-making systems that value not just data presence, but its purpose and relevance.

\subsubsection{From Syntax to Semantics: Meaning in Information}

The transformation of raw symbols into actionable knowledge requires semantic models that comprehend what the data conveys and why it matters. This evolution from syntax to semantics proves fundamental for systems that must adapt to contextual signals and make autonomous decisions.

\paragraph{Knowledge representation}
The progression of knowledge representation techniques shows an expanding ability to encode complex semantic relationships. Early frameworks often relied on rigid data schemas, but recent innovations have introduced more expressive and adaptable models. Wu \textit{et al.}~\cite{Wu2025TII} addressed the challenge of conceptual inconsistency in large-scale industrial systems by developing concept-aware entity alignment networks that enable consistent semantic mapping across varying granularities. Building on this, Liang \textit{et al.}~\cite{Liang2023ICDE} proposed interpretable neural encoders for different knowledge graph modalities, including multimodal and uncertain contexts, which enhanced both scalability and semantic precision. Temporal dimensions were further incorporated by Qian \textit{et al.}~\cite{Qian2025TNNLS}, who introduced distillation-based mutual learning strategies for time-aware knowledge graphs and enabled semantic transfer across models while preserving temporal dependencies. These advancements collectively establish a flexible infrastructure capable of embedding, transferring, and reasoning over rich semantic structures in real-time applications.

\paragraph{Contextual interpretation}
Just as language meaning depends on context, UoI is shaped by the environment in which it is interpreted. Ran \textit{et al.}~\cite{Ran2015ICIP} established that conventional semantic segmentation methods often neglect relational cues, which results in disjointed interpretation. Addressing this limitation, Wang \textit{et al.}~\cite{Wang2024TNNLS} introduced a dual-coupling network that integrates global scene understanding with fine-grained local features and substantially improved semantic continuity and object boundary awareness under weak supervision. This approach mimics human perception by dynamically adjusting interpretations based on multiscale contextual feedback. As intelligent systems operate in increasingly dynamic and heterogeneous environments, such context-aware interpretation mechanisms prove crucial to ensure that semantic understanding remains both accurate and adaptive.

\subsubsection{Semantic Information Theory}

The shift from data fidelity to semantic relevance has motivated new theoretical models that quantify information meaning itself. \emph{Semantic Information Theory} seeks to formalize not just how much information is conveyed, but how valuable that information is for a particular objective.

\paragraph{Formal semantic frameworks}
Traditional information theory fails to distinguish between meaningful content and syntactic noise and assigns equal weight to both if their entropy is the same. In response, Yang \textit{et al.}~\cite{Yang2024TIFS} introduced privacy-utility trade-offs using rate-distortion theory to encode semantic value in a privacy-preserving manner. Building on this idea, Zamani \textit{et al.}~\cite{Zamani2024TIT} developed bounded leakage models that regulate how much semantically sensitive content may be disclosed under utility constraints. Velicheti \textit{et al.}~\cite{Velicheti2025TAC} further advanced this line of inquiry through a semantic game-theoretic framework and showed how sender competition under utility-aware objectives can increase informational value to receivers. Together, these works lay the mathematical foundations for measuring, transmitting, and optimizing meaning in networked decision systems.

\paragraph{Goal-oriented semantics}
Beyond formalism, the true impact of semantics lies in its ability to serve goals. Information becomes valuable only when it contributes to decision-making or action. Fu \textit{et al.}~\cite{Fu2023INFOCOM} introduced task-driven communication frameworks that extract and transmit only the content relevant to the end objective and achieved substantial reductions in overhead. Chen \textit{et al.}~\cite{Chen2024GLOBECOM} established that by focusing on task-critical semantics, communication cost can be cut by up to 44\% in robotic systems without compromising accuracy. Wu \textit{et al.}~\cite{Wu2024TWC} integrated this perspective with timeliness and proposed a hybrid metric that combines semantic relevance with AoI to prioritize urgent and meaningful content. Wang \textit{et al.}~\cite{Wang2023ICDCS} expanded this framework to complex environments such as 6G networks and metaverse platforms and highlighted how semantic prioritization enables more efficient and context-aware system behavior.

\subsubsection{AI-Driven Semantic Understanding}

Recent advances in AI have empowered systems to learn, adapt, and reason about meaning without explicit rules. Semantic extraction is no longer confined to pre-engineered features but emerges through learning from context, data distribution, and task performance.

\paragraph{Deep learning approaches}
Alsmadi \textit{et al.}~\cite{Alsmadi2023TCSS} established how hierarchical neural networks can embed semantic meaning at various levels and improved generalization and robustness in noisy environments. Wang \textit{et al.}~\cite{Wang2025TMM} enhanced semantic segmentation models through novel boundary-refinement modules and global attention layers, which allowed for precise interpretation even in visually ambiguous scenes. Meanwhile, Hu \textit{et al.}~\cite{Hu2024IOTJ} extended these capabilities to edge computing and deployed lightweight convolutional networks capable of real-time semantic inference in bandwidth- and energy-constrained environments. These deep models facilitate adaptive and distributed semantic services and transform how meaning is accessed and applied across devices.

\paragraph{Knowledge graph integration}
Where deep learning captures distributional semantics, knowledge graphs encode structural and relational meaning. Wu \textit{et al.}~\cite{Wu2025TII} advanced the alignment of heterogeneous data sources through concept-aware embeddings and created scalable knowledge representations. Liang \textit{et al.}~\cite{Liang2023ICDE} built interpretable neural graph models that handle uncertainty and multimodality and enabled semantic fusion across diverse information types. Qian \textit{et al.}~\cite{Qian2025TNNLS} contributed distillation-based mutual learning frameworks that allow semantic models to be compressed and transferred without performance degradation and ensured deployment even in resource-constrained nodes. This convergence of symbolic structure and sub-symbolic learning establishes a foundation for intelligent systems capable of robust, real-time semantic reasoning across domains.
\subsection{Cost of Information (CoI)}

The increasing complexity of networked systems has heightened awareness that information acquisition, processing, and utilization consume valuable resources. The CoI framework addresses this fundamental trade-off by quantifying resource expenditures associated with information management throughout its lifecycle. We examine the multidimensional nature of information costs and their implications for next-generation intelligent networks.

\subsubsection{Resource Consumption Models}

As networked systems become more sophisticated, the understanding and optimization of resource consumption has evolved from simple hardware efficiency to comprehensive sustainability frameworks.

\paragraph{Green information processing}
The environmental impact of digital infrastructure has prompted paradigm shifts in how information systems manage their ecological footprint. Early approaches to green computing focused primarily on hardware-level optimizations, but this narrow perspective failed to capture the relationship between information value and energy expenditure. Faist \textit{et al.}~\cite{Faist2015NC} established foundational principles by demonstrating that irreversible information processing inevitably incurs thermodynamical work costs and formulated the minimal work required for any logical process as the entropy of discarded information conditional to computation output. This thermodynamic principle created a crucial bridge between physical energy constraints and information theory. Based on this foundation, Wang \textit{et al.}~\cite{Wang2022TMC} addressed practical implementation challenges in mobile ad hoc networks (MANETs) by developing efficient data acquisition solutions that reduce information retrieval costs while improving success rates through innovative approaches to path continuity. When these principles were extended to wireless networks, Nasir \textit{et al.}~\cite{Nasir2013TWC} introduced time-switching and power-splitting relaying protocols that enable energy-constrained nodes to harvest energy and process information simultaneously and demonstrated how architectural innovations can transform energy constraints into operational advantages.

\paragraph{Network resource utilization}
Beyond processing costs, the transmission and dissemination of information represent significant expenditures in networked systems. The evolution of efficient resource utilization frameworks has progressed from static allocation schemes to dynamic, context-aware approaches. Su \textit{et al.}~\cite{Su2023TNSM} introduced Q-learning-based routing techniques that enable wireless sensors to adaptively select suitable neighboring nodes for energy-efficient information transmission in a decentralized manner. This approach demonstrated how intelligent decision-making can reduce and balance energy consumption while extending network lifetime. To advance resource optimization further, Agarwal \textit{et al.}~\cite{Agarwal2024TGCN} introduced mobile sink-based query processing for IoT-enabled Wireless Sensor Networks and achieved remarkable improvements in average energy consumption (41.26\%), query processing delay (39.84\%), and network lifetime (39.74\%) through optimal rendezvous point identification and mobile sink selection. For more complex energy management scenarios, Li \textit{et al.}~\cite{Li2024TII} developed data-stream-driven distributionally robust control frameworks for hydrogen-based multimicrogrids that accommodate uncertainty streams from renewable energy sources and established how advanced predictive models can enhance resource utilization under fluctuating conditions.

\subsubsection{Energy-Information Trade-offs}

The fundamental tension between information quality and energy expenditure requires sophisticated optimization frameworks that balance these competing objectives.

\paragraph{Green information processing}
The intricate relationship between energy consumption and information processing has evolved from isolated optimization to holistic frameworks that consider both dimensions simultaneously. Wu \textit{et al.}~\cite{Wu2013ECC} established a foundational connection between energy and information by drawing from behavioral ecology and modeled information quality as potential energy within a mechanical framework. This innovative perspective allowed the formulation of optimal path problems that increase information quality while reducing kinetic energy, with solutions that satisfy Euler-Lagrange equations from classical mechanics. This mathematical foundation enabled the development of energy-efficient information gathering strategies for mobile agents in target localization. When this concept was advanced to dynamic environments, Liu \textit{et al.}~\cite{Liu2025TWC} introduced a UAV-assisted integrated sensing, calculation, and communication system that jointly optimizes sensing scheduling, transmit power, and motion parameters to maximize sensing data while minimizing Age of Information under energy constraints. This integration of information freshness metrics with energy considerations demonstrates how advanced multi-objective optimization can balance competing requirements in complex cyber-physical systems.

\paragraph{Energy-aware sampling}
The strategic acquisition of information under energy constraints represents a critical challenge in resource-limited environments. Su \textit{et al.}~\cite{Su2020IOTJ} developed an innovative UAV-assisted wireless charging approach for energy-constrained IoT devices and employed bipartite matching with one-sided preferences to model charging relationships. By formulating the problem through multiple-stage dynamic matching and through Markov decision processes, their solution optimized charging energy transfer while respecting UAV constraints. This approach exemplifies how energy-aware information gathering systems can adapt to dynamic conditions through principled decision-making frameworks. For energy-harvesting networks, Nasir \textit{et al.}~\cite{Nasir2013TWC} demonstrated that time switching-based relaying outperforms power splitting-based approaches at low signal-to-noise ratios and provided essential guidance for protocol selection in energy-constrained scenarios. These innovations together establish a path toward intelligent sampling methods that maximize information value while operating within strict energy budgets.

\subsubsection{Economic Models of Information Acquisition}

Beyond physical resources, information acquisition carries economic costs that must be systematically quantified and optimized for rational decision-making.

\paragraph{Cost-sensitive learning}
The economic dimension of information acquisition is particularly evident in learning systems that must balance information value against acquisition costs. Wang \textit{et al.}~\cite{Wang2023TSMC} introduced a comprehensive cost-sensitive active learning framework for incomplete data that unifies the evaluation of attribute values and labels through softmax regression. Their approach addresses the practical scenario where obtaining certain values and labels incurs varying costs and enables optimal acquisition through permutation and greedy strategies. When these principles were extended to specialized domains, Wan \textit{et al.}~\cite{Wan2024TITS} developed a cost-sensitive graph convolutional network with self-paced learning for hit-and-run crash analysis and demonstrated how semi-supervised approaches can effectively handle both missing label and class imbalance issues. For industrial applications, Deng \textit{et al.}~\cite{Deng2024IOTJ} proposed a knowledge distillation-guided cost-sensitive ensemble learning framework that combines multiple learning paradigms to address imbalanced fault diagnosis. These diverse applications demonstrate how cost-sensitive approaches can enhance learning outcomes across domains by explicitly accounting for the economic dimensions of information acquisition.

\paragraph{Budget-constrained information gathering}
The optimization of information acquisition under budget constraints represents a fundamental challenge for resource-limited systems. Baronov \textit{et al.}~\cite{Baronov2012JPROC} established theoretical foundations for rapid information acquisition in the reconnaissance of random fields and developed a mathematical framework that quantifies the relationship between discovered and undiscovered information. Their approach treats search as an optimal information acquisition problem and enables precise definitions of conservative and aggressive search strategies based on information metric change rates. In networked environments, Khandaker \textit{et al.}~\cite{Khandaker2023ICC} proposed a reverse auction-based dynamic caching and pricing scheme for producer-driven Information-Centric Networks and addressed the challenge of cache allocation when multiple content producers and competing service providers are involved. This market-driven approach maximizes caching benefits for content producers while operating within economic constraints. These frameworks together demonstrate how economic principles can guide efficient information acquisition strategies in diverse application domains, from autonomous exploration to network caching systems.

\subsection{Lessons Learned}

\begin{table*}[!t]
\centering
\caption{Quality/Utility-oriented information metrics: comparative analysis and optimization methods.}
\label{tab:quality_utility_metrics}
\renewcommand{\arraystretch}{1.5}
\begin{tabular}{|p{3cm}|p{4cm}|p{4.2cm}|p{4cm}|}
\hline
\rowcolor[gray]{0.85} \textbf{Key Features} & \textbf{Utility of Information (UoI)} & \textbf{Semantics of Information (Sem-oI)} & \textbf{Cost of Information (CoI)} \\
\hline
Definition & 
Quantifies the practical value and decision-making utility of information for achieving specific application goals \cite{Hu2024TMC, Howard1966TSSC, Yang2024TIFS}. & 
Measures the semantic meaning, contextual relevance, and interpretability of information content \cite{Chaccour2025COMST, Wu2025TII, Wang2024TNNLS}. & 
Evaluates the economic and resource costs associated with information acquisition, processing, and utilization \cite{Agarwal2024TGCN, Wang2023TSMC, Faist2015NC}. \\
\hline
Primary Applications & 
Federated learning systems, autonomous decision-making, and utility-aware resource allocation \cite{Hu2024TMC, NgoICDCS, Zhan2023CAA}. & 
Knowledge graph construction, semantic segmentation, and goal-oriented communication systems \cite{Wu2025TII, Wang2024TNNLS, Fu2023INFOCOM}. & 
Energy-efficient IoT networks, cost-sensitive learning, and resource-constrained wireless systems \cite{Agarwal2024TGCN, Wang2023TSMC, Nasir2013TWC}. \\
\hline
Analytical Techniques & 
Bayesian game theory, reinforcement learning, and utility maximization frameworks \cite{Hu2024TMC, Howard1966TSSC, Bongole2025ICASSP}. & 
Graph neural networks, attention mechanisms, and semantic embedding techniques \cite{Wu2025TII, Liang2023ICDE, Qian2025TNNLS}. & 
Optimization theory, active learning, and energy-information trade-off modeling \cite{Agarwal2024TGCN, Wang2023TSMC, Wu2013ECC}. \\
\hline
Typical Optimization Approaches & 
AutoFL frameworks, contextual bandits, and multi-objective utility optimization \cite{Hu2024TMC, NgoICDCS, Ghosh2024TIT}. & 
Concept-aware entity alignment, weakly-supervised learning, and semantic communication \cite{Wang2024TNNLS, Hu2024TMC}. & 
Mobile sink optimization, cost-sensitive active learning, and energy-delay trade-offs \cite{Agarwal2024TGCN, Wang2023TSMC}. \\
\hline
Main Limitations & 
Subjective utility assessment, context-dependent valuations, and difficulty in standardizing utility functions \cite{Zhan2023CAA,Ferguson2023TCSS, Pan2024TFS}. & 
High computational complexity, semantic ambiguity, and requirement for domain-specific knowledge \cite{Ran2015ICIP, Wu2024TWC, Alsmadi2023TCSS}. & 
Accurate cost modeling challenges, dynamic pricing complexities, and multi-objective trade-off resolution \cite{Faist2015NC, Wang2022TMC}. \\
\hline
\end{tabular}
\end{table*}

The examination of quality/utility-oriented information metrics provides theoretical and practical insights for next-generation networked systems. The evolution from time-focused to value-centered information assessment frameworks shows how evaluation paradigms have matured beyond simplistic availability measures to incorporate contextual relevance and semantic meaning. Howard's foundational work \cite{Howard1966TSSC} established the critical relationship between information quality and decision outcomes, while recent semantic communication frameworks \cite{Getu2024JPROC} have extended this to include goal-oriented information exchange. This progression shows the increasing recognition that the value of information is not derived merely from its timeliness but from its contextual utility within specific operational environments and decision frameworks. The growing integration of economic principles into information valuation, as shown by emerging resource trading frameworks \cite{Abegaz2023TNSM}, also illustrates the multi-dimensional nature of information utility across diverse application domains.

The implementation of quality/utility-oriented metrics faces several practical challenges that require more sophisticated approaches. The need to balance comprehensive semantic understanding with computational efficiency presents a fundamental challenge, particularly in resource-constrained environments such as industrial IoT networks \cite{Karamyshev2024TII}. The integration of contextual awareness mechanisms into existing systems often requires substantial architectural modifications, as legacy infrastructure typically lacks the flexibility to incorporate dynamic quality assessment. In addition, standardization challenges arise when systems attempt to establish consistent utility metrics across heterogeneous networks and applications. Successful implementations have addressed these constraints through adaptive learning approaches that dynamically assess information value relative to operational contexts and use techniques like transformer-based prediction models \cite{Dang2025MCI} that can capture complex temporal relationships in information utility patterns.

Future research in quality/utility-oriented information metrics should focus on several promising directions. The integration of semantic frameworks with economic utility models holds great potential for holistic information assessment that considers both meaning and value dimensions simultaneously. Multi-objective optimization approaches \cite{Dang2025MCI} that balance information utility against resource constraints will be essential for practical deployment in next-generation networks. The development of standardized cross-domain utility metrics would enhance interoperability and facilitate consistent information valuation across diverse application scenarios. As networked systems continue to evolve toward greater autonomy and intelligence, quality/utility-oriented metrics will remain fundamental tools for reliable information exchange that prioritizes value over volume and enables truly efficient and meaningful communication in increasingly complex environments.

\section{Reliability/Robustness-oriented Information Metrics}\label{Reliability}
\begin{figure}[h]
    \centering
    \includegraphics[width=1\linewidth]{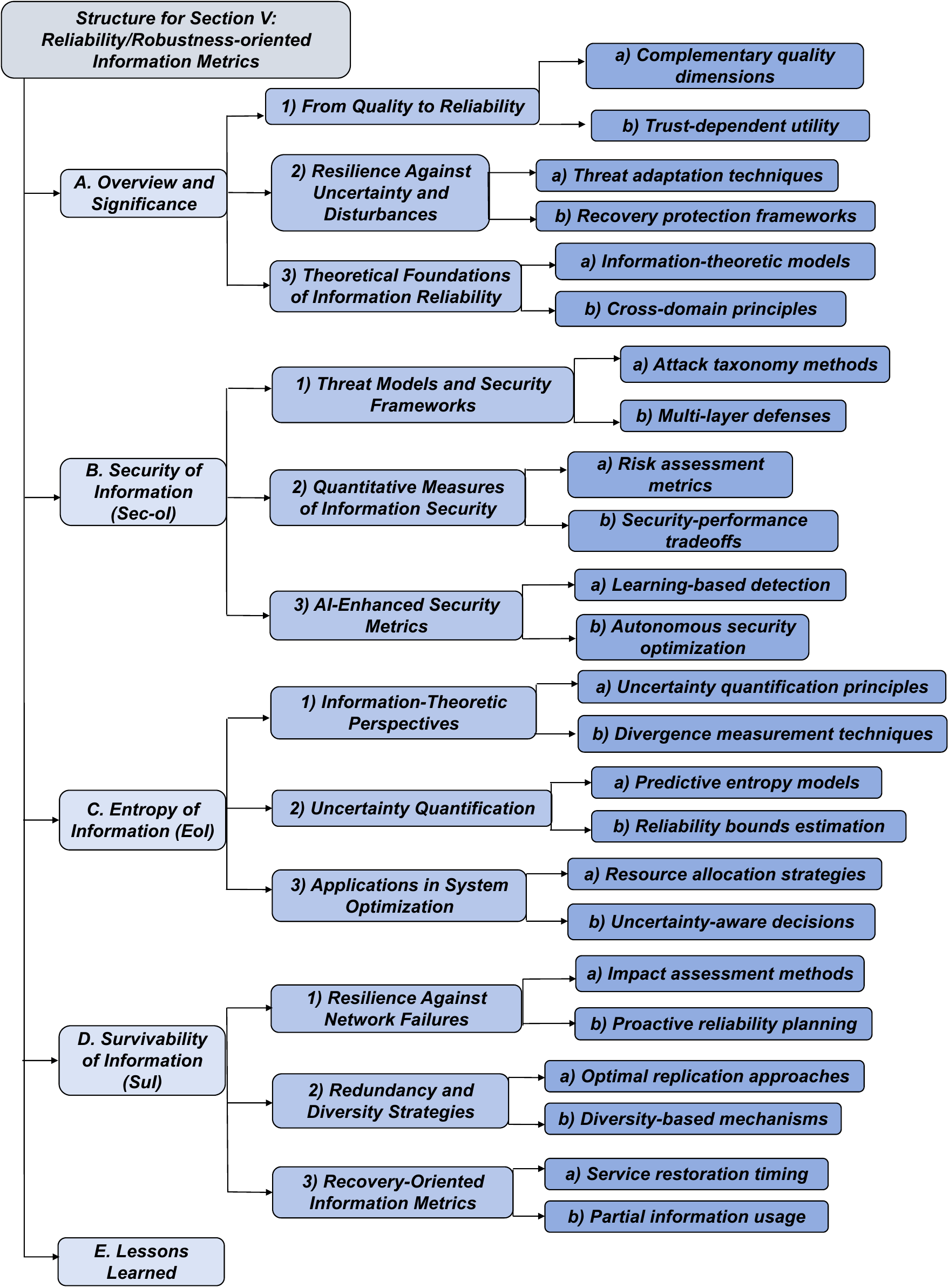}
    \caption{The roadmap of section V.}
    \label{fig:RMV}
\end{figure}

We provide a comprehensive framework for reliability/robustness-oriented information metrics, focusing on their underlying trust mechanisms, resilience modeling under dynamic uncertainty, and theoretical foundations for dependable information evaluation. Metrics such as Security of Information (Sec-oI), Entropy of Information (EoI), and Survivability of Information (SuI) are systematically explored, each contributing uniquely to ensuring information robustness across critical applications with uncertainty, adversarial threats, or system-level failures. The detailed roadmap of Section \ref{Reliability} is illustrated in Fig.~\ref{fig:RMV}.

\subsection{Overview and Significance}

\begin{figure}[ht]
    \centering
    \includegraphics[width=1\linewidth]{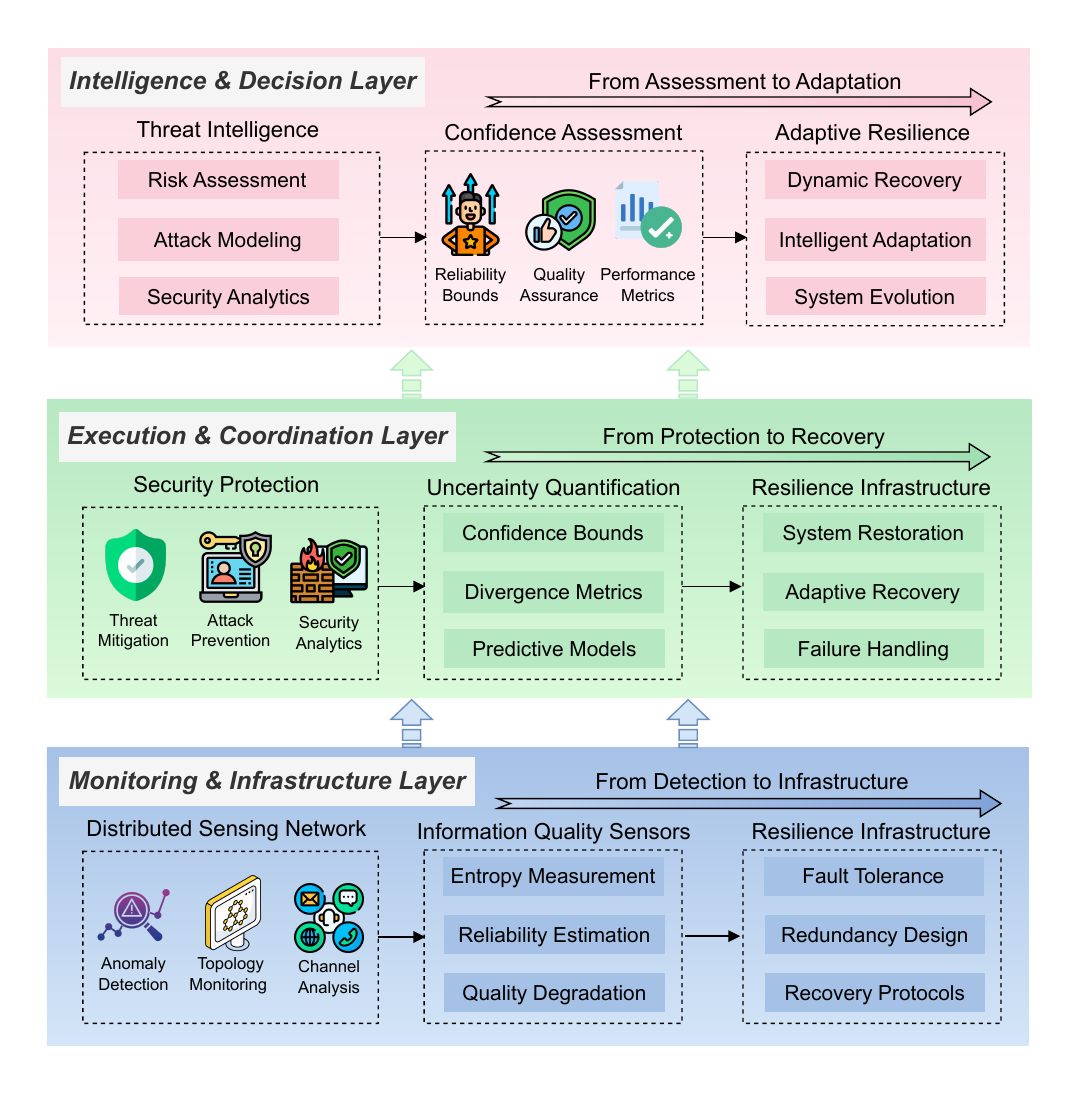}
    \caption{Multi-layered architecture for reliability/robustness-oriented information metrics in next-generation networks.}
    \label{fig:reliability_architecture}
\end{figure}

As illustrated in Fig.~\ref{fig:reliability_architecture}, reliability and robustness-oriented information metrics operate through a progressive three-tier architecture that encompasses monitoring infrastructure, execution coordination, and intelligent decision-making layers. This framework enables systematic evaluation of information trustworthiness, uncertainty quantification, and adaptive resilience in mission-critical networked systems. The monitoring layer continuously assesses information integrity through distributed sensing and anomaly detection. In contrast, the coordination layer implements protective measures and recovery protocols, and the intelligence layer makes strategic decisions about information trustworthiness and system adaptation.

Reliability and robustness-oriented metrics represent a critical evolution in information assessment that moves beyond value-centric and time-based paradigms. In mission-critical systems, information must not only be timely and valuable but also trustworthy, verifiable, and adaptive to disruption. These metrics fundamentally answer the question: Can we trust and depend on the information we receive? Sec-oI measures protection against malicious threats through integrity verification and threat assessment scores. EoI quantifies uncertainty and randomness using information-theoretic calculations and confidence bounds. SuI evaluates system resilience through redundancy analysis and recovery time assessments.

These metrics enable intelligent systems to proactively reason about threats, quantify uncertainty, and recover from disturbances while maintaining consistent decision-making performance. They find critical applications in healthcare systems for protecting patient data integrity, autonomous vehicles for verifying Vehicle-to-Everything (V2X) communication authenticity and Industrial Internet of Things (IoT) networks for maintaining operational safety during component failures. The practical measurement involves deploying security monitoring tools, implementing statistical uncertainty models, and establishing automated failover mechanisms that collectively ensure information systems remain dependable under adverse conditions.

\subsubsection{From Quality to Reliability}
To achieve resilient information systems, quality-oriented paradigms must evolve to address the volatility, trust asymmetry, and adversarial uncertainty present in modern networks.

\paragraph{Complementary quality dimensions}
The concept of trust complements traditional quality evaluation by embedding dependability across semantic, social, and technical layers. Cho \textit{et al.}~\cite{Cho2015ACS} categorized trust into emotional, logical, and relational constructs and mapped them to layered system behaviors. This multi-perspective view set the foundation for reliability-aware evaluation, where correctness is not assumed but continually validated. Premarathne \textit{et al.}~\cite{Premarathne2017TCC} expanded this notion and proposed a federated identity negotiation mechanism in the cloud, where trust is operationalized through fuzzy vulnerability and cooperation scores. Their model showed how system-wide quality hinges on the reliability of authentication and service handoff. These foundational studies demonstrated that quality metrics must be augmented by mechanisms that trace trust propagation through diverse decision contexts.

\paragraph{Trust-dependent utility}
As intelligent systems operate under uncertainty, the utility of information increasingly depends on its trust lineage and interpretability. Lewis \textit{et al.}~\cite{Lewis2023IOTJ} revealed how context-based attacks can compromise trust systems by injecting false evidence into localized IoT environments. Their findings emphasized that trust must be both context-aware and defensively modeled. In such adversarial settings, the perceived utility of information is inseparable from its structural credibility. This reinforces the idea that decision quality must account for system-level dependability, not just content relevance.

\subsubsection{Resilience Against Uncertainty and Disturbances}
Beyond establishing trust, reliable systems must remain operational under unexpected deviations. This resilience is achieved by embedding adaptability into both planning and control layers.

\paragraph{Threat adaptation techniques}
Planning under uncertainty demands a shift from fixed rules to predictive, model-informed responses. Liu \textit{et al.}~\cite{Liu2021TPWRS} addressed this challenge in power distribution networks and proposed a restoration framework that combined model predictive control and hierarchical dispatch. Their system continuously re-optimized restoration actions based on forecast errors and real-time failures. This demonstrated how resilience can be encoded into adaptive infrastructure logic. Similarly, Zhang \textit{et al.}~\cite{Zhang2024TNNLS} presented a neuromorphic solution where spiking neurons autonomously regulate their activation thresholds, which improved inference speed and robustness. These works illustrated distinct yet converging strategies where adaptive thresholds in control logic or biological computing serve as the foundation for reliability.

\paragraph{Recovery protection frameworks}
Effective disturbance recovery requires both localized correction and global coordination. Duan \textit{et al.}~\cite{Duan2021TCNS} proposed distributed estimation algorithms for networked systems under model perturbations. By employing recursive gain updates and node-level covariance bounding, their design avoided centralized bottlenecks while preserving estimation fidelity across uncertain environments. The implication is clear: robust recovery must be decomposable, low-complexity, and inherently aware of its operational envelope.

\subsubsection{Theoretical Foundations of Information Reliability}
To rigorously quantify and reason about robustness, a theoretical framework is needed that unifies probabilistic, evidential, and learning-based perspectives.

\paragraph{Information-theoretic models}
Reliability modeling under epistemic uncertainty often exceeds the capabilities of classical probability. Zhang \textit{et al.}~\cite{Zhang2025TR} responded to this by integrating copula-based parameter correlations into an evidence-theoretic framework for multi-failure systems. Their method yielded belief and plausibility bounds, which enabled bounded estimation when statistical data are sparse. This provides a structured lens through which to understand reliability not just as a scalar value but as a probabilistic interval governed by systemic dependencies.

\paragraph{Cross-domain principles}
In cross-domain environments, robust inference hinges on multi-source data fusion and intelligent filtering. Zhu \textit{et al.}~\cite{Zhu2020IJCNN} proposed a two-stage evidence fusion framework for voltage estimation in power grids that resolved inconsistencies among heterogeneous measurement devices. Their work highlighted the centrality of belief propagation and credibility weighting in ensuring reliable system state estimation. Complementing this, Lv \textit{et al.}~\cite{Lv2021TII} applied AI-driven ranking and spatiotemporal modeling to evaluate CPS trustworthiness in real time. These contributions reinforced that theoretical reliability must embrace not only formal reasoning but also intelligent adaptation across layers and modalities.

\subsection{Security of Information (Sec-oI)}
Sec-oI plays a foundational role in establishing robustness to ensure that information remains trustworthy not only under uncertainty but also in the presence of adversarial threats. We examine how reliability is maintained in systems that must actively defend against intrusions, adapt to evolving threat models, and strike a balance between security and operational performance. The analysis proceeds from foundational threat characterizations to quantitative security metrics and culminates with AI-enhanced defense mechanisms that enable autonomous adaptation.

\subsubsection{Threat Models and Security Frameworks}
To effectively protect against cyber threats, it is essential to understand how attacks are classified, structured, and modeled across multiple domains of system operation.

\paragraph{Attack taxonomy methods}
The rapid integration of intelligent agents, wireless nodes, and IoT devices has exposed emerging systems to increasingly diverse and stealthy attacks. Adil \textit{et al.}~\cite{Adil2023TITS} provided a comprehensive survey of UAV-aided IoT systems and demonstrated how unstructured deployments and wireless dependencies create vulnerabilities across the communication stack. Their taxonomy emphasized the importance of viewing system threats through both physical and protocol layers. Sadeghi \textit{et al.}~\cite{Sadeghi2017TSE} complemented this by presenting a formal taxonomy of Android software threats, based on a rigorous synthesis of over 300 studies, which identified long-term shifts in exploit strategies and the evolution of code-level vulnerabilities. In software development contexts, Barabanov \textit{et al.}~\cite{Barabanov2018ICAICT} expanded this scope and introduced a life-cycle-oriented threat model that mapped 35 discrete threats to attacker profiles and ISO/IEC 12207 processes, thereby formalising the emergence of threats across system engineering phases. These works collectively laid the groundwork for robust security assessments by framing threats not as isolated anomalies but as system-wide phenomena embedded in architecture, behaviour, and deployment strategies.

\paragraph{Multi-layer defenses}
Once threats are modeled, effective protection must span physical, cyber, and control layers. Upadhyay \textit{et al.}~\cite{Upadhyay2022TNSM} proposed a multi-layered security framework for SCADA systems and integrated lightweight symmetric and asymmetric cryptographic mechanisms to satisfy real-time and broadcast requirements under hardware constraints. Their design demonstrated how resource-aware defense mechanisms can maintain integrity and confidentiality without exceeding system capacity. He \textit{et al.}~\cite{He2024TYCB} shifted focus to cascading failures in cyber–physical systems and showed how initial attacks can propagate via tightly coupled interdependencies. Their modeling approach, based on historical case studies, revealed that vulnerabilities are not isolated but systemic, which reinforced the value of defense-in-depth designs. Kumar \textit{et al.}~\cite{Kumar2023TITS} further strengthened this view by introducing deep-learning-powered cyber threat intelligence for maritime IoT and used detection and classification schemes to anticipate attacks and provide early warning. Together, these studies validated a layered security philosophy, where each abstraction, from microcontroller encryption to high-level system analytics, contributes to end-to-end robustness.

\subsubsection{Quantitative Measures of Information Security}
To design defensible systems, it is not enough to model threats qualitatively; security must be measured, scored, and weighed against system constraints in real time.

\paragraph{Risk assessment metrics}
Quantifying security risk transforms abstract vulnerabilities into actionable system metrics. Zeng \textit{et al.}~\cite{Zeng2024TSG} developed a real-time cyberattack risk assessment framework for distribution networks and embedded feeder automation behavior into cyber-physical state deterioration models. They introduced an entropy-like metric that captures both the likelihood of an attack and its system impact. In a healthcare context, Coronato \textit{et al.}~\cite{Coronato2022TKDE} built a Markov model-based dynamic probabilistic risk assessment methodology compliant with ISO 14971. Their approach not only supported pre-deployment certification but also enabled adaptive post-market surveillance using real-world data. These risk models demonstrated how temporal, probabilistic reasoning can bridge the gap between perceived threats and measured impact, which reinforced the integration of reliability assurance into critical system domains.

\paragraph{Security-performance tradeoffs}
Maximizing security often introduces overhead, latency, or energy costs, which makes tradeoff analysis essential. Zhao \textit{et al.}~\cite{Zhao2022TITS} proposed a hybrid accident-risk assessment model for crowd flow management based on entropy and energy distributions and revealed how dynamic risk control can optimize safety while preserving spatial efficiency. Ramos \textit{et al.}~\cite{Ramos2017COMST} provided a sweeping survey of model-based network security metrics and illustrated how tradeoffs are mathematically formalized to support decision-making. This aligned with Sanders~\cite{Sanders2014MSP}, who challenged the notion that perfect metrics are required for valuable insight and argued that even imperfect quantification enables strategic improvements. On the physical layer, Bao \textit{et al.}~\cite{Bao2018VTC} analyzed multiple-input single-output secure transmission and illustrated the impact of artificial noise and unitary beamforming on secrecy outage probability. Fu \textit{et al.}~\cite{Fu2024WCNC} further explored this tradeoff in V2X networks and modeled how hybrid automatic repeat request (HARQ) schemes improve reliability but can leak information, which required optimal parameter tuning to balance secrecy and success probability. These works collectively underlined that robust systems must quantify not only what is protected, but at what cost.

\subsubsection{AI-Enhanced Security Metrics}
In highly dynamic threat environments, static policies fall short. Learning-based approaches enable adaptive security evaluation and improve both detection and response over time.
\paragraph{Learning-based detection}
In dynamic and heterogeneous environments, reliable intrusion detection depends on the system's ability to learn from evolving behaviors. Rather than focusing on fixed rule sets, recent research has emphasized adaptive models that generalize across domains while capturing localized anomalies. For instance, Ferrag \textit{et al.}~\cite{Ferrag2022JAS} identified key ML-based design patterns in agricultural systems. In contrast, Otoum \textit{et al.}~\cite{Otoum2019LNET} contrasted deep and hybrid detection methods in sensor networks and revealed tradeoffs between inference speed and generalizability. Beyond algorithmic design, Grosse \textit{et al.}~\cite{Grosse2023TIFS} provided field evidence that practical deployment depends as much on organizational readiness and perceived exposure as on technical capability. Together, these studies highlighted a transition in metric thinking: from predefined thresholds toward behavior-aware, data-driven evaluation strategies, where detection reliability is shaped by the model's adaptability and the ecosystem's evolving threat surface.

\paragraph{Autonomous security optimization}
Detection alone is insufficient in environments where security, performance, and privacy are in constant tension. Intelligent systems must autonomously optimize their defensive strategies and account for conflicting objectives such as latency, interpretability, and resource consumption. This need is evident in mobility-driven systems, where security decisions must adapt to spatiotemporal dynamics. Xiu \textit{et al.}~\cite{Xiu2025TVT} tackled this by jointly optimizing trajectory and beamforming for autonomous aerial vehicles and improved secrecy rates while preserving sensing accuracy. Qin \textit{et al.}~\cite{Qin2022TVT} proposed a task offloading scheme for C-V2X that balanced physical-layer noise injection with latency and energy goals and captured the tradeoff between proactive defense and real-time performance. At the architectural level, Namakshenas \textit{et al.}~\cite{Namakshenas2024TICPS} addressed the privacy-security dichotomy in federated learning for industrial CPS and integrated homomorphic encryption with Shapley-based interpretability to enable secure yet explainable AI. In vehicular networks, Kamal \textit{et al.}~\cite{Kamal2021TITS} proposed optimized fingerprinting algorithms that minimize computational overhead without sacrificing authentication fidelity. Finally, Jahan \textit{et al.}~\cite{Jahan2019ACS} provided a modeling framework for security adaptation in autonomous systems and emphasized the need to formalize system behavior under evolving threat conditions. Together, these works represented a shift toward security as a dynamic, multi-objective optimization problem where robustness emerges not from fixed policies but from continual tradeoff navigation and structural self-awareness.

\subsection{Entropy of Information (EoI)}
Entropy-based metrics provide a foundational lens for characterizing information uncertainty, a crucial aspect of robustness in decision-making systems. We explore how entropy and divergence support reliability modeling from core theoretical formulations to predictive confidence estimation and system-level optimization under incomplete information.

\subsubsection{Information-Theoretic Perspectives}
At the foundation of uncertainty modeling lies the need to quantify the amount of disorder or unpredictability in system states and probabilistic outcomes.

\paragraph{Uncertainty quantification principles}
Entropy measures formalize how much information is needed to describe a system's behavior and offer early insight into reliability even before empirical failures occur \cite{Brown1987TR,Lutwak2005TIT,Perrone2024TIT}. For example, entropy-based reliability estimators can fuse prior belief with limited development data to support early-stage assessments \cite{Brown1987TR}. More broadly, generalizations of Shannon entropy, such as Rényi entropy, provided extended inequalities for measuring tail behavior and extremal uncertainty under moment constraints \cite{Lutwak2005TIT}. Theoretical abstractions, such as Markov categories, go further and frame entropy as a measure of deviation from determinism across probabilistic structures \cite{Perrone2024TIT}. Across these approaches, entropy emerges not just as a static property, but as a flexible tool for shaping probabilistic insight during system design and modeling.

\paragraph{Divergence measurement techniques}
While entropy quantifies uncertainty within a distribution, divergence metrics enable the comparison of distributions and provide a way to reason about confidence shifts, update consistency, or deviation from expected behavior \cite{Liese2006TIT,Heidari2025TIT}. In robust statistical inference, $f$-divergences encapsulated a spectrum of uncertainty-aware comparisons, from Kullback-Leibler to total variation, all of which inform posterior reasoning and model credibility \cite{Liese2006TIT}. In communication systems, these divergence tools underlie variable-length reliability bounds in channels with feedback, where information evolves and must be evaluated dynamically \cite{Heidari2025TIT}. Ultimately, divergences serve as operational metrics to track and control informational drift in uncertain or interactive systems.

\subsubsection{Uncertainty Quantification}
Moving beyond static theory, modern systems require tools to estimate and act on uncertainty during prediction, learning, and control, particularly in cases of data sparsity or environmental shifts.

\paragraph{Predictive entropy models}
Reliable predictions must come not only with answers, but with measures of confidence. In perceptual quality modeling, discrete entropy distributions helped quantify variability in human ratings and enabled more precise quality of experience estimates \cite{Saupe2024LSP}. In meta-learning and continual adaptation, models like neural processes balanced average prediction with variance estimation, which allowed them to react effectively to changing data or limited supervision \cite{Wang2023TNNLS}. These trends reflect a shift from treating uncertainty as post-hoc error to integrating it into the model's primary reasoning process, especially in semi-supervised or incremental scenarios where data quality is heterogeneous \cite{Cui2024TNNLS}.

\paragraph{Reliability bounds estimation}
In complex control systems, it is not enough to estimate uncertainty; bounds must be placed on system behavior to guarantee stability under uncertainty. Data-driven $\mathcal{H}_{\infty}$-norm estimation provided such bounds for model errors in vibration isolation systems and supported high-confidence robust design without complete plant modeling \cite{Oomen2014TCST}. In power grids and multi-agent systems, reinforcement learning and stochastic stability techniques were employed to maintain control performance despite uncertainties in cyber and communication systems \cite{Duan2018YSG,Haddad2020TAC}. Across these domains, the common thread is clear: reliability is enabled when uncertainty is not only measured, but bounded and planned for in real-time system dynamics.

\subsubsection{Applications in System Optimization}
Entropy and uncertainty metrics are increasingly informing resource and policy optimization in systems that must act reliably under ambiguity.

\paragraph{Resource allocation strategies}
From industrial control to cloud services, uncertainty-aware resource allocation has become a key strategy for achieving stability and efficiency. Entropy-based clustering and complexity metrics helped match resources to tasks under preference ambiguity or cloud-side variability and improved throughput and fairness \cite{Zhao2016KIS,Chen2018TETCI}. In offline optimization contexts, generative uncertainty models (e.g., cGAN ensembles) can be used to simulate policy outcomes and reduce the need for costly online experimentation in sensitive environments \cite{Feng2024IJCNN}. These methods highlight how entropy-based reasoning can help bridge the gap between data-driven planning and performance-aware control.

\paragraph{Uncertainty-aware decisions}
Decision-making frameworks that actively adapt to uncertainty show promise in high-dimensional, dynamic environments. In transfer learning, entropy-based reweighting aligned feature learning with domain reliability and guided better generalization across mismatched conditions \cite{Pan2024ICASSP}. In energy-constrained IoT networks, control decisions were co-optimized with communication strategies to adapt under dropouts and delays \cite{Lu2022ICCC}. In graph-based learning, distributionally robust optimization yielded uncertainty-aware embeddings that defend against structural noise and missing data \cite{Chen2025ICASSP}. These examples reflect a broader evolution: reliable systems increasingly embed entropy not as a diagnostic metric, but as a driving signal for adaptive behavior.
\subsection{Survivability of Information (SuI)}
In adversarial, failure-prone, or resource-constrained environments, system robustness must extend beyond reliability toward survivability, which is the ability to remain functional despite partial degradation.  We explore key strategies to sustain information operations under failures, including network resilience, redundancy and diversity, as well as recovery-aware metric design.

\subsubsection{Resilience Against Network Failures}
Information survivability begins with anticipating how failures propagate and how systems can absorb their impact without total collapse.

\paragraph{Impact assessment methods}
Survivability assessment requires more than binary availability checks. It demands models that reflect how cascading disruptions unfold in topologically complex systems \cite{Deng2023TII,Lenders2008ToN,Mahmoud2019TSTE}. For instance, Deng \textit{et al.}~\cite{Deng2023TII} proposed a Bayesian impact model for cyber-physical systems that combines epidemic spreading dynamics with load-loss ratios and enabled quantifiable estimates of cyberattack severity. In ad hoc routing, Lenders \textit{et al.}~\cite{Lenders2008ToN} introduced a density-aware anycast paradigm and demonstrated how delivery reliability improves when routing incorporates not just proximity, but the redundancy of reachable nodes. These approaches reflect a growing consensus: survivability is best assessed by modeling both structural fragility and functional interdependence.

\paragraph{Proactive reliability planning}
Proactive survivability shifts the focus from damage assessment to failure anticipation, often through reconfigurable routing or structural redundancy \cite{Alam2016TNSM,Li2011ICSPCC,Rubin2009ADH,Misra2009TVT}. In virtualized infrastructures, Alam \textit{et al.}~\cite{Alam2016TNSM} designed a multi-path link embedding scheme that guarantees survivability under single or multiple link failures with minimal redundancy. Li \textit{et al.}~\cite{Li2011ICSPCC} modeled geographic region failures and formulated region-disjoint routing optimizations to maintain network throughput under disaster conditions. Meanwhile, Rubin \textit{et al.}~\cite{Rubin2009ADH} and Misra \textit{et al.}~\cite{Misra2009TVT} emphasized robustness in mobile ad hoc and localization networks and showed how route lifetime, trust, and anchor honesty affect overall survivability. These studies collectively advocate for forward-looking strategies that anticipate failure paths and encode redundancy where it matters most.

\subsubsection{Redundancy and Diversity Strategies}
Survivability is strengthened not only by backup routes or plans, but by the strategic use of multiplicity across space, channel, and logic.

\paragraph{Optimal replication approaches}
Replication ensures continued access to critical services and data, but it incurs additional resource costs. Hence, survivability planning must optimize for availability under constraints \cite{Gupta2006TPDS,Tornatore2021TNSM,Jiang2023TCOM}. In peer-to-peer multicast networks, Gupta \textit{et al.}~\cite{Gupta2006TPDS} developed hierarchical gossip protocols that adapt replication overhead to network domain boundaries and preserve scalability without sacrificing delivery. Tornatore \textit{et al.}~\cite{Tornatore2021TNSM} surveyed reliability strategies for 5G/6G networks and highlighted service replication at the edge and network slicing as tools to ensure ultra-reliable low-latency service continuity. Jiang \textit{et al.}~\cite{Jiang2023TCOM} extended these ideas to satellite networks and combined multipath scheduling and redundant coding for deterministic transmission under dynamic topologies. Across these examples, optimal replication is not about brute force. It is about embedding recoverability in the system fabric.

\paragraph{Diversity-based mechanisms}
Where replication seeks to copy, diversity seeks to hedge and leverages differences in routing paths, hardware components, or physical channels to mitigate correlated failures \cite{Liu2022IOTJ,Liu2024TMC,Asad2024IOTJ}. Liu \textit{et al.}~\cite{Liu2022IOTJ} proposed a self-healing vehicular routing algorithm using ant colony optimization and fuzzy link metrics to maintain multi-hop connectivity under frequent path breakages. In contact-free health monitoring, Liu \textit{et al.}~\cite{Liu2024TMC} enhanced mmWave sensing with diversity across antennas, chirps, and spatial channels to maintain signal strength in obstructed environments. Similarly, Asad \textit{et al.}~\cite{Asad2024IOTJ} explored RF energy harvesting diversity and optimized transmission and harvesting tradeoffs using deep reinforcement learning. These studies highlight how diversity transforms fragile systems into adaptive ones by dispersing dependency and increasing resilience margins.

\subsubsection{Recovery-Oriented Information Metrics}
True survivability involves not only withstanding failures, but restoring operations and extracting value from degraded conditions.

\paragraph{Service restoration timing}
Resilient systems must adapt to both known and unknown disruptions through fault-tolerant control and dynamic reconfiguration \cite{Sharida2023JIES,Hou2024ICSE,Zhong2023TYCB,Chu2024TITS}. For example, Sharida \textit{et al.}~\cite{Sharida2023JIES} designed a master–slave coordination scheme for power rectifiers that automatically reassigns roles under communication faults. In contrast, Hou \textit{et al.}~\cite{Hou2024ICSE} analyzed ransomware disruption phases and proposed phase-specific detection techniques to minimize service loss. In multi-agent systems, Zhong \textit{et al.}~\cite{Zhong2023TYCB} and Chu \textit{et al.}~\cite{Chu2024TITS} developed distributed consensus protocols with adaptive event-triggered control to ensure re-synchronization after link or node faults. Together, these works reveal that restoration timing must account for fault origin, propagation path, and control loop feedback to enable graceful recovery.

\paragraph{Partial information usage}
When full recovery is not immediately possible, systems must make the best of what remains and leverage partial information to preserve basic functionality \cite{Zhang2021TASE,Pang2022TCSII,Zhang2023TAES}. In infrastructure monitoring, Zhang \textit{et al.}~\cite{Zhang2021TASE} developed a reconstruction method for magnetic flux leakage data using multifeature conditional risk and recovered critical safety indicators despite missing sensor measurements. Pang \textit{et al.}~\cite{Pang2022TCSII} integrated observer-based compensation into predictive fault-tolerant controllers to maintain tracking accuracy under sensor and actuator faults. Meanwhile, Zhang \textit{et al.}~\cite{Zhang2023TAES} proposed a resilient adaptive controller for malicious attack scenarios and combined sliding-mode observers and hybrid actuator fault models to achieve convergence even when network and actuation layers are compromised. These methods exemplify how intelligent inference, fault estimation, and model redundancy can extract value from incomplete states and turn information decay into graceful degradation.
\subsection{Lessons Learned}

\begin{table*}[!t]
\centering
\caption{Reliability/Robustness-oriented information metrics: comparative analysis and optimization methods.}
\label{tab:reliability_robustness_metrics}
\renewcommand{\arraystretch}{1.5}
\begin{tabular}{|p{2.5cm}|p{4.2cm}|p{4cm}|p{4.4cm}|}
\hline
\rowcolor[gray]{0.85} \textbf{Key Features} & \textbf{Security of Information (Sec-oI)} & \textbf{Entropy of Information (EoI)} & \textbf{Survivability of Information (SuI)} \\
\hline
Definition & 
Measures information trustworthiness, integrity, and protection against adversarial threats and cyber attacks \cite{Adil2023TITS, Kumar2023TITS, Ramos2017COMST}. & 
Quantifies information uncertainty, randomness, and confidence bounds for probabilistic decision-making \cite{Chen2024TIT, Brown1987TR}. & 
Evaluates system resilience, fault tolerance, and ability to maintain functionality under failures \cite{Alam2016TNSM, Rubin2009ADH, He2024TYCB}. \\
\hline
Primary Applications & 
UAV-aided IoT security, maritime transportation systems, and cyber-physical system protection \cite{Adil2023TITS, Kumar2023TITS, Upadhyay2022TNSM}. & 
Markov source scheduling, CNN model optimization, and uncertainty-aware decision systems \cite{Chen2024TIT, Li2024Neuro, Wang2023TNNLS}. & 
Network virtualization, mobile ad hoc networks, and fault-tolerant system design \cite{Alam2016TNSM, Rubin2009ADH}. \\
\hline
Analytical Techniques & 
Threat modeling, multi-layer defense frameworks, and machine learning-based detection \cite{Adil2023TITS, Kumar2023TITS, Ferrag2022JAS}. & 
Information theory, statistical modeling, and entropy-based optimization \cite{Chen2024TIT, Lutwak2005TIT, Liese2006TIT}. & 
Graph-based analysis, redundancy modeling, and failure propagation assessment \cite{Alam2016TNSM, He2024TYCB, Deng2023TII}. \\
\hline
Typical Optimization Approaches & 
DLTIF frameworks, comprehensive security surveys, and adaptive threat intelligence \cite{Kumar2023TITS, Adil2023TITS, Otoum2019LNET}. & 
Restless bandit policies, graph-based pruning, and uncertainty quantification \cite{Chen2024TIT, Li2024Neuro, Wang2023TNNLS}. & 
Multi-path link embedding, robust routing algorithms, and survivability optimization \cite{Alam2016TNSM, Li2011ICSPCC}. \\
\hline
Main Limitations & 
High computational overhead, evolving threat landscapes, and trade-offs with system performance \cite{Zhao2022TITS, Sanders2014MSP, Bao2018VTC}. & 
Complex probability modeling, computational intensity, and difficulty in real-time uncertainty estimation \cite{Saupe2024LSP, Cui2024TNNLS, Oomen2014TCST}. & 
Resource redundancy requirements, scalability challenges, and complexity in multi-failure scenarios \cite{Gupta2006TPDS}. \\
\hline
\end{tabular}
\end{table*}

Reliability and robustness-oriented metrics provide a crucial extension to conventional information evaluation by addressing the ability of systems to function under threat, uncertainty, and degradation. Across this section, a clear thematic shift emerges: from static assessments of information quality to dynamic modeling of system resilience. Security-oriented metrics defend against adversarial threats; entropy-based measures quantify uncertainty and probabilistic robustness; survivability metrics ensure continued operation through redundancy, diversity, and recovery strategies. These dimensions are not isolated; they are complementary. Each tackles a different vulnerability class while reinforcing the shared objective of trustworthy, durable decision-making.

Also, the role of reliability metrics is evolving from descriptive indicators to operational enablers. Rather than passively monitoring system states, these metrics now inform adaptive control, reconfiguration, and learning-driven defenses. Whether through attack-aware detection thresholds, entropy-guided optimization, or event-triggered recovery protocols, robustness is increasingly embedded into system logic. The integration of reliability dimensions through unified models, real-time metric fusion, and cross-layer feedback will be essential for dependable intelligence in next-generation networks.

\section{Network/Communication-oriented Information Metrics}\label{Network}
\begin{figure}[ht]
    \centering
    \includegraphics[width=1\linewidth]{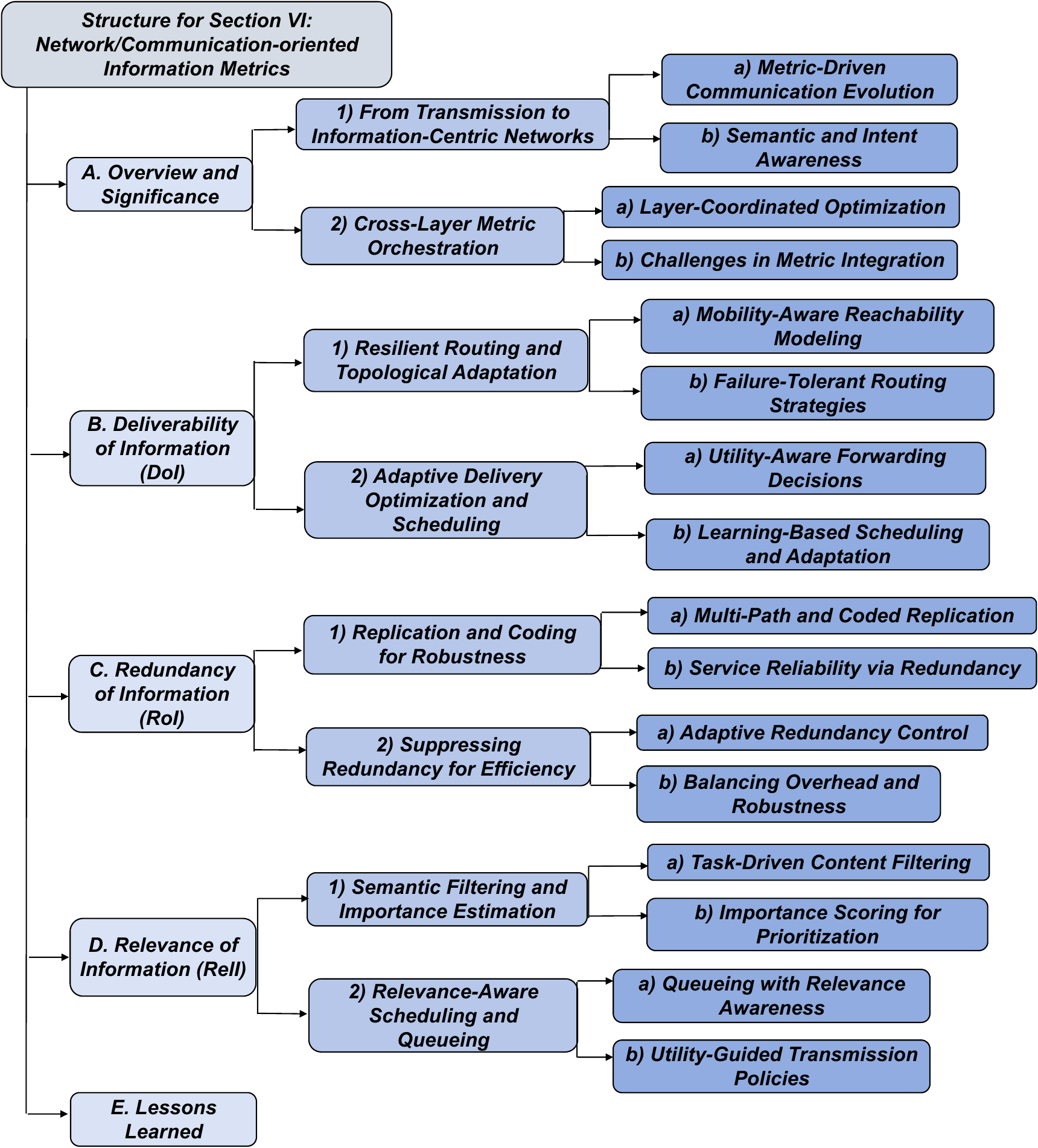}
    \caption{The roadmap of section VI.}
    \label{fig:RMVI}
\end{figure}

We provide a comprehensive framework for network/communication-oriented information metrics that focuses on the evolution of the network as an intelligent information substrate. Beyond traditional throughput-centric paradigms, emerging architectures prioritize the value, relevance, and robustness of information delivery. Metrics such as Deliverability of Information (DoI), Redundancy of Information (RoI), and Relevance of Information (RelI) are systematically explored, each contributing uniquely to optimizing information utility across dynamic, resource-constrained communication environments. The detailed roadmap of Section \ref{Network} is illustrated in Fig.~\ref{fig:RMVI}.

\subsection{Overview and Significance}
\begin{figure}[ht]
    \centering
    \includegraphics[width=1\linewidth]{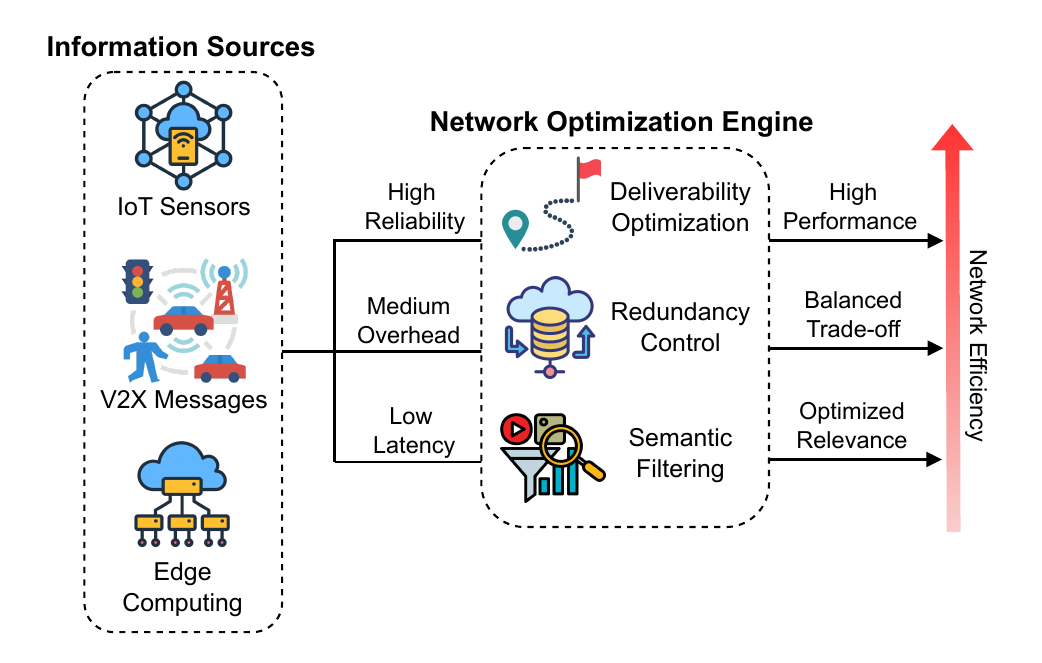}
    \caption{End-to-end information flow optimization framework for network/communication-oriented metrics.}
    \label{fig:network_framework}
\end{figure}

As illustrated in Fig.~\ref{fig:network_framework}, network and communication-oriented information metrics operate through an integrated optimization framework that transforms diverse information sources into performance-optimized applications via intelligent deliverability, redundancy, and semantic filtering mechanisms. This framework encompasses information sources from IoT sensors and V2X messages to edge computing nodes, applying multi-layered optimization that balances high reliability, controlled overhead, and low latency requirements. The framework dynamically adjusts deliverability optimization, redundancy control, and semantic filtering to achieve optimal network efficiency while maintaining service quality.

Network and communication-oriented metrics represent a pivotal rethinking of system design, where delivering the right information under dynamic conditions becomes as critical as delivering it quickly. These metrics enable networks to reason about data value, context, and resilience under disruptions and lay the foundation for intelligent, utility-aware communication systems. The core metrics address distinct communication challenges: DoI measures the network's ability to transmit information despite topology changes and failures successfully. RoI quantifies optimal replication strategies that balance robustness against resource overhead. RelI evaluates semantic importance and contextual utility to prioritize critical data transmission.

These metrics transform networks from simple data pipes into intelligent information substrates that adapt to application needs and environmental conditions. They find essential applications in autonomous vehicle networks, where V2X communications must reliably deliver safety-critical information; UAV systems, which require adaptive routing under dynamic topologies; Industrial IoT networks, which manage sensor data with varying importance levels; and healthcare monitoring systems, which ensure that critical patient information reaches medical staff promptly. Practical measurement involves implementing mobility-aware routing protocols, deploying semantic filtering algorithms, and establishing utility-driven scheduling mechanisms that collectively optimize information flow based on content value rather than mere transmission speed.

\subsubsection{From Transmission to Information-Centric Networks}
The transformation from data pipes to intelligent information substrates has redefined the mission of network architectures.

\paragraph{Metric-Driven Communication Evolution}
Traditional communication systems were optimized primarily for throughput and delay, and assumed that all transmitted data had equal value. However, as information-centric networking (ICN) paradigms emerge, networks are increasingly evaluated by their ability to deliver valuable and trustworthy information under dynamic conditions \cite{Anisetti2022TNSM, Safitri2024TITS, Tan2024TWC}. Anisetti \textit{et al.}~\cite{Anisetti2022TNSM} highlighted the need for continuous security verification in ICN environments and emphasized that metrics must now include trustworthiness and content integrity, not just delivery success. Similarly, Safitri \textit{et al.}~\cite{Safitri2024TITS} surveyed the use of ICN in railway communication systems and identified how quality of service must be redefined in environments where seamless, content-driven delivery outpaces traditional connection-oriented goals. Together, these shifts illustrate that metric frameworks must evolve beyond throughput to capture information reliability, security, and relevance.

\paragraph{Semantic and Intent Awareness}
Networks must also become aware of the meaning and purpose of the data transmitted, which builds on metric-driven communication. Tan \textit{et al.}~\cite{Tan2024TWC} proposed a hybrid-coding access control framework for ICN that dynamically adjusts security and content retrieval based on user authorization and reflected an early step toward semantic-aware networking. Zhang \textit{et al.}~\cite{Zhang2023IOTJ} extended this view and integrated intelligent caching schemes for IoT networks using deep reinforcement learning to predict content popularity under dynamic conditions. These developments indicate that future networks will increasingly prioritize data not solely by destination address, but by content utility, contextual importance, and user intent. This requires a new generation of semantic-aware metrics to guide forwarding and caching decisions.

\subsubsection{Cross-Layer Metric Orchestration}
As information-centric metrics evolve, they must be operationalized coherently across communication layers.

\paragraph{Layer-Coordinated Optimization}
Optimizing deliverability, redundancy, and relevance simultaneously demands cross-layer metric coordination. Nguyen \textit{et al.}~\cite{Nguyen2022TNSE} proposed a Green ICN design where power adaptation, caching behavior, and link rate adjustment are jointly optimized based on dynamic content popularity and demonstrated how physical, MAC, and network layers must collaboratively manage resource efficiency and information utility. Qiao \textit{et al.}~\cite{Qiao2024TGCN} introduced a multipath TCP scheduler enhanced by transformer-based learning and predicted future throughput patterns in open radio access networks (O-RAN) to guide link adaptation decisions. These examples highlight that no single layer can independently guarantee information robustness, and that metric-driven co-optimization must permeate the stack.

\paragraph{Challenges in Metric Integration}
Despite its promise, cross-layer orchestration introduces significant challenges in metric compatibility, feedback timing, and tradeoff resolution. Polese \textit{et al.}~\cite{Polese2023TMC} demonstrated through the ColO-RAN framework that while deep reinforcement learning can dynamically adapt scheduling, slicing, and model training in O-RANs, practical deployment must address the lack of standardized metric interfaces and the difficulty of integrating real-time feedback. Kuo \textit{et al.}~\cite{Kuo2012SECON} showed that balancing multiplexing gain, diversity gain, and energy efficiency in underwater acoustic sensor networks requires highly specialized, cross-layer metric tuning, rather than generic optimization strategies. Finally, Li \textit{et al.}~\cite{Li2025IOTJ} highlighted that minimizing the Age of Information while maximizing throughput necessitates complex metric arbitration across scheduling, power control, and resource allocation, with learning-based adaptation serving as a critical enabler. These studies collectively affirm that future communication systems must embrace metric heterogeneity and intelligent orchestration to achieve real-time, information-value-aware optimization.
\subsection{Deliverability of Information (DoI)}
Deliverability metrics address the need to maintain reliable information transmission across dynamic, heterogeneous, and failure-prone communication environments. This section explores routing resilience strategies against mobility-induced disruptions, followed by adaptive scheduling mechanisms that prioritize information delivery based on utility and resource optimization.

\subsubsection{Resilient Routing and Topological Adaptation}
In highly dynamic networks, achieving sustained deliverability requires routing protocols that can anticipate topology changes and efficiently recover from failures.

\paragraph{Mobility-Aware Reachability Modeling}
Mobility patterns have a significant impact on link stability and routing decisions in dynamic networks. Arafat \textit{et al.}~\cite{ArafatJSEN} proposed an adaptive reinforcement learning-based mobility-aware routing (ARMR) protocol for heterogeneous wireless body area networks and employed link lifetime estimation to select nodes with more extended connection periods and enhance communication reliability under high mobility. Zhang \textit{et al.}~\cite{Zhang2024JOE} developed a robust graph-based bathymetric simultaneous localization and mapping approach for autonomous underwater vehicles and utilized seabed topography and multibeam sonar data to improve position estimates and support navigation resilience against uncertainties. Masip-Bruin \textit{et al.}~\cite{Masip2010LCOMM} introduced a constraint-based routing strategy that integrates prediction mechanisms and link-state cost innovations to mitigate the effects of routing inaccuracy under topology dynamics. These contributions demonstrate that integrating mobility prediction and environmental modeling into routing design enhances the robustness of reachability against network volatility.

\paragraph{Failure-Tolerant Routing Strategies}
Beyond anticipating mobility impacts, networks must also be able to tolerate unexpected link and node failures to maintain information flow. Sourlas \textit{et al.}~\cite{Sourlas2018TNSM} enhanced information resilience in disruptive information-centric networks and introduced satisfied interest tables that enable redirection of interests toward cached content after failures. Naday \textit{et al.}~\cite{Naday2014MNET} addressed source recovery in ICN environments and utilized multiple publishers to recover delivery processes when sources fail. Kini \textit{et al.}~\cite{Kini2010TNET} proposed a proactive backup path calculation and tunneling mechanism that enables fast recovery from dual-link or single-node failures in IP networks. Liu \textit{et al.}~\cite{Liu2009TNET} analyzed the scalability of network resilience using probabilistic graphical models and revealed how resilience properties vary with network size and failure probabilities. These strategies collectively highlight that failure-tolerant deliverability requires proactive protection mechanisms, multipath redundancy, and information-centric recovery frameworks.

\subsubsection{Adaptive Delivery Optimization and Scheduling}
Beyond maintaining paths, efficient information delivery under dynamic conditions demands intelligent scheduling mechanisms that adapt to changing network states and prioritize critical information.

\paragraph{Utility-Aware Forwarding Decisions}
Effective forwarding decisions must optimize not only for connectivity but also for the utility associated with transmitted information. Cheng \textit{et al.}~\cite{cheng2017SIGMOD} formulated utility-aware ridesharing scheduling on road networks and prioritized rider assignments to maximize aggregate utility under spatial-temporal constraints. Song \textit{et al.}~\cite{Song2025MNET} designed Smart-Flycast, an AI-driven multipath transmission system for UAV-based live streaming, which dynamically adjusts multipath transmissions based on network conditions and service requirements to improve Quality of Experience. Aboagye \textit{et al.}~\cite{Aboagye2025TCOM} proposed a mobility-aware resource allocation framework for hybrid THz/VLC wireless networks and optimized user-AP assignments and resource allocation with handoff-aware sum rate and energy efficiency metrics. These studies demonstrate that utility-driven transmission policies, when aligned with network conditions and service semantics, significantly enhance deliverability outcomes.

\paragraph{Learning-Based Scheduling and Adaptation}
Learning-based scheduling approaches enable real-time adaptation to the dynamic requirements of delay sensitivity, resource limitations, and application-driven priorities. Yang \textit{et al.}~\cite{Yang2024IOTJ} proposed a multipolicy deep reinforcement learning framework for multi-objective joint routing and scheduling in deterministic networks, utilising an A3C-based multi-strategy optimisation algorithm and graph convolutional networks to improve schedulability and resource utilisation. Jayanetti \textit{et al.}~\cite{Jayanetti2024TPDS} introduced a multi-agent reinforcement learning framework for renewable energy-aware workflow scheduling across distributed cloud datacenters and minimized energy consumption while maintaining workflow deadlines. Lu \textit{et al.}~\cite{Lu2023TPDS} designed RLPTO, a reinforcement learning-based mechanism for performance-time optimized task and resource scheduling in distributed machine learning environments that balances resource utilization and training convergence time. Papandriopoulos and Evans~\cite{Papandriopoulos2008TNET} developed distributed protocols for optimal cross-layer design of physical and transport layers in mobile ad hoc networks and achieved global optimality through convex optimization techniques. Akan~\cite{Akan2004JSAC} proposed an adaptive transport layer suite for next-generation wireless Internet that dynamically adjusts congestion control mechanisms to heterogeneous wireless architectures. Collectively, these contributions demonstrate that reinforcement learning, multiagent coordination, and adaptive congestion control strategies are central to future deliverability optimization under complex and dynamic conditions.
\subsection{Redundancy of Information (RoI)}
Redundancy-oriented information metrics aim to enhance data delivery robustness through replication and coding, while carefully balancing the associated overhead to maintain efficiency. This section discusses both the application of redundancy for resilience and strategies for suppressing unnecessary redundancy to optimize system performance.

\subsubsection{Replication and Coding for Robustness}
Robust communication in dynamic and failure-prone environments benefits significantly from redundancy in transmission paths and coding strategies that improve data survivability.

\paragraph{Multi-Path and Coded Replication}
Multipath forwarding and coding techniques provide strong resilience against network disruptions. Zhang \textit{et al.}~\cite{Zhang2015ICCC} proposed an LT code-aided multipath forwarding strategy for content-centric networking and transmitted coded content over multiple disjoint paths to reduce download time and enhance load balancing. Fagundes \textit{et al.}~\cite{Fagundes2024TIE} designed a redundancy-based microgrid optimization strategy and integrated intelligent algorithms to coordinate redundant battery and fuel cell resources under real-time pricing constraints, which improved system reliability and operational efficiency. Chen \textit{et al.}~\cite{Chen2015TCC} introduced a k-out-of-n computing framework for mobile cloud environments and allowed mobile devices to retrieve or process data from any subset of available servers to achieve fault tolerance and energy efficiency. These works highlight how redundancy, whether through network coding or system-level resource allocation, serves as a core mechanism for enhancing robustness in communication and computation systems.

\paragraph{Service Reliability via Redundancy}
Targeted redundancy mechanisms further ensure service-level reliability in challenging environments. Ramasubramonian \textit{et al.}~\cite{Ramasubramonian2010LSP} combined practical network coding with multiple description coding for video multicast and enabled efficient quality-of-service control across heterogeneous user conditions. Peng \textit{et al.}~\cite{Peng2025TNET} proposed redundancy management strategies for edge computing systems and analyzed optimal replication factors that balance latency reduction and resource availability. Li \textit{et al.}~\cite{Li2024TAC} developed a secure distributed state estimation framework that exploits observability redundancy to achieve robust sensor network performance under sparse integrity attacks. These studies illustrate that well-designed redundancy schemes are crucial for maintaining service continuity, even under failure or adversarial conditions.

\subsubsection{Suppressing Redundancy for Efficiency}
While redundancy improves resilience, excessive replication can strain bandwidth, energy, and storage resources. Efficient redundancy management thus becomes critical.

\paragraph{Adaptive Redundancy Control}
Adaptive mechanisms dynamically regulate redundancy in response to network conditions and task requirements. Hu \textit{et al.}~\cite{Hu2021ACCESS} proposed a coding-based distributed congestion-aware packet spraying method for data center networks and adjusted coding redundancy in response to asymmetry among paths to eliminate packet reordering and minimize flow completion times. Wang \textit{et al.}~\cite{Wang2021CVPR} developed a structural redundancy reduction approach for convolutional neural network pruning and statistically modeled network pruning to identify and remove structural redundancies for compact and efficient architectures. Pei \textit{et al.}~\cite{Pei2024TITS} presented a dual-model pruning method for federated learning in intelligent transportation systems and reduced communication overhead and maintained model robustness through two-stage pruning at client and server levels. These approaches demonstrate that adaptive redundancy control must be context-aware and balance performance goals against resource limitations.

\paragraph{Balancing Overhead and Robustness}
Redundancy must be carefully calibrated to optimize both robustness and efficiency. Li \textit{et al.}~\cite{Li2024Neuro} introduced a redundancy graph decomposition method for CNN filter pruning and achieved high compression rates with minimal accuracy loss by systematically removing redundant filters. Park \textit{et al.}~\cite{Park2018TCOM} designed low-density parity-check codes for distributed storage systems and minimized repair bandwidth while enhancing reliability through tailored graph constructions. Distler \textit{et al.}~\cite{Distler2016TC} proposed resource-efficient Byzantine fault-tolerant systems that reduce resource usage during normal operations by passivating a subset of replicas and reactivate them only upon fault detection. These studies underscore that the tradeoff between redundancy and resource consumption must be dynamically managed to achieve resilient yet efficient system designs.
\subsection{Relevance of Information (RelI)}
Relevance-oriented information metrics focus on optimizing the delivery and processing of information by dynamically evaluating its semantic importance and task-specific utility. Rather than treating all data equally, relevance-aware systems prioritize content based on contextual value and significantly enhance efficiency in resource-constrained and dynamic environments.

\subsubsection{Semantic Filtering and Importance Estimation}
Managing the overwhelming volume of information requires filtering mechanisms that intelligently assess semantic relevance and assign importance to different data streams.

\paragraph{Task-Driven Content Filtering}
To enable selective information delivery, systems must first be capable of discerning content relevance based on task requirements and environmental context. One emerging solution involves augmenting metadata to enrich open IoT data descriptions, as demonstrated by Montori \textit{et al.}~\cite{Montori2023IOTJ}, who proposed a cascading ensemble classification framework that autonomously annotates heterogeneous datasets for more effective integration and retrieval. Complementing metadata-based approaches, Sun \textit{et al.}~\cite{Sun2022TASE} integrated location-aware graph embeddings into collaborative filtering models and enabled intelligent IoT service discovery by simultaneously considering service quality and geographic context. In mobile environments such as delay-tolerant networks, Chang \textit{et al.}~\cite{Chang2013WCNC} emphasized the importance of social and community structures and introduced the CROP protocol, which dynamically routes messages based on community relevance rather than simple path metrics. Together, these advances reveal that task-driven semantic filtering must incorporate multiple dimensions of context, including metadata, spatial proximity, and social ties, to maximize relevance in large-scale, decentralized systems.

\paragraph{Importance Scoring for Prioritization}
Beyond filtering, systems require mechanisms to quantify the relative importance of information to prioritize delivery when resources are constrained. A foundational step in this direction is the analysis of semantic error decay, as addressed by Shi \textit{et al.}~\cite{Shi2023TWC}, who derived theoretical bounds for excess distortion in semantic-aware MIMO systems and offered a new metric for evaluating transmission urgency. Moving toward practical communication system design, Xu \textit{et al.}~\cite{Xu2025JSAC} introduced semantic values derived from generative foundation models and linked perceptual similarity with transmission fidelity and enabled power allocation strategies that inherently prioritize high-importance data. In publish/subscribe networks, Zhang \textit{et al.}~\cite{Zhang2017ICDCS} extended traditional covering optimization to relevance-based top-k filtering and allowed scalable systems to deliver only the most semantically critical updates to subscribers. These developments collectively suggest that accurate, application-driven importance scoring is essential to guide information prioritization across diverse networked systems.

\subsubsection{Relevance-Aware Scheduling and Queueing}
Even after filtering and importance assessment, systems must actively schedule transmissions to ensure that critical information is delivered promptly and reliably under dynamic conditions.

\paragraph{Queueing with Relevance Awareness}
Effective queue management requires policies that dynamically adjust based on the current relevance and urgency of queued data. Zhao \textit{et al.}~\cite{Zhao2024TASE} proposed a deep reinforcement learning-based reactive scheduling framework for dynamic job shops, where attention-based networks prioritize pending tasks not based on fixed dispatching rules but directly on relevance to delivery goals. In wireless networks, Liu \textit{et al.}~\cite{Liu2006TIT} showed that scheduling based on utility-gradient principles, where packet transmission decisions incorporate both channel quality and delay-sensitive utility functions, can significantly optimize overall system performance. Extending relevance-aware scheduling into softwarized 5G environments, Petrov \textit{et al.}~\cite{Petrov2018JSAC} developed dynamic orchestration strategies that prioritize mission-critical traffic across radio access and core networks and ensured end-to-end reliability even under resource contention. These examples demonstrate that relevance-aware queue management, which blends semantic priority and real-time adaptability, is crucial for achieving timely and efficient information delivery in modern networks.

\paragraph{Utility-Guided Transmission Policies}
The final step in relevance-driven information management involves designing transmission policies that adaptively balance utility maximization with resource limitations. Recent surveys by Aceto \textit{et al.}~\cite{Aceto2014MCOM} highlighted the role of AI in enabling future traffic classification and management systems that can dynamically prioritize flows based on learned utility patterns. In the context of energy-constrained scheduling, Guan \textit{et al.}~\cite{Guan2000TPS} modeled the realizability of energy delivery schedules and emphasized the need to align workload scheduling with system constraints, such as ramp rates. Fu \textit{et al.}~\cite{Fu2022TNET} further advanced the field and introduced Learning-NUM. This dynamic network utility maximization framework can learn user utility functions online despite delayed feedback, which enabled adaptive rate control. At the edge, Dong \textit{et al.}~\cite{Dong2023TWC} proposed LOMES, a Lyapunov-based workload scheduling algorithm that balances system utility and queue stability across heterogeneous servers. In cloud environments, Balla and Dutta~\cite{Balla2018ICDIS} integrated reinforcement learning with queuing theory to enhance task reliability under dynamic conditions. These works collectively show that utility-guided transmission policies, when informed by relevance and dynamically optimized, are pivotal for realizing efficient, robust, and user-centric networked systems.
\subsection{Lessons Learned}

\begin{table*}[!t]
\centering
\caption{Network/Communication-oriented information metrics: comparative analysis and optimization methods.}
\label{tab:network_communication_metrics}
\renewcommand{\arraystretch}{1.5}
\begin{tabular}{|p{2.4cm}|p{4.3cm}|p{4.4cm}|p{4cm}|}
\hline
\rowcolor[gray]{0.85} \textbf{Key Features} & \textbf{Deliverability of Information (DoI)} & \textbf{Redundancy of Information (RoI)} & \textbf{Relevance of Information (RelI)} \\
\hline
Definition & 
Measures information reachability and reliable transmission across dynamic, heterogeneous, and failure-prone communication environments \cite{ArafatJSEN, Sourlas2018TNSM, Zhang2024JOE}. & 
Quantifies data delivery robustness through replication and coding strategies while balancing associated overhead costs \cite{Zhang2015ICCC, Fagundes2024TIE, Chen2015TCC}. & 
Evaluates information semantic importance and task-specific utility for optimized delivery and processing \cite{Montori2023IOTJ, Sun2022TASE, Sun2022TASE}. \\
\hline
Primary Applications & 
Heterogeneous wireless body area networks, disruptive information-centric networks, and autonomous underwater vehicles \cite{ArafatJSEN, Sourlas2018TNSM, Zhang2024JOE}. & 
Content-centric networking, redundancy-based microgrids, mobile cloud environments, and video multicast systems \cite{Zhang2015ICCC, Fagundes2024TIE, Chen2015TCC}. & 
IoT data annotation, collaborative filtering systems, delay-tolerant networks, and semantic communication \cite{Montori2023IOTJ, Sun2022TASE, Chang2013WCNC}. \\
\hline
Analytical Techniques & 
Reinforcement learning-based routing, graph-based SLAM approaches, and information resilience modeling \cite{ArafatJSEN, Zhang2024JOE, Sourlas2018TNSM}. & 
Multi-path forwarding strategies, economic operation optimization, and energy-efficient fault-tolerant frameworks \cite{Zhang2015ICCC, Fagundes2024TIE, Chen2015TCC}. & 
Cascading ensemble classification, location-aware graph embeddings, and community-relevance based routing \cite{Montori2023IOTJ,Sun2022TASE, Chang2013WCNC}. \\
\hline
Typical Optimization Approaches & 
ARMR protocols with Q-learning, link lifetime estimation, and satisfied interest tables for failure recovery \cite{ArafatJSEN, Masip2010LCOMM, Naday2014MNET}. & 
Multi-path coding strategies, intelligent algorithm coordination, and k-out-of-n computing frameworks \cite{Zhang2015ICCC, Fagundes2024TIE}. & 
Metadata-assisted frameworks, collaborative filtering with graph embeddings, and opportunistic routing protocols \cite{Montori2023IOTJ, Sun2022TASE}. \\
\hline
Main Limitations & 
Dynamic topology challenges, mobility-induced disruptions, and scalability in heterogeneous environments \cite{Masip2010LCOMM, Kini2010TNET, Liu2009TNET}. & 
Resource overhead vs. robustness trade-offs, computational complexity, and energy consumption optimization \cite{Hu2021ACCESS, Wang2021CVPR, Pei2024TITS}. & 
Semantic understanding complexity, context adaptation challenges, and computational overhead in real-time processing \cite{Shi2023TWC, Zhang2017ICDCS, Petrov2018JSAC}. \\
\hline
\end{tabular}
\end{table*}

Network and communication-oriented information metrics mark a profound shift from conventional throughput and delay-based evaluations toward intelligent, relevance-driven optimization of information flow. Throughout this section, a cohesive narrative emerges: information systems are evolving from raw data transmission platforms to context-aware, utility-maximizing infrastructures. Deliverability metrics prioritize resilient and adaptive delivery under dynamic topology changes; redundancy metrics balance robustness and efficiency through intelligent replication control; relevance metrics refine transmission decisions by embedding semantic filtering and task-driven prioritization into scheduling frameworks.

This evolution signals a transformation in the operational logic of networked systems. Information relevance, semantic significance, and utility tradeoffs are becoming primary optimization objectives, which influence routing, coding, resource allocation, and queue management in real time. Next-generation networks will increasingly require cross-layer coordination of these metrics, unified semantic-aware frameworks for information valuation, and AI-driven adaptation mechanisms that dynamically align communication behavior with task-critical objectives. This vision will demand the seamless integration of semantic modeling, learning-based scheduling, and utility-guided redundancy across heterogeneous and mission-critical environments.

\section{Application Scenarios and Case Studies}\label{Application}
\subsection{Autonomous Vehicles and Intelligent Transportation}
Autonomous Vehicles (AVs) and Intelligent Transportation Systems (ITS) demand exceptional information quality across multiple dimensions for safe and efficient operation. These complex systems must process sensor data, V2X messages, and control commands with stringent requirements for timeliness, reliability, relevance, and security \cite{Guo2023MNET}. Managing the coexistence of heterogeneous services, which range from safety-critical status updates focused on freshness to data-intensive communications that require low latency and high reliability, presents significant challenges \cite{Guo2023MNET}. We illustrate how multi-dimensional information metrics, optimized via AI, address the critical needs of AVs and ITS.

\subsubsection{Criticality of Timeliness: AoI and PAoI in AV Perception and Control}
Safe navigation hinges on real-time environmental perception and control. Stale sensor data or delayed commands can lead to catastrophic failures. Time-oriented metrics, such as Age of Information (AoI) and Peak Age of Information (PAoI), are fundamental \cite{Guo2023MNET}. AoI minimization is crucial for scheduling interdependent tasks within an ADS, as it directly impacts response time and throughput. AI-based schedulers showed promise in outperforming traditional methods, even with fewer resources \cite{Xu2022INFOCOM}. In cooperative driving scenarios that use MEC, freshness optimization requires joint scheduling of vehicle transmissions and edge processing to minimize AoI for all relevant information flows \cite{Sorkhoh2022TITS}. This complex optimization used advanced techniques, such as Benders decomposition to find near-optimal solutions efficiently, which ensured timely decisions based on fresh data \cite{Sorkhoh2022TITS}. Low AoI and bounded PAoI are central themes in AV design, particularly for safety-critical functions.

\subsubsection{Connectivity and Relevance: Network \& Quality Metrics in V2X}
V2X communication extends AV awareness and enables cooperation and safety alerts, but requires reliable and relevant information exchange in dynamic networks. Metrics like Deliverability of Information (DoI), Relevance of Information (RelI), and Semantics of Information (Sem-oI) are vital. Context-aware architectures dynamically adapt communication profiles across ITS-G5 and C-V2X based on application needs and environmental conditions, which enhances quality of service (QoS) and DoI \cite{Sepulcre2018CN}. However, large volumes of data exchange, such as for collaborative perception, can overload V2X networks. SemCom offered a solution by extracting and transmitting only task-relevant semantic features \cite{Feng2024MNET, Gimenez2024MNET}. AI-driven techniques can prioritize transmissions based on information relevance or importance \cite{Fang2025TON}, use semantic understanding to filter messages \cite{Feng2024MNET}, and adapt compression or transmission strategies to channel conditions \cite{Fang2025TON}. Focusing on semantic meaning rather than raw data significantly reduced network load and latency, which made future ITS applications viable despite spectrum constraints \cite{Feng2024MNET, Gimenez2024MNET,Huang2025TMC}.

\subsubsection{Safety and Trust: Reliability \& Security Metrics in ITS}
The safety requirements of AVs and ITS require robust mechanisms for information reliability and security. Metrics such as Security of Information (Sec-oI), Survivability of Information (SuI), and Age of Incorrect Information (AoII) are crucial. Trust in vehicular networks, particularly for crowdsourced services, can use decentralized blockchain frameworks that transparently manage trust scores related to Sec-oI and reliability \cite{Wang2020Trustcom}. System resilience relies on fault tolerance; scalable consensus protocols tailored for vehicular networks' localized interactions enhanced reliability against node failures compared to traditional BFT algorithms \cite{Deshmukh2025TVT,Wu2025WASA,Wu2025ARXIV}. Security remained a significant concern, as critical components, such as multi-sensor fusion algorithms were vulnerable to false data injection attacks, which required thorough security analysis and adaptive defense strategies \cite{Hu2025IF,Wu2023MPE}. The concept of trustworthiness, potentially integrating AoII to account for both timeliness and correctness in the face of potential attacks or errors, has become increasingly important, especially for advanced 6G applications that involve ITS \cite{Chen2025MNET}.
\subsection{Wireless Sensor Networks and Industrial IoT}
Wireless Sensor Networks (WSNs) and Industrial Internet of Things (IIoT) underpin numerous monitoring and automation applications, often deploying vast numbers of nodes under strict resource constraints, particularly energy. Unlike the high-bandwidth, low-latency requirements often seen elsewhere, WSNs/IIoT frequently prioritize network lifetime, data utility, and operational robustness. Information metrics are essential for optimizing performance, managing resources effectively, and ensuring the delivery of valuable and reliable data from potentially constrained or inaccessible devices. This section examines the application of information metrics tailored to the unique challenges of WSNs and IIoT.

\subsubsection{Timeliness vs. Cost in Resource-Constrained WSNs}
A central challenge in WSNs is balancing the need for timely information, often measured by AoI, against the imperative to conserve energy— a key aspect of the Cost of Information (CoI) , and extend network lifetime. Frequent transmissions minimize AoI but rapidly deplete node batteries. This trade-off is particularly relevant in systems that incorporate energy harvesting, where adaptive scheduling is necessary to manage stochastic energy availability and sampling instants to optimize freshness metrics like PAoI while minimizing power consumption \cite{Fang2022JSAC}. Techniques such as adaptive sleep-scheduling and active queue management reduced energy costs while maintaining acceptable information freshness \cite{Fang2022JSAC, Wu2023access, Fang2022TVT,Pan2023SCIS}. The broader economic perspective, which encompasses the Value of Information (VoI) derived from sensor data versus the Cost of Processing (CoP) associated with its collection, is also crucial for designing sustainable and user-accepted Internet of Things (IoT) systems \cite{Turgut2017MCOM}. AI techniques, such as deep reinforcement learning, found optimal resource management strategies that navigate these complex trade-offs, for instance, minimizing average PAoI while considering power dissipation in innovative health systems \cite{Wu2023access}.

\subsubsection{Scalability, Reliability, and Security in Large-Scale Deployments}
WSNs/IIoT deployment often involves numerous devices spread over large areas, which requires scalable, reliable, and secure operation. Information DoI and network SuI in the face of node failures, unreliable channels, or attacks are critical. Scalability challenges are addressed through techniques like network clustering and the use of mobile chargers/data collectors, which can efficiently recharge nodes and gather data via optimized itineraries. These approaches improved both energy efficiency and data delivery timeliness compared to static approaches \cite{Wedaj2023TGCN}. Efficient consensus protocols, like Byzantine Fault Tolerant (BFT) mechanisms tailored for wireless environments, are crucial for maintaining data integrity and network reliability, especially when using blockchain for secure service provision \cite{Zhang2024TWC}. Randomized proposer selection and efficient signature schemes enhanced throughput and resilience against attacks in such protocols \cite{Zhang2024TWC}. Additionally, simulation tools are crucial for evaluating the performance of large-scale, heterogeneous IoT applications that involve cloud/edge integration prior to deployment. Scalable simulators, such as TinySim, enable developers to assess performance metrics like energy and latency with high fidelity and interactivity \cite{Chen2023IOTJ}. Sec-oI is a paramount concern, particularly for low-cost bare-metal IoT devices often managed via companion apps. Systematic analysis frameworks identified vulnerabilities in crucial processes like Device Firmware Updates across authentication, acquisition, and verification stages, as flaws can expose systems to significant threats \cite{Xue2024TDSC}.

\subsubsection{Information Metrics for Industrial Control and Monitoring}
In IIoT, particularly for real-time wireless feedback control systems, information quality directly impacts control performance, stability, and safety. While AoI is relevant, the study found that minimizing it may not always optimize control performance \cite{Tong2024IOTJ}. Metrics like AoII or VoI can better capture the timeliness requirements for control stability. These metrics led to information update policies that optimize control metrics while significantly reducing communication resource consumption compared to purely AoI-minimal approaches \cite{Tong2024IOTJ}. The quality and reliability of sensed data are also vital for industrial applications, such as food quality assessment or power device health monitoring. Machine learning-based multi-sensor fusion techniques provided accurate quality assessments and predictions \cite{Song2024TII}, while effective data processing suppressed environmental noise and enabled reliable health indicators even under dynamic conditions \cite{Peng2024IOTJ}. In collaborative perception tasks, common in multi-robot or vehicular IIoT settings, metrics that capture the timeliness of perceived targets (Age of Perceived Targets, AoPT) are used alongside task-relevance-guided adaptive communication strategies to balance inference accuracy, communication overhead, and robustness to packet corruption \cite{fang2024ARXIV}. Also, collecting valuable key-value data from numerous IoT devices for service improvement must contend with privacy concerns. Utility-optimized Local Differential Privacy (LDP) mechanisms were designed to provide tailored privacy protection based on data sensitivity, which enhanced the Utility of Information (UoI) of collected data compared to standard LDP methods while preserving user privacy \cite{Wang2025IOTJ}.
\subsection{Real-time Healthcare and Digital Twins}
Networked healthcare systems, which encompass real-time remote patient monitoring, telehealth services, and the sophisticated paradigm of Healthcare Digital Twins (HDTs), impose stringent requirements on the quality of information. Clinical efficacy and patient safety often depend on the timeliness, correctness, reliability, and security of data exchanged between sensors, actuators, diagnostic systems, and virtual models. This section examines how various information metrics are leveraged, often in conjunction with AI, to address the unique demands of modern digital health applications.

\subsubsection{Timeliness and Reliability for Critical Monitoring and Intervention}
In critical healthcare scenarios, such as remote monitoring of vital signs or time-sensitive diagnostic processes, the freshness of information is paramount. AoI is a key metric used to characterize the timeliness of physiological status updates transmitted from healthcare IoT devices \cite{Ling2022IOTJ}. AoI minimization often involves trade-offs, for instance, with energy consumption in energy-harvesting sensor devices. Distributionally robust optimization techniques, potentially using Conditional Value-at-Risk (CVaR) methods, minimized average AoI under energy and transmission constraints, even with imperfect channel knowledge \cite{Ling2022IOTJ}. Low latency and high Quality of Service (QoS) for real-time data processing is vital, particularly in large-scale systems. AI-driven resource management, potentially utilizing aerial computing platforms like Unmanned Aerial Vehicles (UAVs) within a Software-Defined Network (SDN)-IoT architecture, can dynamically allocate bandwidth and processing resources to reduce latency and energy consumption for critical applications like telemedicine \cite{Lv2025}. Also, the integration of AoI optimization directly into diagnostic frameworks, such as federated digital twins that use split learning and AI-Generated Content (AIGC), has significantly enhanced the timeliness of medical responses. Deep reinforcement learning optimized data and model updates, which reduced AoI and improved the potential for prompt, accurate diagnostics \cite{Wu2025MNET}.

\subsubsection{Information Quality and Consistency for Healthcare Digital Twins}
HDTs, which create dynamic virtual replicas of patients or healthcare processes, offer transformative potential for personalized medicine, treatment planning, and system optimization \cite{Benedictis2023JBHI, Chen2024IOTJ}. The value of an HDT hinges on the quality, accuracy, and consistency of the underlying information used to build and update it. HDT establishment required the integration of diverse data sources, management of high-fidelity virtual modeling, and assurance of strong bidirectional information interaction between the physical and digital realms, often challenged by scarce or noisy data \cite{Chen2024IOTJ}. Generative AI (GAI) emerged as a promising tool to address these data challenges and empower HDT creation and application across the healthcare spectrum, from personalized monitoring to rehabilitation \cite{Chen2024IOTJ}. Generalized Digital Twin (DT) frameworks helped identify the key functional components needed for effective data exchange and virtualization \cite{Benedictis2023JBHI}. The vision extended towards a collaborative Virtual Human Twin (VHT) infrastructure, intended as a shared resource of data, models, and standards to accelerate the development, validation, and clinical adoption of digital twins in healthcare. This infrastructure supported researchers, industry, clinicians, and patients \cite{Viceconti2024JBHI}. This requires robust metrics for data quality (UoI) and semantic consistency (Sem-oI) to ensure the twins are valid and useful representations.

\subsubsection{Security, Privacy, and Ethics in Networked Healthcare}
The digitization of healthcare and the use of HDTs introduce significant security and privacy challenges due to the highly sensitive nature of patient data. Compromising a healthcare DT could have severe real-world consequences, which makes Sec-oI a critical concern \cite{KarakraAICCSA}. Attackers may target various components of the DT ecosystem, which requires comprehensive threat modeling that considers the interplay between IoT, cloud computing, and AI technologies involved \cite{KarakraAICCSA}. Medical record protection, a core component of HDTs, requires advanced security measures. Storing sensitive records directly in the cloud, even encrypted, posed risks; solutions that combined post-quantum searchable encryption with blockchain for verification offered enhanced security and privacy for cloud-based medical record systems that support DTs \cite{Yi2023JCC}. Beyond reactive defense, DTs themselves were leveraged to bolster cybersecurity. By creating virtual replicas of healthcare IoT systems, security professionals simulated cyber-attacks, identified vulnerabilities, and tested security mechanisms in a safe environment without disrupting real-time services or compromising patient safety, which enabled dynamic and adaptive security solutions \cite{Pirbhulal2022VTC}. Addressing these Sec-oI and privacy requirements is fundamental for building trustworthy digital health systems.
\subsection{UAV-assisted Communication Systems}
UAVs, or drones, introduce flexibility to wireless networks and can act as mobile base stations, relays, or data collectors, especially in areas that lack terrestrial infrastructure or during emergency responses \cite{Xing2023IOTJ}. Their mobility and potential for establishing favorable line-of-sight links are key advantages. However, UAV-assisted system optimization requires addressing challenges unique to aerial platforms, which include limited energy, highly dynamic network topologies, and the need to ensure the quality and timeliness of collected or relayed information. This section examines how various information metrics guide the design and operation of these dynamic aerial communication systems.

\subsubsection{Information Freshness Optimization with Mobile UAVs}
When UAVs collect data, particularly from IoT devices or sensors that monitor time-varying phenomena, maintaining information freshness, measured by AoI, is often critical \cite{Zhang2025TWC, Yin2021VTC}. AoI minimization requires joint optimization of UAV trajectories, data collection schedules, and potentially task completion times, which considers complex factors like realistic air-to-ground channel models \cite{Zhang2025TWC}. This optimization is often non-convex and complex, which makes traditional methods difficult \cite{Zhang2025TWC, Yin2021VTC}. AI approaches like Deep Reinforcement Learning (DRL) were often used to find effective UAV trajectories and scheduling policies that minimize AoI or guarantee maximum AoI thresholds \cite{Yin2021VTC, Yang2025TWC}. Federated learning enhanced DRL training across multiple UAVs while preserving privacy \cite{Zhang2025TWC}. A major constraint is the UAV's limited energy; therefore, visiting planning must balance AoI minimization with maximizing the number of visited nodes or collected data volume within the UAV's energy budget \cite{Hong2023VTC}. Optimization techniques like the Reformulation-Linearization Technique (RLT) and Alternating Direction Method of Multipliers (ADMM) \cite{Hong2023VTC}, or hierarchical DRL combined with optimization methods like Semidefinite Relaxation (SDR) \cite{Yang2025TWC}, tackled these constrained AoI-energy trade-offs efficiently.

\subsubsection{Reliable Connectivity in Dynamic UAV Networks}
The inherent mobility of UAVs leads to dynamic network topologies, which poses challenges for maintaining reliable communication links (DoI) and overall network resilience (SuI). In UAV mesh networks managed using SDN, dynamic topology changes increased control overhead and reduced routing reliability \cite{Wang2023ICCC}. Mechanisms like Directed Acyclic Graph (DAG)-based reliable routing (DBRR), which consider link quality and expiration time to select optimal next hops and set timeouts, improved routing reliability and reduced overhead in these dynamic scenarios \cite{Wang2023ICCC}. Understanding the unique characteristics of air-to-ground links is also crucial; studies showed that UAV-assisted WSN links may differ significantly from terrestrial WSN links, with compressed connectivity regions and potentially deceptive link quality indicators (Received Signal Strength Indicator, RSSI; Signal-to-Noise Ratio, SNR), which required careful protocol design \cite{Xia2021WCNC}. Since UAVs are often deployed for mission-critical tasks, rigorous reliability modeling, analysis, and design that consider various failure causes are essential \cite{Xing2023IOTJ}. Multi-UAV cooperation offered a promising approach to enhance power efficiency and potentially communication reliability by forming large virtual antenna arrays, which allowed joint optimization of trajectories, cooperation decisions, and beamforming to overcome individual UAV Size, Weight, and Power (SWAP) constraints \cite{Xiang2020WCNC}.

\subsubsection{Intelligent Data Management and Task Offloading via UAVs}
UAVs can serve beyond simple relays; they can act as intelligent nodes for data processing and task execution at the network edge. Their operation optimization involves considering the value or relevance of the information they handle. In dual-functional radar-communication multi-UAV networks, joint optimization of UAV location, user association, and power control aimed to maximize network utility (UoI) under sensing accuracy constraints, which used spectral clustering and game theory \cite{Wang2021TCOM}. When UAVs performed tasks like object detection, semantic communication that leveraged knowledge graphs significantly reduced data transmission volume compared to transmitting raw images, which improved communication robustness and computational efficiency even under poor channel conditions or high compression rates \cite{Song2024ICC}. UAVs are increasingly used as Mobile Edge Computing (MEC) platforms to assist IoT devices. Adaptive learning frameworks that integrated Multi-Objective Reinforcement Learning (MORL) and optimization techniques like Lyapunov optimization managed UAV trajectories and resource allocation to balance conflicting objectives such as task completion rates, energy efficiency, and system stability in dynamic environments \cite{Bakhrani2025IOTJ}. Also, UAVs supported joint MEC and data collection (DC) tasks, which required sophisticated DRL-based approaches to optimize UAV movement, user power, and association for minimizing latency and maximizing collected data volume simultaneously \cite{Wang2025IOTJ1}.
\subsection{Agent-to-Agent (A2A) Communication in LLM Ecosystems}
Large Language Models (LLMs) have spurred the development of sophisticated AI agents capable of complex reasoning and task execution. Effective communication between these agents (A2A) is crucial for collective intelligence, which enables them to collaborate, share knowledge, and coordinate actions in distributed ecosystems. Unlike traditional data communication, A2A interactions are driven by intent, semantics, and the utility of information to achieve shared or individual goals. This section examines how various information metrics are essential for optimizing these communication processes.

\subsubsection{Semantics, Relevance, and Utility in Goal-Driven A2A Dialogue}
In LLM-based multi-agent systems, communication is the cornerstone for learning, reasoning, and achieving specific objectives. The quality of information exchanged, in terms of its semantic meaning (Sem-oI), contextual relevance (RelI), and contribution to task goals (UoI), dictates the overall system performance. LLM agents were trained and adapted their policies through communication that integrated diverse feedback, both linguistic and non-linguistic, which used universal buffers and iterative learning pipelines to enhance their versatility in various environments \cite{Wang2025ACL}. To overcome the limitations of individual LLMs in complex reasoning, collaborative multi-agent frameworks leveraged efficient knowledge-sharing and communication-driven networks, which enabled synergistic reasoning capabilities \cite{Das2023CIC}. Multi-agent systems like CommLLM were designed to enhance LLM potential in specific domains such as 6G communications by employing specialized agents for data retrieval, collaborative planning, and solution evaluation/refinement, which ensured the communication content is highly relevant and useful for the task at hand \cite{Jiang2024MWC}. The adaptability of A2A dialogue was also highlighted in frameworks where multiple LLM agents, each with a distinct persona, engaged in role-playing communication to autonomously tackle novel challenges and problem-solving scenarios, which demonstrated the utility of nuanced interaction \cite{Rasal2024Arxiv}. Also, LLM-powered multi-agent frameworks facilitated highly targeted goal-oriented learning, for example, in intelligent tutoring systems by accurately mapping learner objectives to required skills and personalizing learning paths and content, which showcased a clear UoI for educational outcomes \cite{Wang2025Arxiv}.

\subsubsection{Efficiency, Coordination, and Trust in Multi-Agent LLM Systems}
The overall efficiency, coordination, and trustworthiness of communication are critical for scalable and reliable multi-agent LLM systems. A significant challenge is the substantial token overhead and associated economic costs (CoI) inherent in many multi-agent communication pipelines \cite{Zhang2024Arxiv}. Frameworks like \texttt{AgentPrune} addressed this by identifying and pruning redundant or even malicious messages from the inter-agent communication graph, which achieved comparable performance to state-of-the-art systems at a fraction of the token cost and also enhanced robustness against certain adversarial agent behaviors \cite{Zhang2024Arxiv}. Efficient coordination also benefited from dynamic agent collaboration, where frameworks like Dynamic LLM-Powered Agent Network (\textbf{DyLAN}) automatically selected optimal teams of agents for specific tasks and allowed them to collaborate within dynamic communication structures, which improved performance on various reasoning and decision-making benchmarks \cite{Liu2024Arxiv}. The CoI also extends to the foundational processes of training and inferring from LLMs. Systems like Lins optimized the Zero Redundancy Optimizer (ZeRO) for LLM training by employing flexible sharding strategies, which significantly reduced communication overhead and improved training throughput at scale \cite{Chen2024IWQoS}. Similarly, for LLM inference, Optimal Sample Compute Allocation (OSCA) by finding an optimal mix of inference configurations yielded better accuracy with substantially less compute, particularly crucial for agentic workflows \cite{Zhang2025scaling}. However, the increasing sophistication of LLMs brings forth significant security and privacy challenges (Sec-oI). LLMs were vulnerable to attacks such as jailbreak, data poisoning, and leakage of personally identifiable information, which can be exploited in A2A settings \cite{Das2025ACS}. LLM-based Code Completion Tools (LCCTs), with their unique workflows and reliance on proprietary datasets, presented distinct security risks, which included successful jailbreak and training data extraction attacks \cite{Cheng2025AAAI}. Addressing these vulnerabilities and optimizing the broader economic costs of LLM usage by predicting output quality and selecting appropriate models to balance cost, quality, and latency are essential for building efficient, reliable, and trustworthy A2A ecosystems \cite{Shekhar2024Arxiv}.
\subsection{Metaverse and Persistent Virtual Environments}
The Metaverse, envisioned as an immersive and persistent amalgamation of interconnected virtual worlds, presents challenges for information management. Its goal of enabling seamless, real-time interaction among numerous users and digital objects requires rethinking how information is generated, transmitted, processed, and secured. The quality, consistency, and timeliness of information are paramount for enhancing user experience, maintaining the integrity of virtual economies, and ensuring the overall viability of these complex ecosystems. This section examines the role of various information metrics in addressing these unique demands.

\subsubsection{Information Consistency and Real-time Interaction}
A fundamental requirement for the Metaverse is the consistent and synchronized perception of the virtual environment across all participants. Delays or discrepancies in state updates can severely degrade immersion and utility. Adaptive sampling and transmission schemes were crucial for managing the massive information exchange needed for frequent updates of panoramic information, which minimized statistical AoI by considering user attention and varying wireless channel conditions \cite{Xiao2024JSAC}. In specialized domains like healthcare metaverses, data freshness (AoI) for sensitive information was critical, with decentralized approaches like blockchain-empowered federated learning explored for privacy-preserving and incentivized data sharing \cite{Kang2024TCCN}. The computational demands of rendering and processing millions of virtual objects for immersive experiences often require distributed edge computing. To ensure low latency and resilience against stragglers or malicious servers in such setups, Resilient, Secure, and Private Coded Distributed Computing (RSPCDC) schemes for tasks like matrix-vector multiplication were proposed to optimize computation latency and reliability \cite{Qiu2024TCCN}. Also, the industrial metaverse, which aimed for real-time interaction with physical counterparts, highlighted the coupled importance of AoI, latency, and reliability, often that leveraged 6G Ultra-Reliable and Low-Latency Communication (URLLC) with optimized short-packet structures and co-design of sensing, communication, control, and computing \cite{Cao2023TWC}.

\subsubsection{Information Quality Optimization for Immersive Experiences}
The quality of immersive experiences in the Metaverse and Extended Reality (XR) applications extends beyond timeliness and encompasses the correctness, relevance, and utility of presented information. For XR applications, the AoII was a vital metric, as it captured both the error and AoI-based performance of semantic communication systems, which ensured that the information is not only fresh but also semantically accurate \cite{Chen2023JSTSP}. Semantic communication was also key in UAV-assisted Metaverse scenarios for building and updating DTs; by encoding data as semantic information (e.g., scene graphs), communication costs were significantly reduced \cite{Liew2024TVT}. Incentive mechanisms, such as combinatorial auctions, facilitated the trading of such semantic data based on its relatedness, freshness (AoI), and richness \cite{Liew2024TVT}. Quality of Information (QoI), which can be a composite of throughput and AoI, was crucial for mobile crowdsensing campaigns that support Metaverse applications. Multi-Agent Deep Reinforcement Learning (MADRL) frameworks, potentially that incorporated user mobility prediction and relational graph learning, optimized UAV trajectories and communication partner selection to maximize overall QoI \cite{Ye2023JSAC}. Also, in the context of semantic information marketing, where Virtual Service Providers (VSPs) procured sensing data from IoT devices to create digital copies of the physical world, learning-based contract theory frameworks were proposed to design incentive mechanisms that addressed information asymmetry and maximized VSP profit while ensuring truthful participation from sensing devices \cite{Lotfi2024JSAC}.

\subsubsection{Digital Asset Management and Security in Virtual Economies}
Persistent virtual environments feature sophisticated digital economies built around virtual assets, user-generated content, and services. The security and integrity of these assets and transactions are paramount. Frameworks like MetaFusion aimed to provide secure and scalable Metaverse environments by integrating IoT technologies with hyperdimensional computing for data encoding and advanced cryptography (e.g., homomorphic computation, secure multi-party computation) to ensure low-latency data synchronization with high security resilience \cite{Zhang2025TCE}. When Metaverse Terminal Devices (MTDs) outsourced complex tasks like graphic rendering to MEC servers, both privacy and accountability were critical. Blockchain-based frameworks (e.g., Meta-BMEOC) facilitated privacy-preserving and accountable outsourced computing that used smart contracts, threshold secret sharing,  trusted execution environments, and incentive mechanisms to deter malicious behavior \cite{Li2024TGCN}. However, as Metaverse applications proliferated, they encountered security and privacy threats, which included personal data breaches \cite{Kang2025MCOM}. A thorough understanding of these vulnerabilities, informed by existing standards and application characteristics, was necessary to propose and implement effective security and privacy requirements and countermeasures for a trustworthy Metaverse \cite{Kang2025MCOM}.

\section{Challenges and Future Directions}\label{future}
\subsection{Integration and Fusion of Multiple Information Metrics}
The integration of multiple information metrics represents a fundamental paradigm shift towards holistic system optimization in next-generation networked systems. While individual metrics such as Age of Information (AoI), Utility of Information (UoI), and Security of Information (Sec-oI) provide valuable insights into specific aspects of information quality, the complex interdependencies and trade-offs among these dimensions present significant challenges \cite{Mazloomi2022AHN,Fei2017COMST}. Current approaches suffer from computational complexity when attempting to jointly optimize multiple conflicting objectives across diverse network scenarios and application requirements \cite{Zeng2023CVPR,Iyer2022TIT}. Furthermore, the lack of unified theoretical frameworks that can capture the fundamental relationships between different metric dimensions hampers the development of systematic integration methodologies \cite{Thi2017arxiv}. Advanced Artificial Intelligence (AI)-driven approaches, particularly reinforcement learning and neural optimization techniques, offer promising solutions for addressing these challenges through adaptive metric fusion, intelligent orchestration, and context-aware decision making.
\begin{itemize}
\item \textbf{Theoretical Frameworks for Metric Composition:} Mathematical foundations that capture the fundamental relationships between different information metric dimensions require novel approaches to correlation analysis, dependency modeling, and unified bounds derivation. This includes establishing information-theoretic limits for joint metric optimization, deriving composition rules that preserve individual metric properties while enabling meaningful aggregation, and creating mathematical models that can express complex trade-offs between temporal freshness, semantic utility, security robustness, and network efficiency \cite{Natarajan2005POTICOML}.
\item \textbf{AI-Enhanced Dynamic Metric Weighting:} Machine learning approaches that adaptively balance multiple metrics based on application context, network conditions, and user requirements present opportunities for context-sensitive optimization strategies. This involves developing neural network architectures that can learn optimal metric weighting schemes from historical performance data, implementing reinforcement learning agents that continuously adjust metric priorities based on environmental feedback, and creating adaptive algorithms that can handle dynamic trade-offs between competing objectives \cite{Ji2014VTCSpring}.
\end{itemize}
\subsection{Information-driven Resource Allocation and Optimization}
Future network architectures are expected to transition from traditional bandwidth-centric allocation towards intelligent, information-aware resource management paradigms. Unlike conventional approaches that prioritize throughput maximization and latency minimization, information-driven systems must consider the actual utility, relevance, and quality of transmitted data when making allocation decisions \cite{Imran2024JIOT,Wang2025JIOT}. This fundamental shift requires sophisticated understanding of information value dynamics and their relationship to network resource consumption \cite{Wu2024TCOMM,Wang2025TVT}. The heterogeneous nature of modern applications, which range from real-time control systems to content delivery services, creates complex optimization landscapes where traditional metrics fail to capture actual system performance \cite{Liao2024ICC}. Advanced Artificial Intelligence (AI) techniques and game-theoretic frameworks offer promising solutions for creating resource allocation mechanisms that truly optimize for information value rather than merely data quantity.
\begin{itemize}
\item \textbf{Game-Theoretic Information Marketplace Design:} Economic models that treat information as a tradeable commodity with dynamic pricing based on quality metrics, utility, and scarcity require development. This involves auction mechanisms where network resources are allocated based on information value propositions, blockchain-based smart contracts for automated information-resource exchanges, and incentive structures that encourage high-quality information sharing while penalizing low-utility transmissions \cite{Xie2024TNSE}.
\item \textbf{Hierarchical Information-Aware Network Slicing:} Multi-dimensional network slicing that considers information characteristics alongside traditional Quality of Service (QoS) requirements enables dynamic slice reconfiguration based on real-time information quality assessment. This encompasses AI-driven slice orchestrators that can predict information utility trends, adaptive isolation mechanisms that maintain slice performance while optimizing global information value, and cross-slice information sharing protocols that maximize network-wide utility \cite{Sun2025OJCOMS}.
\end{itemize}
\subsection{Semantic and Goal-oriented Communications}
Semantic and goal-oriented communications face three fundamental challenges that hinder the transition beyond traditional Shannon-based paradigms: the absence of universal semantic representation standards that can operate across heterogeneous platforms and applications \cite{Shao2024TMC,KurisummoottilThomas2023JSAIT}, the computational complexity of achieving real-time semantic interpretation in resource-constrained environments \cite{DiLorenzo2023IOTM,FernandezVillamor2014TSMCC}, and the lack of standardized metrics for evaluating semantic fidelity and goal achievement in communication systems \cite{Chaccour2025COMST}. Foundation models and large-scale AI architectures offer transformative opportunities for developing communication systems that understand content meaning and user intent.
\begin{itemize}
\item \textbf{Cognitive Communication Architectures:} Systems that possess human-like understanding of content semantics through integration of multimodal foundation models \cite{Li2025MCOM} and cognitive reasoning frameworks require development. This involves neural-symbolic hybrid systems that combine deep learning with logical reasoning, self-evolving protocols that improve through continuous interaction, and meta-learning approaches for rapid domain adaptation.
\item \textbf{Proactive Intent-Driven Networking:} Infrastructures that anticipate user needs before explicit requests through predictive analytics \cite{Leivadeas2023COMST} and behavioral modeling need establishment. This encompasses temporal intention modeling systems, collaborative intelligence networks for global resource optimization, and anticipatory content positioning based on predicted semantic relevance.
\end{itemize}
\subsection{Information Metrics in Federated Learning and Edge Intelligence}
Federated learning and edge intelligence environments present unique complexities where traditional centralized approaches to information quality assessment become inadequate due to distributed data ownership and varying local processing capabilities \cite{Tariq2025TCE,Zhou2024AHN}. The fundamental tension between enabling collaborative information sharing for global optimization and preserving individual privacy creates intricate trade-offs that require novel metric design frameworks \cite{Wang2024JIOT,Gao2021CVPR}. Additionally, the dynamic nature of edge environments, characterized by intermittent connectivity, heterogeneous computational resources, and non-uniform data distributions, demands adaptive strategies that can maintain consistent information quality standards across diverse operational contexts \cite{Chen2019TOIT}. Advanced privacy-preserving techniques and distributed intelligence frameworks provide promising pathways for reconciling these competing requirements.
\begin{itemize}
\item \textbf{Privacy-Preserving Information Quality Assessment:} Cryptographic and statistical techniques that enable distributed evaluation of information metrics without revealing sensitive data require development. This involves secure multi-party computation protocols \cite{Zhou2024ACCESS} for joint metric calculation, differential privacy mechanisms that add controlled noise while preserving metric utility, and homomorphic encryption schemes that enable computation over encrypted information quality indicators.
\item \textbf{Hierarchical Edge-Cloud Information Orchestration:} Multi-tier intelligence frameworks that coordinate information processing across edge devices, regional servers, and centralized cloud infrastructure need creation. This encompasses adaptive workload distribution algorithms \cite{Gort2025TMLCN} that consider both computational constraints and information quality requirements, intelligent caching strategies for high-utility information at edge locations, and seamless handoff mechanisms that maintain quality continuity during device mobility.
\end{itemize}
\subsection{Green and Sustainable Information-centric Networks}
The exponential growth in data traffic and computational demands has created an urgent need to address the environmental impact of information-centric systems while maintaining service quality standards. Current energy optimization approaches primarily focus on hardware efficiency and load balancing, often neglecting the fundamental relationship between information value and energy expenditure \cite{Zhong2024MNET}. The challenge intensifies when considering the carbon footprint of distributed computing infrastructures, where optimizing for information quality may conflict with sustainability objectives \cite{Cao2024TSUSC,Bashroush2018TSUSC}. Furthermore, the lack of comprehensive frameworks for quantifying the environmental cost of information operations hampers the development of genuinely sustainable network architectures \cite{Allu2024TGCN}. Renewable energy integration and carbon-aware computing paradigms offer transformative opportunities for creating environmentally responsible information systems.
\begin{itemize}
\item \textbf{Carbon-Conscious Information Processing Frameworks:} Algorithms that explicitly consider the carbon footprint of information operations enable real-time optimization between service quality and environmental impact and require development. This involves carbon accounting models \cite{Nguyen2015TSC} for distributed computing tasks, renewable energy-aware scheduling algorithms that prioritize clean energy sources, and adaptive service degradation strategies that maintain essential functionality while minimizing emissions during peak demand periods.
\item \textbf{Sustainable Information Lifecycle Management:} Comprehensive frameworks for managing information from creation to disposal with explicit environmental considerations need to be established. This encompasses intelligent data retention policies \cite{Thalor2024ICBDS} that balance storage costs with information utility, energy-efficient compression strategies that optimize long-term storage sustainability, and circular information economy principles that promote the reuse and recycling of processed information.
\end{itemize}
\subsection{Security, Privacy, and Ethical Considerations}

The proliferation of AI-driven information systems has intensified concerns about adversarial manipulation, where sophisticated attacks target information quality metrics to compromise system integrity and deceive automated decision-making processes \cite{Radanliev2023CSUR}. Information quality assessments often require detailed analysis of user behavior patterns and system operations, creating inherent tensions between comprehensive quality evaluation and privacy preservation \cite{Qian2017TC,Zhao2021TMC}. The automated nature of these systems also raises fundamental questions about algorithmic fairness, as AI-driven quality determinations may inadvertently perpetuate or amplify existing social biases in information access and prioritization \cite{Ramachandranpillai2025TNNLS}. Emerging cryptographic techniques and ethical AI frameworks offer pathways for developing trustworthy information systems that strike a balance between security, privacy, and fairness requirements.

\begin{itemize}

\item \textbf{Zero-Trust Information Quality Architectures:} Implementing security frameworks that assume no inherent trust in information sources or processing components, requiring continuous verification of quality claims. This involves developing cryptographic proof systems \cite{Alevizos2022SP} for information quality assertions, creating distributed consensus mechanisms for quality verification, and implementing behavioral analysis systems that detect anomalous information quality patterns indicative of attacks or manipulation.

\item \textbf{Fairness-Aware Information Metric Design:} Establishing principles and methodologies for ensuring that information quality assessments do not introduce or amplify social biases and inequality. This encompasses developing bias detection algorithms \cite{Mehrabi2021CSUR} for information quality metrics, creating inclusive evaluation frameworks that consider diverse user populations, and implementing algorithmic auditing tools that identify and mitigate discriminatory outcomes in information access decisions.

\end{itemize}

\section{Conclusion}\label{conclus}

In this survey, we have given prominence to the meticulous study of multi-dimensional information metrics for future-generation networked systems, which lays the groundwork for the visionary ``X of Information'' continuum beyond traditional throughput-based paradigms. We have developed a four-dimensional taxonomic framework consisting of temporal, quality/utility, reliability/robustness, and network/communication-centered metrics, and revealed the progressive interdependencies and subtle trade-offs among these dimensions. We have also examined artificial intelligence technologies such as deep reinforcement learning, multi-agent systems, and neural optimization paradigms, which support adaptive and context-aware optimization across competing information quality objectives. The practical effectiveness of these multi-dimensional metrics is demonstrated through detailed case studies covering six critical application domains including autonomous vehicles and intelligent infrastructures, wireless sensor networks and industrial IoT, real-time healthcare and digital twins, UAV-assisted communication networks, agent-to-agent communication in LLM ecosystems, and metaverse infrastructure, which together illustrate their transformative value in meeting diverse operational demands. We have highlighted implementation challenges such as integration and fusion of metrics, information-driven resource management, semantic and goal-oriented communications, federated learning optimization, and sustainable network design. These explorations collectively reflect the paradigm shift toward intelligent, value-sensitive communication systems that emphasize information utility over raw data volume. The survey concludes by identifying key research directions in unified theoretical frameworks, AI-driven dynamic optimization, and cross-layer orchestration mechanisms, encouraging continued investigation in this field that will shape the future of intelligent networked systems.


\end{document}